\documentclass[%
 reprint,
 amsmath,amssymb,
prc,
]{revtex4-1}

\usepackage{graphicx}% Include figure files
\usepackage{multirow}
\usepackage{dcolumn}% Align table columns on decimal point
\usepackage{bm}% bold math
\usepackage{caption}
\usepackage{subcaption}
\usepackage{makecell}
\usepackage{hyperref}
\hypersetup{
   colorlinks=true,
    linkcolor=blue,
    filecolor=magenta,      
    urlcolor=blue,
    pdftitle={Sharelatex Example},
    bookmarks=true,
    citecolor=blue,
    pdfpagemode=FullScreen,
}

\begin{document}

\preprint{APS/123-QED}

\title{Interaction and Identification of the Meson-Baryon molecules}% Force line breaks with \\

\author{D. P. Rathaud}
 \altaffiliation[Also at ]{Department of Applied Physics, Sardar Vallabhbhai National Institute of Technology, Surat, Gujarat-395007, India.}%Lines break automatically or can be forced with \\
\author{Ajay kumar Rai}%
 \email{dharmeshphy@gmail.com}
\affiliation{%
 Department of Applied Physics, Sardar Vallabhbhai National Institute of Technology, Surat, Gujarat-395007, India.
}%


%	
%\affiliation{
%\affiliation{%

\date{\today}% It is always \today, today,

\begin{abstract}
The challenges with the molecular model of the multiquark systems are the identification of the hadronic molecules and the interaction between two color neutral hadrons. We study the di-hadronic molecular systems with proposed interaction potential as s-wave one boson exchange potential along with Screen Yukawa-like potential, and arrived with the proposal that within hadronic molecule the two color neutral hadrons experience the dipole-like interaction. The present study is the continuation  of our previous study \cite{arxiv-Rathaud-penta}. With the proposed interaction potential, the mass spectra of  $\Sigma_{s}K^{*}$, $\Sigma_{c}K^{*}$, $\Sigma_{b}K^{*}$, $\Sigma_{s}D^{*}$, $\Sigma_{c}D^{*}$, $\Sigma_{b}D^{*}$, $\Sigma_{s}B^{*}$, $\Sigma_{c}B^{*}$, $\Sigma_{b}B^{*}$, $\Xi_{s}K^{*}$, $\Xi_{c}K^{*}$, $\Xi_{b}K^{*}$, $\Xi_{s}D^{*}$, $\Xi_{c}D^{*}$, $\Xi_{b}D^{*}$, $\Xi_{s}B^{*}$, $\Xi_{c}B^{*}$, $\Xi_{b}B^{*}$  meson-baryon molecules are predicted. The Weinberg compositeness theorem which provides clue for the compositeness of the state is used for determination of the scattering length and effective range. The present study predict  $P_{c}(4450)$ pentaquark  sate as $\Sigma_{c}D^{*}$ molecule with $I(J^{P})=\frac{1}{2}(\frac{3}{2}^{-})$. The formalism also predicts some very interesting open as well as hidden flavour near threshold  molecular pentaquark states.
\end{abstract}

\pacs{Valid PACS appear here}% PACS, the Physics and Astronomy
\maketitle

%\tableofcontents

\section{Introduction}
\label{Introduction}
The multiquark states comes again in the focus of  interest, with the discovery of the two new observations $P_{c}(4380)^{+}$ and $P_{c}(4450)^{+}$, reported by LHCb Collaboration in 2015 \cite{Aaij-prl2015-115-pentaquark}. Both of these states could be potential candidate of the pentaquark state. The subject of the non conventional hadrons (exotic hadrons) has been speculated since the beginning of the quark model \cite{Gell-Mann-physLett1964-8,Jaffe-prd1977-15}. The idea of non conventional baryon composed of four quark and an anti-quark was introduced in refs. \cite{Strottman-prd1979-20,Lipkin-plb1987-195}, while the name pentaquark was devised by Lipkin \cite{Lipkin-plb1987-195}. 

The first claim of the pentaquark was in 2003 by LEPS Collaboration, as a $\Theta_{s}^{+}$ state, with suggesting the minimal quark content $uudd\overline{s}$, having the strangeness S=+1 \cite{Nakano-prl2003-91}. After this claim, two other experimental groups DIANA  \cite{Barmin-phyAtnucl2003-66} and CLAS  \cite{Stepanyan-prl2003-91} were found positive signatures for $\Theta_{s}^{+}$. With following these positive motivation, charm and bottom analogue of  $\Theta_{s}^{+}$ were predicted as $\Theta_{c}$ and $\Theta_{b}$. Meanwhile in 2004, H1 collaboration at HERA  reported the observation of the $\Theta_{c}$ \cite{Aktas-plb2004-588}. However, some experiments reported negative results regarding the status of $\Theta_{s}^{+}$, $\Theta_{c}$ and $\Theta_{b}$ \cite{Abt-prl2004-93,Knöpfle-jpg2004-30,Bai-prd2004-70,Aubert-prd2007-76,Link-plb2005-622}. Even, some recent experimental efforts, such as E19 Collaboration at J-PARC for the search of $\Theta^{+}$ \cite{Moritsu-prc2014-90} and ALICE collaboration for the $\phi(1869)$ pentaquark \cite{Abelev-epjc2015-75}, reported negative results. Within this situation, LHCb in 2015 has boosted the interest for the search of the multiquark states and their studies by reported promising hidden charm pentaquark states, named as $P_{c}(4380)^{+}$ with mass $4380\pm8\pm29$ MeV and width  $205\pm18\pm86$ MeV, respectively along with  another state $P_{c}(4450)^{+}$ with mass $4449.8\pm1.7\pm2.5$ MeV and a narrow width $39\pm5\pm19$ MeV. The spin-parity of these two states are $\frac{3}{2}$ are $\frac{5}{2}$ with opposite party have been proposed \cite{Aaij-prl2015-115-pentaquark}. The definite  spin and parity for both of these states has not yet been defined. 

These controversial history regarding the search and study of multiquark states, for instant pentaquark states, made the subject more fascinating from both theoretical and experimental point of view. Since various theoretical efforts have been made to understand the properties of these states, such as compact pentaquarks model \cite{Li-jhep2015-2015,Lebed-plb2015-749,Maiani-plb2015-749,Wang-epjc2016-76}, meson-baryon molecules \cite{Rui-Chen-npa2016-954-penta,Rui-Chen-cpc2017-41-penta,Azizi-prd2017-96,Azizi-prd2017-95,Shen-arxiv2017}, topological soliton model \cite{Scoccola-prd2015-92}, kinematic rescattering effects \cite{Guo-prd2015-92,Guo-epjc2016-52,Meißner-plb2015-751,Xiao-Hai-plb2016-757}, cups like effect \cite{}, hadroquarkonia model \cite{Eides-epjc2018-78} etc. One can find some brief review articles on multiquark states including pentaquarks in Refs. \cite{Chen-physreport2016-639,Swanson-PhysRep2006-429,Esposito-phys.Rep2017-668,Ali-Prog.Part.Phys2017-97}, for the review on hadronic molecules, one should see ref. \cite{Guo-RevModPhys2018-90} and references therein. These theoretical efforts suggested that the study of the multiquark states required more attention to reveal the information of their  substructure and nature.

The present article focus on the molecular picture of the pentaquark (meson-baryon bound states) states. The work  of this article is the continuation of our previous study presented in ref. \cite{arxiv-Rathaud-penta}. In \cite{arxiv-Rathaud-penta}, the results of the dimesonic systems have been presented. 

The aim of our study is to attempt two challenges of the molecular model:(i) interaction between two color neutral hadrons (ii) identification of the hadronic molecules from confined states. The interaction between tow color neutral states is largely unknown. The various interaction potentials have been used to explain such hadronic molecular structure \cite{Machleidt-PhysRep1987-149,Lee-prd2011-84,Ding-prd2009,Rai-epjc2015,Rathaud-epjp2017-132,Rathaud-IJP2016-90}. The hdronic molecules should close to the s-wave. Therefore, we proposed the interaction potential as s-wave One Boson Exchange Potential where the range of the force is proportional to the inverse mass of the exchange mesons. We notice that the s-wave OBE potential could not gain sufficient attractive strength for bound state, thus, we have added a screen Yukawa-like potential for additional attractive strength. This additional potential is added with a proposal        that the two color neutral states experienced dipole-like interaction. The potential parameters of the s-wave OBE and screen Yukawa-like potential are fitted to get experimental binding energy of the deuteron. The screen parameter 'c' of screen Yukawa-like potential is the only free parameter of the model and only fitted for deuteron, while it is taken as a fixed parameter for all di-hadronic calculation. With this model, the results and analysis of the meson-baryon molecular systems are presented in this article.  The characteristic contributory nature of the individual s-wave meson exchange and effective s-wave OBE potential are presented in result and discussion  section. The second prerequisite of hadronic molecule model, identification of the hadronic molecule, we have adopted the Weinberg's compositeness theorem and results of scattering length ($a_{s}$) and effective range ($r_{e}$) are extracted for attempted meson-baryon molecular states. 

The article is organized as follows: after the brief introduction, theoretical framework and the model is introduce in the section-\ref{Theoretical framework}, after that the results of the mass spectra of meson-baryon molecular systems are presented in section-\ref{Mass spectra of meson-baryon systems}. The compositeness theorem and the results of the scattering length and effective range for attempted systems are presented in section-\ref{Results of as and re from compositeness theorem}, finally the summary and conclusion is presented in the last section-\ref{Discussion and summary}.  

%\begin{eqnarray}
%\label{as and re equation}
%&&a_{s}=\left[2(1-Z)/(2-Z)\right]R+{\cal{O}}(1/\beta) \nonumber \\
%&&r_{e}=\left[-Z/(1-Z)\right]R+{\cal{O}}(1/\beta)
%\end{eqnarray}
%where  $R\equiv \sqrt{2\mu \epsilon}$, $\epsilon$ is the binding energy and $\mu$ is the reduced mass of the composite system. The ${\cal{O}}(1/\beta)$ is the range of the force and could be calculated if one know the information of the interaction. In order to determine the state of the particle as in a bare elementary or in a composite state, he argued that the renormalization constant Z takes the value  $0 \leq Z \leq 1$. If Z=0 then the particle is in a pure composite state, while for Z=1 it becomes a purely elementary. This argument have previously discussed by other authors \cite{Vaughn-PhesRev1961,Acharya1962} and followed by Weinberg\cite{Weinberg-pr1965-137}. For the case Z=0 (the deuteron as a composite particle) the Eq(\ref{as and re equation}) becomes $a_{s}=R$ and $r_{e}={\cal{O}}(1/\beta)$ which is in agreement with the experimental vales : $a_{s}=+5.41$ fm, $r_{e}=+1.75$ fm. 

\section{Theoretical framework}
\label{Theoretical framework}
The Hamiltonian of di-hadronic molecule is given by
\begin{equation}
\label{Hamiltonian}
H=\sqrt{P^2+m_{d}^{2}}+\sqrt{P^2+m_{b}^2}+V_{hh}(r_{db})
\end{equation}
$m_{d}$ and $m_{b}$ are the masses of constituents and P is the relative momentum of two hadrons while the $V_{hh}(r_{db})$ is the inter-hadronic interaction potential. In the variational scheme, we have used the hydrogenic trial wave function to determine the expectation value of the Hamiltonian. 

%We propose the dipole-dipole like interaction between two color neutral hadrons, could be either permanent dipole or induced dipole in which the latter one is weakest. The uneven color charge field distribution between two hadrons creates a dipole effect while in other case the color charge field of two hadron get influence to each other and leads to temporary dipole. The screening of color charge field tends to vacuum polarization and creates quark anti quark pairs. 

%Hence, let us describe the interaction potential in terms of the One Boson Exchange (OBE) potential plus phenomenological attractive screen Yukawa-like potential. 

The di-hadronic interaction potential is given by
\begin{equation}
\label{Interaction di-hadronic potential}
V_{hh}(r_{db})=V_{OBE}(r_{db})+V_{Y}(r_{db})
\end{equation}
the term $V_{Y}(r_{db})$ is screen Yukawa-like potential and $V_{OBE}$ is the s-wave One Boson Exchange (OBE) potential. 

The additional phenomenological screen Yukawa-like potential is used along with s-wave OBE potential to get sufficient attractive strength for bound state.  For use of such potential, we have taken an approximation that two color neutral hadrons experienced the dipole-dipole like interaction, where it could be either permanent dipole or induced dipole in which the latter one is weakest. The screen Yukawa-like potential  expressed as  
\begin{eqnarray}
V_{Y}&=& -\frac{k_{mol}}{r_{db}} e^{\frac{{-c^{2}r_{db}^{2}}}{2}}
\end{eqnarray}

here, c is a screen fitting parameter of the potential while $K_{mol}$ is the residual running coupling constant, namely

\begin{eqnarray}
\label{Kmol}
k_{mol}(M^{2}) = \frac{4\pi}{(11-\frac{2}{3}n_{f})ln\frac{M^{2}+ {M_{B}}^{2}}{\Lambda_{Q}^{2}}}
\end{eqnarray}

where M=2$m_{d}$ $m_{b}$/ ($m_{d}$+$m_{b}$), $m_{d}$ and $m_{b}$ are constituent masses, $M_{B}$=1 GeV, $\Lambda_{Q}$ is QCD scale parameter, respectively. The term $n_{f}$ is number of flavour \cite{Ebert-prd2009,Badalian-prd2004}. This effective coupling constant has introduced to incorporate the asymptotic behavior at short distance as well as to reduce the free parameter of the model.

The light mesons under consideration for the OBE Potential are as follows \cite{Machleidt-PhysRep1987-149}: 

Pseudoscalar meson $(ps)=\pi$, $\eta$ 

Scalar meson $(s)=\sigma,\delta$ (also known as $a_{0}$)

Vector meson $(v)=\omega,\rho$. 

The OBE potential is sum of the all one meson exchange, namely 
\begin{eqnarray}
V_{OBE}=V_{ps}+V_{s}+V_{v}
\end{eqnarray}
The OBE potential with finite size effect due to extended structure of the hadrons can be expressed as \cite{Machleidt-PhysRep1987-149}
\begin{eqnarray}
V_{\alpha}(r_{db})=V_{\alpha}(m_{\alpha},r_{db})- &&F_{\alpha 2}V_{\alpha}(\Lambda_{\alpha 1},r_{db})\nonumber \\ && +F_{\alpha 1}V_{\alpha}(\Lambda_{\alpha 2},r_{db})
\end{eqnarray}
where $\alpha$ = $\pi$, $\eta$, $\sigma$, $\delta$, $\omega$ and $\rho$ mesons, while  
\begin{eqnarray}
\Lambda_{\alpha 1}=\Lambda_{\alpha}+\epsilon \hspace{0.99cm} and \hspace{0.99cm} \Lambda_{\alpha 2}=\Lambda_{\alpha}-\epsilon \nonumber \\ 
F_{\alpha 1}=\frac{\Lambda_{\alpha 1}^{2}-m_{\alpha}^{2}}{\Lambda_{\alpha 2}^{2}-\Lambda_{\alpha 1}^{2}} \hspace{0.5cm} and \hspace{0.5cm} F_{\alpha 2}=\frac{\Lambda_{\alpha 2}^{2}-m_{\alpha}^{2}}{\Lambda_{\alpha 2}^{2}-\Lambda_{\alpha 1}^{2}}\\ \nonumber
\end{eqnarray}
the subscript $\alpha$ tends for mesons ($\pi$, $\eta$, $\sigma$, $\delta$, $\omega$ and $\rho$)  $\epsilon/\Lambda_{\alpha}\ll1$, thus $\epsilon$=10 MeV is an appropriate choice. \\
Hence, the individual meson exchange potential with finite size effect can be expressed as \cite{Machleidt-PhysRep1987-149}

\begin{widetext}
\begin{eqnarray}
V_{ps}(r_{db})_{F}&&=\frac{1}{12}\left[\frac{g^{2}_{\pi qq}}{4\pi}\left \{ \left(\frac{m_{\pi}}{m}\right)^{2} \frac{e^{-m_{\pi}r_{db}}}{r_{db}} -(F_{\pi 2})\left(\frac{\Lambda_{\pi 1}}{m}\right)^{2} \frac{e^{-\Lambda_{\pi 1}r_{db}}}{r_{db}} +  \left. (F_{\pi 1})\left(\frac{\Lambda_{\pi 2}}{m}\right)^{2} \frac{e^{-\Lambda_{\pi 2}r_{db}}}{r_{db}} \right \} \left(\tau_{d}\cdot\tau_{b}\right) \right. \right. \nonumber \\ &&  \left.  + \frac{g^{2}_{\eta qq}}{4\pi}\left \{ \left(\frac{m_{\eta}}{m}\right)^{2} \frac{e^{-m_{\eta}r_{db}}}{r_{db}} - (F_{\eta 2})\left(\frac{\Lambda_{\eta 1}}{m}\right)^{2} \frac{e^{-\Lambda_{\eta 1}r_{db}}}{r_{db}} +  (F_{\eta 1})\left(\frac{\Lambda_{\eta 2}}{m}\right)^{2} \frac{e^{-\Lambda_{\eta 2} r_{db}}}{r_{db}} \right \}\right]\left(\sigma_{d}\cdot\sigma_{b}\right)
\end{eqnarray}
\begin{eqnarray}
V_{s}(r_{db})_{F}&&=-\frac{g_{\sigma qq}^{2}}{4\pi} \left \{ m_{\sigma}\left[1-\frac{1}{4}\left(\frac{m_{\sigma}}{m}\right)^{2}\right]\frac{e^{-m_{\sigma}r_{db}}}{m_{\sigma}r_{db}}- F_{\sigma 2}\Lambda_{\sigma 1}\left[1-\frac{1}{4}\left(\frac{\Lambda_{\sigma 1}}{m}\right)^{2}\right]\frac{e^{-\Lambda_{\sigma 1} r_{db}}}{\Lambda_{\sigma 1} r_{db}}  +F_{\sigma 1}\Lambda_{\sigma 2} \right.  \nonumber \\ &&\left. \left[1-\frac{1}{4}\left(\frac{\Lambda_{\sigma 2}}{m}\right)^{2}\right] \frac{e^{-\Lambda_{\sigma 2} r_{db}}}{\Lambda_{\sigma 2} r_{db}} \right  \} + \frac{g_{\delta qq}^{2}}{4\pi} \left \{ m_{\delta}\left[1-\frac{1}{4}\left(\frac{m_{\delta}}{m}\right)^{2}\right]\frac{e^{-m_{\delta}r_{db}}}{m_{\delta}r_{db}}  - F_{\delta 2}\Lambda_{\delta 1}\left[1-\frac{1}{4}\left(\frac{\Lambda_{\delta 1}}{m}\right)^{2}\right]\right. \\ \nonumber  &&\left. \frac{e^{-\Lambda_{\delta 1} r_{db}}}{\Lambda_{\delta 1} r_{db}}  +F_{\delta 1}\Lambda_{\delta 2}\left[1-\frac{1}{4}\left(\frac{\Lambda_{\delta 2}}{m}\right)^{2}\right]\frac{e^{-\Lambda_{\delta 2} r_{db}}}{\Lambda_{\delta 2} r_{db}} \right \}\left(\tau_{d}\cdot\tau_{b}\right)
\end{eqnarray}
\begin{eqnarray}
V_{v}(r_{db})_{F}&&=\frac{g_{\omega qq}^{2}}{4\pi}\left \{ \left(\frac{e^{-m_{\omega}r_{db}}}{r_{db}}\right)- F_{\omega 2}\left(\frac{e^{-\Lambda_{\omega 1} r_{db}}}{r_{db}}\right)+F_{\omega 1}\left(\frac{e^{-\Lambda_{\omega 2} r_{db}}}{r_{db}}\right) \right \}+\frac{1}{6}\frac{g_{\rho qq}^{2}}{4\pi}\frac{1}{m^{2}}  \left \{ \left(\frac{e^{-m_{\rho}r_{db}}}{r_{db}}\right)-\right. \\ \nonumber  &&\left. F_{\rho 2}\left(\frac{e^{-\Lambda_{\rho 1} r_{db}}}{r_{db}}\right)+F_{\rho 1}\left(\frac{e^{-\Lambda_{\rho 2}r_{db}}}{r_{db}}\right) \right \}\left(\tau_{d}\cdot\tau_{b}\right)\left(\sigma_{d}\cdot\sigma_{b}\right)
\end{eqnarray}
The net s-wave OBE potential with finite size effect can be expressed as 
\begin{eqnarray}
V_{OBE}=V_{ps}(r_{db})_{F}+V_{s}(r_{db})_{F}+V_{v}(r_{db})_{F}
\end{eqnarray}
\begin{eqnarray}
\label{OBEpotential}
&&V_{OBE}(r_{db})_{F}=\frac{1}{12}\left[\frac{g^{2}_{\pi qq}}{4\pi}\left \{ \left(\frac{m_{\pi}}{m}\right)^{2} \frac{e^{-m_{\pi}r_{db}}}{r_{db}} -(F_{\pi 2})\left(\frac{\Lambda_{\pi 1}}{m}\right)^{2} \frac{e^{-\Lambda_{\pi 1}r_{db}}}{r_{db}} +  (F_{\pi 1})\left(\frac{\Lambda_{\pi 2}}{m}\right)^{2} \right. \right. \nonumber \\ && \left.  \left.\frac{e^{-\Lambda_{\pi 2}r_{db}}}{r_{db}} \right \} \left(\tau_{d}\cdot\tau_{b}\right) + \frac{g^{2}_{\eta qq}}{4\pi}\left \{ \left(\frac{m_{\eta}}{m}\right)^{2} \frac{e^{-m_{\eta}r_{db}}}{r_{db}} - (F_{\eta 2})\left(\frac{\Lambda_{\eta 1}}{m}\right)^{2} \frac{e^{-\Lambda_{\eta 1}r_{db}}}{r_{db}} +  (F_{\eta 1})\left(\frac{\Lambda_{\eta 2}}{m}\right)^{2} \right. \right. \nonumber \\  && \left. \left. \frac{e^{-\Lambda_{\eta 2} r_{db}}}{r_{db}} \right \}\right]\left(\sigma_{d}\cdot\sigma_{b}\right)  -\frac{g_{\sigma qq}'^{2}}{4\pi} \left \{ m_{\sigma}\left[1-\frac{1}{4}\left(\frac{m_{\sigma}}{m}\right)^{2}\right]\frac{e^{-m_{\sigma}r_{db}}}{m_{\sigma}r_{db}}- F_{\sigma 2}\Lambda_{\sigma 1}\left[1-\frac{1}{4}\left(\frac{\Lambda_{\sigma 1}}{m}\right)^{2}\right]\right.  \nonumber \\ &&\left.\frac{e^{-\Lambda_{\sigma 1} r_{db}}}{\Lambda_{\sigma 1} r_{db}}  +F_{\sigma 1}\Lambda_{\sigma 2}\left[1-\frac{1}{4}\left(\frac{\Lambda_{\sigma 2}}{m}\right)^{2}\right]\frac{e^{-\Lambda_{\sigma 2} r_{db}}}{\Lambda_{\sigma 2} r_{db}} \right  \} +
\frac{g_{\delta qq}^{2}}{4\pi} \left \{ m_{\delta}\left[1-\frac{1}{4}\left(\frac{m_{\delta}}{m}\right)^{2}\right]\frac{e^{-m_{\delta}r_{db}}}{m_{\delta}r_{db}} \right. \nonumber \\   &&\left. - F_{\delta 2}\Lambda_{\delta 1}\left[1-\frac{1}{4}\left(\frac{\Lambda_{\delta 1}}{m}\right)^{2}\right]\frac{e^{-\Lambda_{\delta 1} r_{db}}}{\Lambda_{\delta 1} r_{db}}  +F_{\delta 1}\Lambda_{\delta 2}\left[1-\frac{1}{4}\left(\frac{\Lambda_{\delta 2}}{m}\right)^{2}\right]\frac{e^{-\Lambda_{\delta 2} r_{db}}}{\Lambda_{\delta 2} r_{db}} \right \}\left(\tau_{d}\cdot\tau_{b}\right)+\nonumber \\ && \frac{g_{\omega qq}^{2}}{4\pi}\left \{ \left(\frac{e^{-m_{\omega}r_{db}}}{r_{db}}\right)- F_{\omega 2}\left(\frac{e^{-\Lambda_{\omega 1} r_{db}}}{r_{db}}\right)+F_{\omega 1}\left(\frac{e^{-\Lambda_{\omega 2} r_{db}}}{r_{db}}\right) \right \}+\frac{1}{6}\frac{g_{\rho qq}^{2}}{4\pi}\frac{1}{m^{2}}  \left \{ \left(\frac{e^{-m_{\rho}r_{db}}}{r_{db}}\right) \right. \nonumber \\  && \left. -F_{\rho 2}\left(\frac{e^{-\Lambda_{\rho 1} r_{db}}}{r_{db}}\right)+F_{\rho 1}\left(\frac{e^{-\Lambda_{\rho 2}r_{db}}}{r_{db}}\right) \right \}\left(\tau_{d}\cdot\tau_{b}\right)\left(\sigma_{d}\cdot\sigma_{b}\right)
\end{eqnarray}
\end{widetext}

The overall contribution form OBE is very less due to its delicate cancellation with each other. However, two points should be noted  on the overall contribution (attraction/repulsion) of the OBE potential: (i) its contribution is strongly related to the coupling constant of the each individual meson exchange and (ii) it depends on the spin-isospin channels.

%The estimates of the coupling constants of OBE potential are given in the most of the realistic potentials \cite{Machleidt-PhysRep1987-149,Machleidt-prc2001-63,Paris-prc2000-62,Naghdi-PhysParNucl2014-45} which are developed to reproduce NN-phase shift data and to explain the deuteron properties. 

%It is reasonable to take the quark-meson coupling in the present interaction scheme. The quark-meson coupling constant with respect to nucleon-meson coupling constant can be derive by Goldberger-Treiman relation \cite{Riska-npa2001} 
%\begin{equation}
%g_{\pi qq}=\frac{3m_{q}}{5m_{N}}g_{\pi NN}
%\end{equation}
%where $g_{\pi qq}$ and $g_{\pi NN}$ are pion-quark and pion-nucleon coupling constant,  respectively.
 
\begin{table}[]
\begin{center}
\caption{OBE potential parameters, this parameters are taken from \cite{Machleidt-PhysRep1987-149,Machleidt-prc2001-63}}
\label{OBEP parameters}
\scalebox{1}{
\begin{tabular}{ccccccc}
\hline
\hline
Mesons & $\pi$ & $\eta$ & $\sigma$  & $a_{0}(\delta)$ & $\omega$ & $\rho$ \\
\\
$\frac{g^{2}_{\alpha NN}}{4\pi}$ & 13.6 & 3 & 7.7823 $^{*}$ & 2.6713 & 20 & 0.85 \\
\\
$\Lambda_{\alpha}$ & 1.3 & 1.5 & 2.0 & 2.0 & 1.5 & 1.3\\
\\
Mass (in MeV) & 134.9 & 548.8 & 710 & 983 & 782.6 & 775.4 \\
\\
\hline
\hline
\end{tabular}
}
\end{center}

\begin{center}
{\tiny($^{*}$The  $\frac{g^{2}_{\sigma NN}}{4\pi}$ for the $\sigma$-exchange given in the table is used for total isospin $I_{T}$=1. Whereas for $I_{T}$=0, $\frac{g^{2}_{\sigma NN}}{4\pi}$=16.2061 have been used.)}
\end{center}
\end{table}

\begin{table}
\caption{The threshold mass, reduced mass and residual running coupling constant($k_{mol}$) of the di-hadronic systems. In the calculation of the $k_{mol}$ the $\Lambda_{Q}$=0.150 GeV is taken while $n_{f}$=2,3,4 is taken as per involvement of the light quarks, charm quark and bottom quark, respectively. Here, $\Sigma^{*}$ and $\Xi^{*}$ are $J^{P}$=$\frac{3}{2}^{+}$ constituents.}
\label{kmol and threshold mass}
\begin{center}
\scalebox{0.8}{
\begin{minipage}[]{0.6\linewidth}
\begin{tabular}{cccc}
\hline
System & Threshold & Reduce & $k_{mol}$  \\
& mass & mass &\\
& GeV & GeV &\\
\hline
\hline 
\\
$\Sigma_{s}$ $K^{*}$  & 2.088 & 0.511 &  0.2882  \\
$\Sigma_{s}^{*}$  $K^{*}$   & 2.278 & 0.543 &  0.2841   \\
$\Sigma_{c}$ $K^{*}$  & 3.349 & 0.656 &  0.2911   \\
$\Sigma_{c}^{*}$ $K^{*}$  & 3.414 & 0.660 &  0.2906   \\
$\Sigma_{b}$ $K^{*}$  & 6.708 & 0.776 &  0.3003   \\
$\Sigma_{b}^{*}$ $K^{*}$  & 6.728 & 0.776 &  0.3002  \\
\\
$\Sigma_{s}$ $D^{*}$  & 3.199 & 0.748 &  0.2809  \\ 
$\Sigma_{s}^{*}$ $D^{*}$  & 3.389 & 0.818 &  0.2739 \\
$\Sigma_{c}$ $D^{*}$  & 4.460 & 1.103 &  0.2509  \\
$\Sigma_{c}^{*}$ $D^{*}$  & 4.525 & 1.116 &  0.2500 \\
$\Sigma_{b}$ $D^{*}$  & 7.819 & 1.491 &  0.2477 \\
$\Sigma_{b}^{*}$  $D^{*}$  & 7.839 & 1.493 &  0.2476 \\
\\
$\Sigma_{s}$  $B^{*}$  & 6.517 & 0.974 &  0.2812   \\
$\Sigma_{s}^{*}$  $B^{*}$  & 6.708 & 1.097 &  0.2714  \\
$\Sigma_{c}$  $B^{*}$  & 7.778 & 1.679 &  0.2392   \\
$\Sigma_{c}^{*}$ $B^{*}$  & 7.844 & 1.709 &  0.2380    \\
$\Sigma_{b}$  $B^{*}$  & 11.138 & 2.779 &  0.2078   \\
$\Sigma_{b}^{*}$  $B^{*}$  & 11.157 & 2.783 &  0.2077    \\

\\
\hline
\hline
\end{tabular}
\end{minipage}

\begin{minipage}[]{0.7\linewidth}
\begin{tabular}{cccc}
\hline
System & Threshold & Reduce & $k_{mol}$  \\
& mass & mass &\\
& GeV & GeV &\\
\hline
\hline 
\\
$\Xi_{s}$ $K^{*}$  & 2.210 & 0.532 &  0.2855 \\
$\Xi_{s}^{*}$ $K^{*}$  & 2.427 & 0.565 &  0.2815 \\
$\Xi_{c}$ $K^{*}$  & 3.366 & 0.657 &  0.2909 \\
$\Xi_{c}^{*}$ $K^{*}$  & 3.541 & 0.669 &  0.2896 \\
$\Xi_{b}$ $K^{*}$  & 6.683 & 0.775 &  0.3003 \\
\\
$\Xi_{s}$ $D^{*}$  & 3.321 & 0.794 &  0.2762 \\
$\Xi_{s}^{*}$ $D^{*}$  & 3.538 & 0.868 &  0.2692 \\
$\Xi_{c}$ $D^{*}$  & 4.477 & 1.107 &  0.2506 \\
$\Xi_{c}^{*}$ $D^{*}$  & 4.652 & 1.141 &  0.2484 \\
$\Xi_{b}$ $D^{*}$  & 7.794 & 1.49 &  0.2478 \\
\\
$\Xi_{s}$ $B^{*}$  & 6.64 & 1.054 &  0.2747 \\
$\Xi_{s}^{*}$ $B^{*}$  & 6.857 & 1.189 &  0.2649 \\
$\Xi_{c}$ $B^{*}$  & 7.796 & 1.687 &  0.2389 \\
$\Xi_{c}^{*}$ $B^{*}$  & 7.971 & 1.767 &  0.2357 \\
$\Xi_{b}$ $B^{*}$  & 11.113 & 2.773 &  0.2079 \\
\\
\hline
\hline
&&&\\
&&&\\
&&&\\
\end{tabular}
 \end{minipage}
 
%\begin{minipage}[]{0.2\linewidth}
%\begin{tabular}{cccc}
%\end{tabular}
%\end{minipage}
}
\end{center}
\end{table}  
The parameters of the model are as follows: (i) hadron masses (ii) coupling constant of the exchange mesons (iii) regularization parameter $\Lambda_{\alpha}$ (iv) residual running coupling constant $k_{mol}$ (v) color screening parameter c. 

The masses of the hadrons  and exchange mesons are taken from the PDG \cite{Patrignani-PDG2016}. The coupling constant of the exchange mesons and regularization parameter $\Lambda_{\alpha}$ are obtained from refs.\cite{Machleidt-PhysRep1987-149,Machleidt-prc2001-63,Paris-prc2000-62,Naghdi-PhysParNucl2014-45}, also tabulated in Table-\ref{OBEP parameters}. The estimates of the coupling constants of OBE potential are given in the most of the realistic potentials \cite{Machleidt-PhysRep1987-149,Machleidt-prc2001-63,Paris-prc2000-62,Naghdi-PhysParNucl2014-45} which were developed to reproduce NN-phase shift data and to explain the deuteron properties. Hence, we have taken them same as estimated in Refs. \cite{Machleidt-PhysRep1987-149,Machleidt-prc2001-63} and approximated the meson-hadron coupling constant for other hadronic molecular cases as 
\begin{equation}
g_{\alpha hh} \simeq g_{\alpha NN}
\end{equation}
where $g_{\alpha hh}$ and $g_{\alpha NN}$ are the meson-hadron and meson-nucleon coupling constants,  respectively. Thus, the masses, exchange meson coupling constant and $\Lambda_{\alpha}$ are the fixed parameters  and obtained from Refs. \cite{Patrignani-PDG2016,Machleidt-PhysRep1987-149,Machleidt-prc2001-63}, also tabulated in Table-\ref{HadronMaases-PDG} $\&$ \ref{OBEP parameters}. Apart from these, the residual running coupling constant $k_{mol}$ is calculated by using Eq.(\ref{Kmol}), and calculated values for attempted di-hadronic systems are tabulated in Table-\ref{kmol and threshold mass}. The color screening parameter 'c' is the only free parameter of the model and we fitted it to get the empirical value of binding energy of the deuteron. For c=0.0686 GeV, we obtained the binding energy of the deuteron. Hence, we took it as a constant and have not changed for any further calculations of the di-hadronic molecules.

%We fit c=0.065 GeV and $\alpha_{s}$=0.29  strength of the OBE Potential where coupling constants of OBE are calculated as discuss in the previous section. Hence, we have fixed c=0.065 GeV for further calculation and taken $\alpha_{s}$=0.25, 0.24, 0.20  according to involvement of double charm, charm-bottom or double bottom quarks, respectively, in the composition of di-hadronic molecules.

\begin{table}[]
\begin{center}
\caption{Mass spectra of proton-neutron (deuteron) and neutron-neutron bound system.}
\label{deuteron-table from OBE}
\scalebox{0.83}{
\begin{tabular}{cccccccc}
\hline
\hline
\\
$[I_{1}(J_{1}^{P_{1}})]^{Q_{1}}$ - $[I_{2}(J_{2}^{P_{2}})]^{Q_{2}}$ & System & $I(J^{P})$ & $\mu$ & B.E. & Mass & $\sqrt{r^{2}}$ \\
& & & GeV & MeV & MeV & fm \\
\\
\multirow{1}{*}{$[\frac{1}{2}$($\frac{1}{2}^{+}$)$]^{+}$-$[\frac{1}{2}$($\frac{1}{2}^{+}$)$]^{0}$}
&\multirow{1}{*}{$p-n$}
& $0(1^{+})$ & 0.1090	& -2.2211 & 1.875 & 03.13\\
\\
\multirow{1}{*}{$[\frac{1}{2}$($\frac{1}{2}^{+}$)$]^{0}$-$[\frac{1}{2}$($\frac{1}{2}^{+}$)$]^{0}$}
&\multirow{1}{*}{$n-n$}
& $1(0^{+})$ & 0.1004	& +0.9528 & 1.880 & 03.40\\
\\
\hline
\hline
\end{tabular}
}
\end{center}
\end{table}
\begin{figure}[!h]
\caption{The characteristic nature of the s-wave one meson exchange potential (a) in case of proton-neutron (deuteron) system (b) in case of neutron-neutron system}
\label{p-n OBE and net plot}
\includegraphics[scale=0.255]{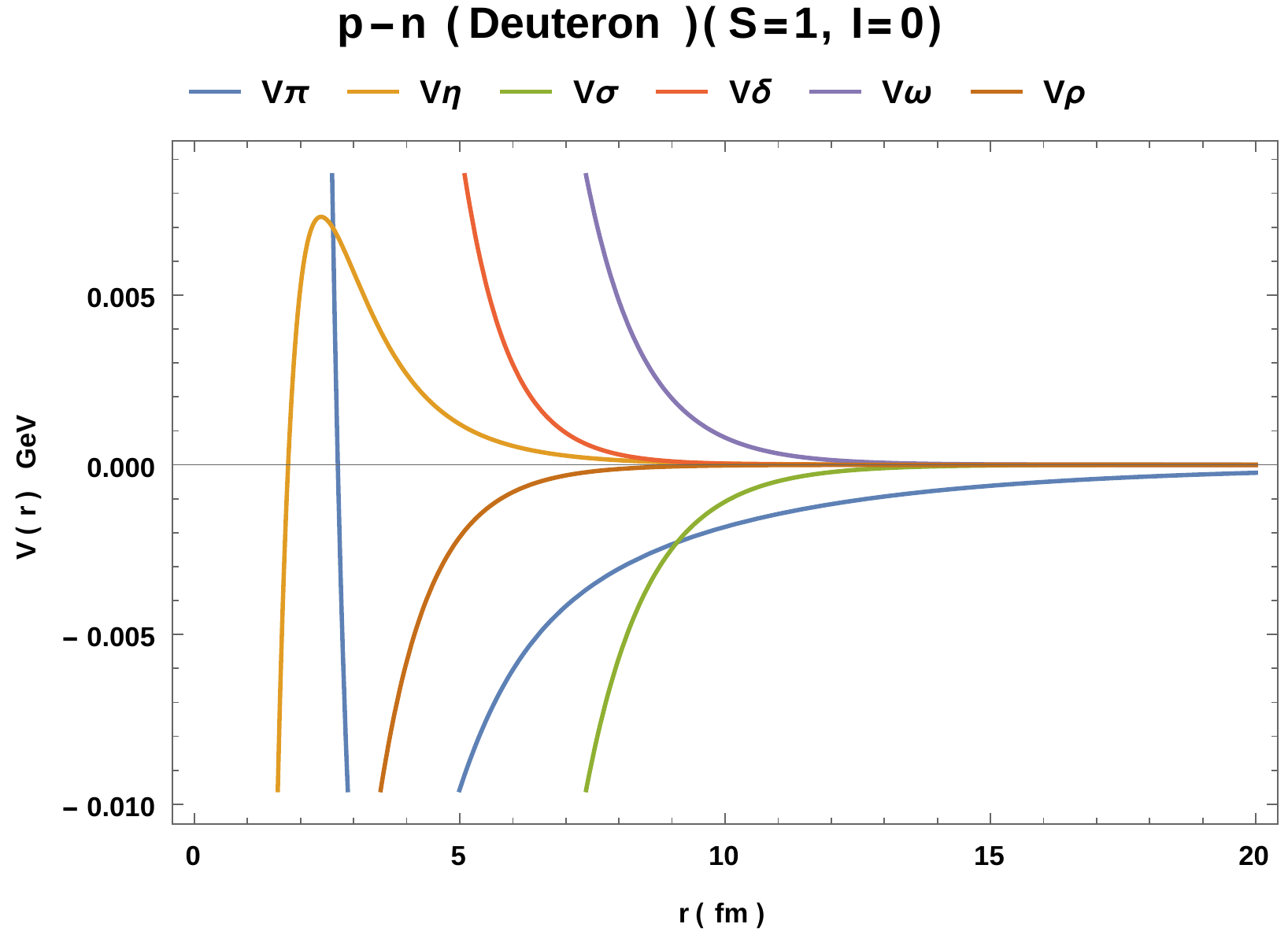}
\includegraphics[scale=0.255]{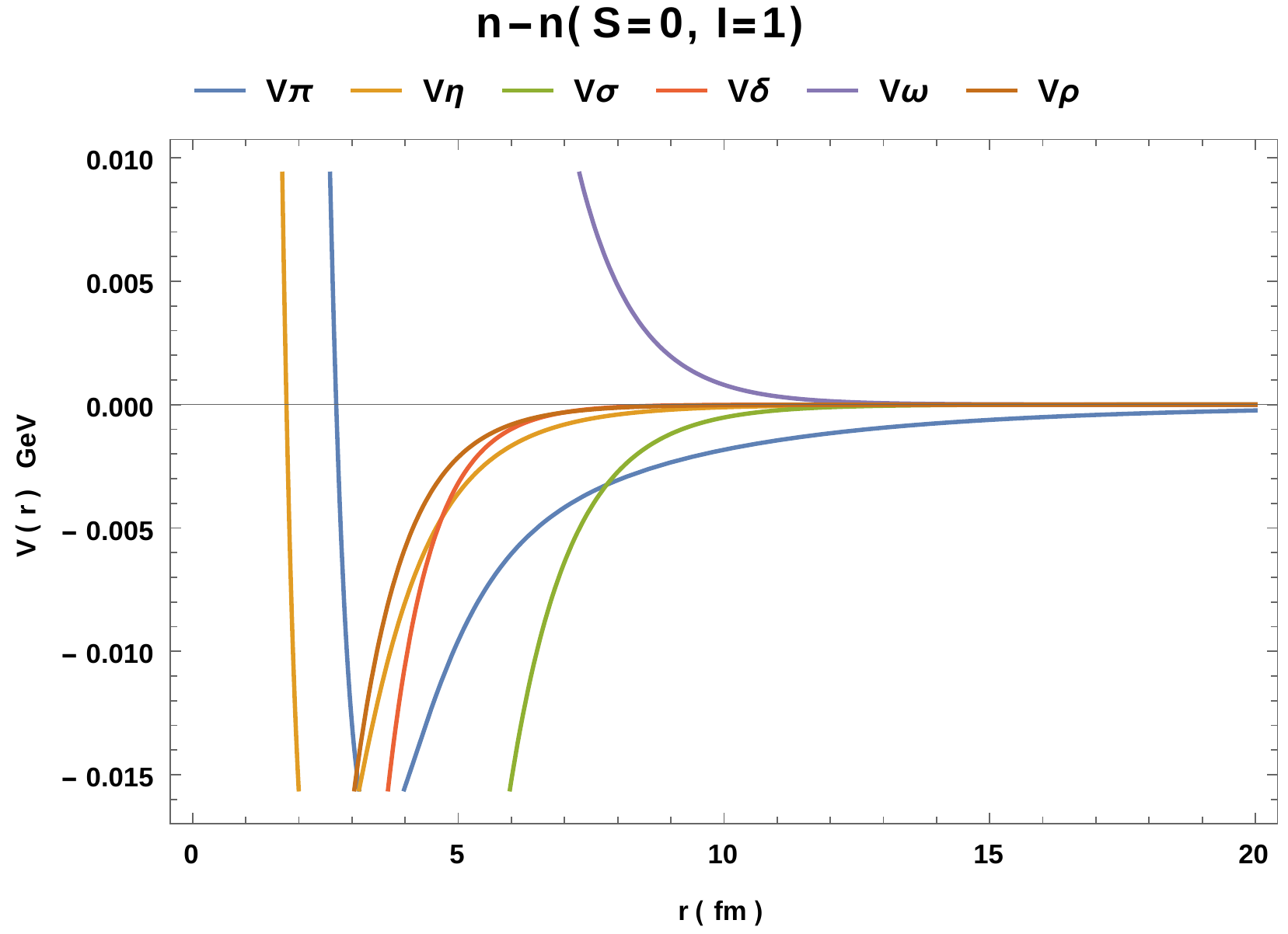}
\caption*{\hspace{1cm}(a)\hspace{4cm}(b)}
\caption{The characteristic nature of the net s-wave One Boson Exchange (OBE) potential (a) in case of p-n (deuteron) system (b) in case of neutron-neutron system}
\label{OBE net plot}
\includegraphics[scale=0.32]{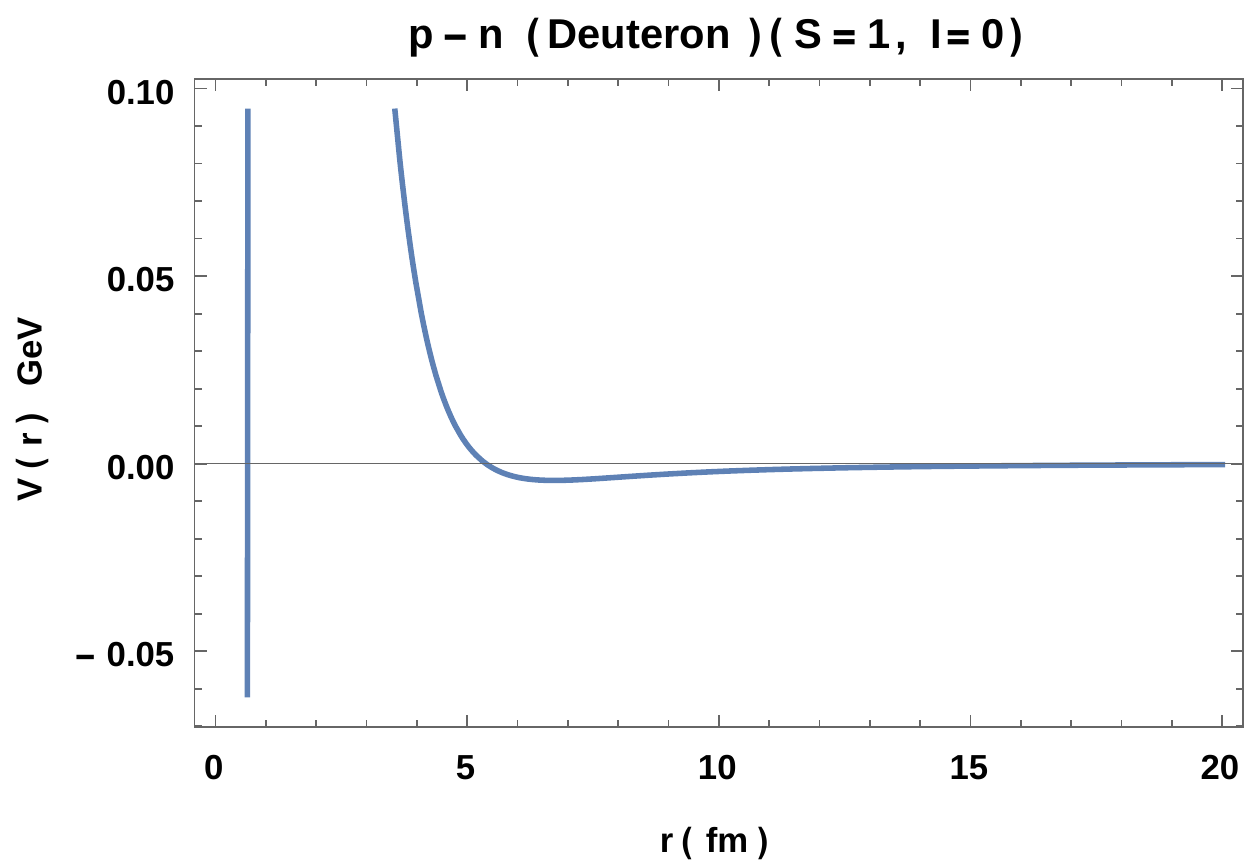}
\includegraphics[scale=0.32]{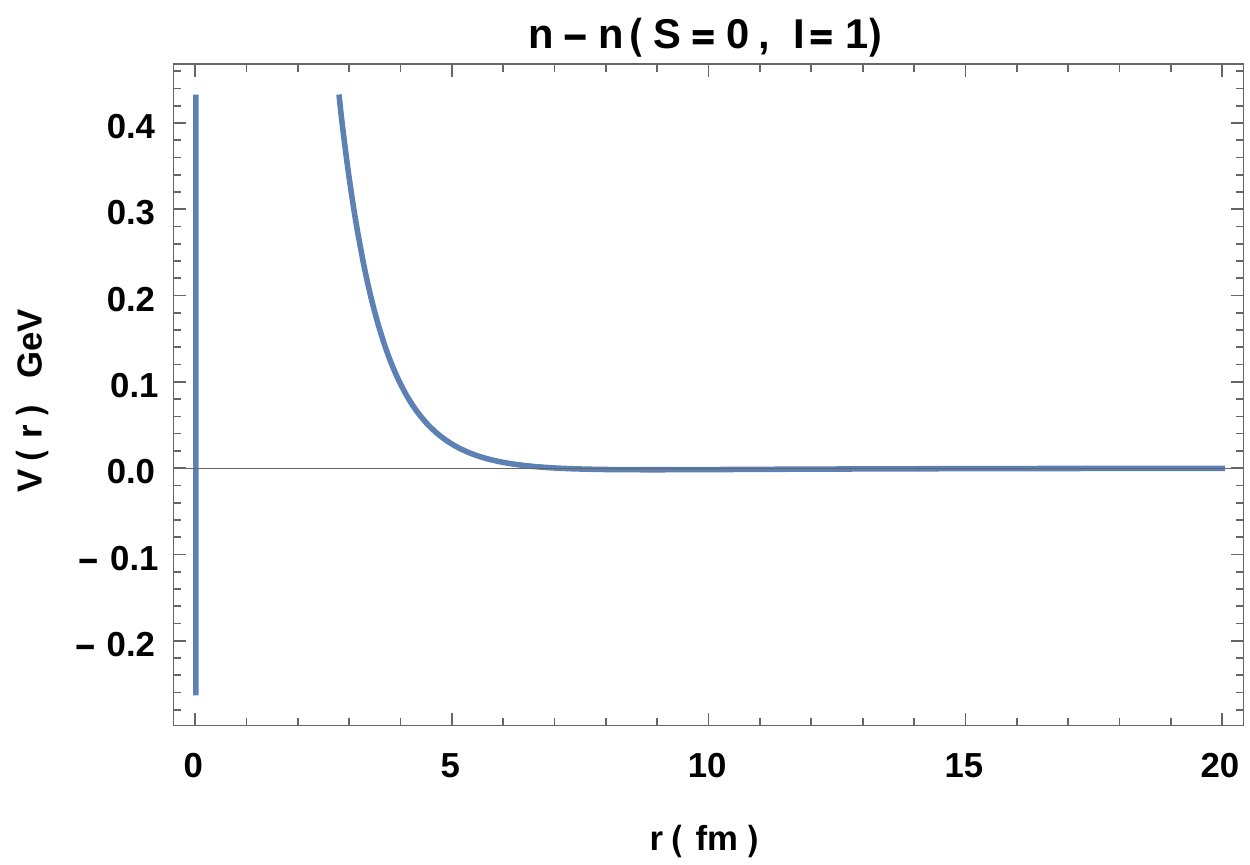}
\caption*{\hspace{1cm}(a)\hspace{4cm}(b)}
\caption{The characteristic nature of the s-wave OBE potential, screen Yukawa-like potential and net effective potential (a) in case of p-n (deuteron) system (b) in case of neutron-neutron system}
\label{n-n OBE and net plot}
\includegraphics[scale=0.255]{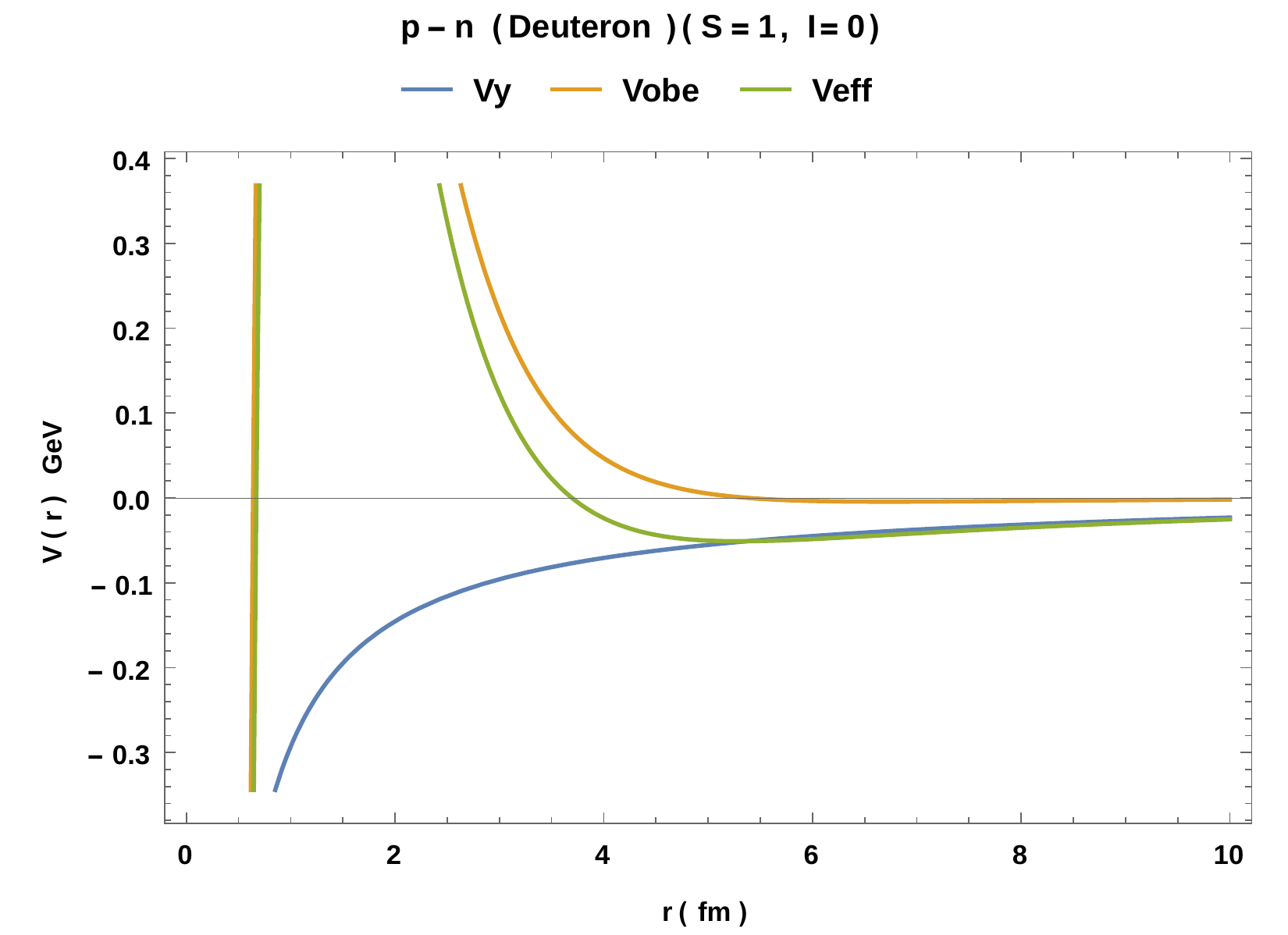}
\includegraphics[scale=0.255]{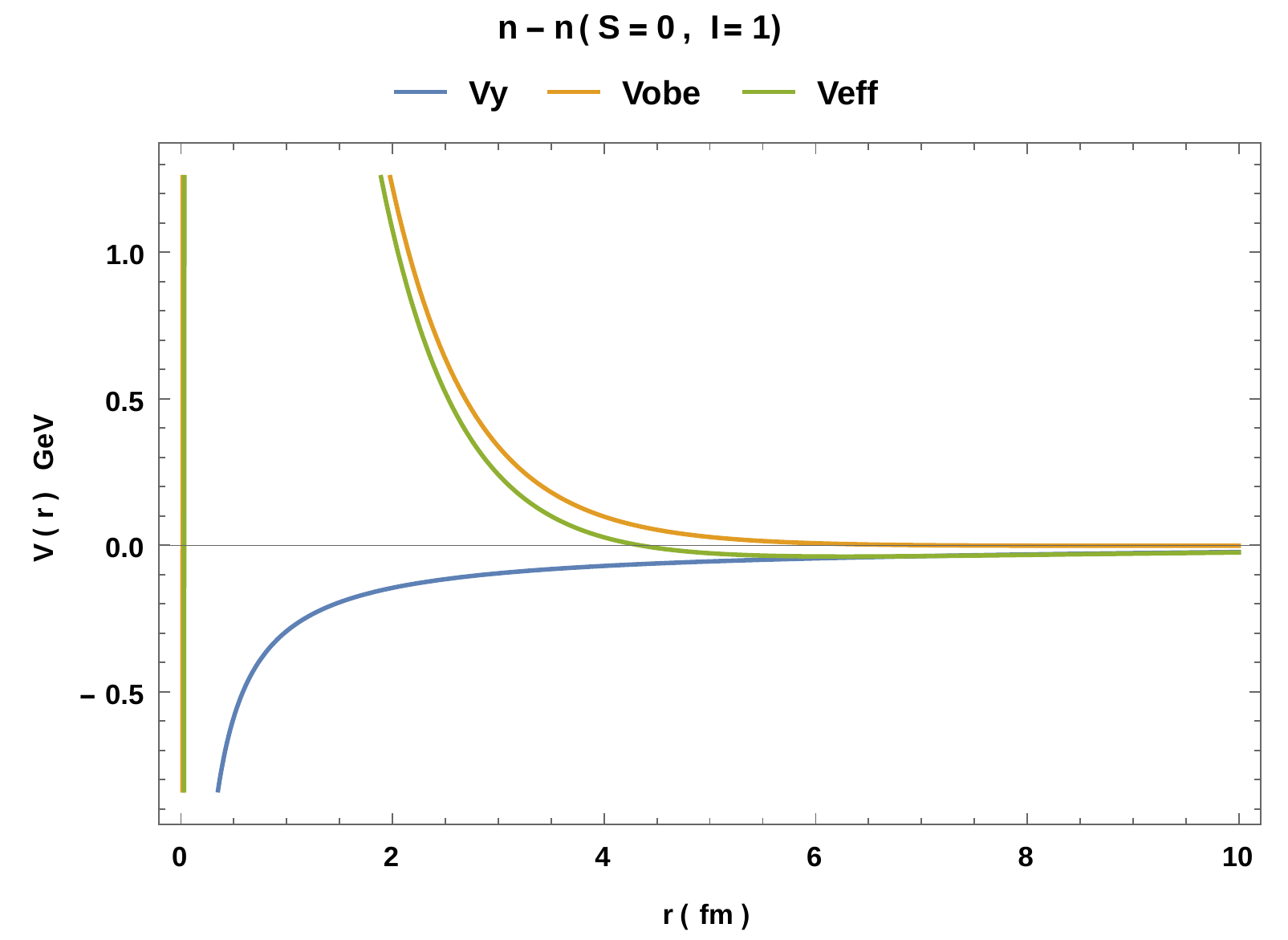}
\caption*{\hspace{1cm}(a)\hspace{4cm}(b)}
\end{figure}

\begin{figure*}[!ht]
\caption{The characteristic nature of the s-wave OBE potential for the case of $\Sigma_{c}D^{*}$ and $\Sigma_{c}^{*}D^{*}$, for all possible spin-isospin channels}
\label{SigmaD* net OBE}
\includegraphics[scale=0.335]{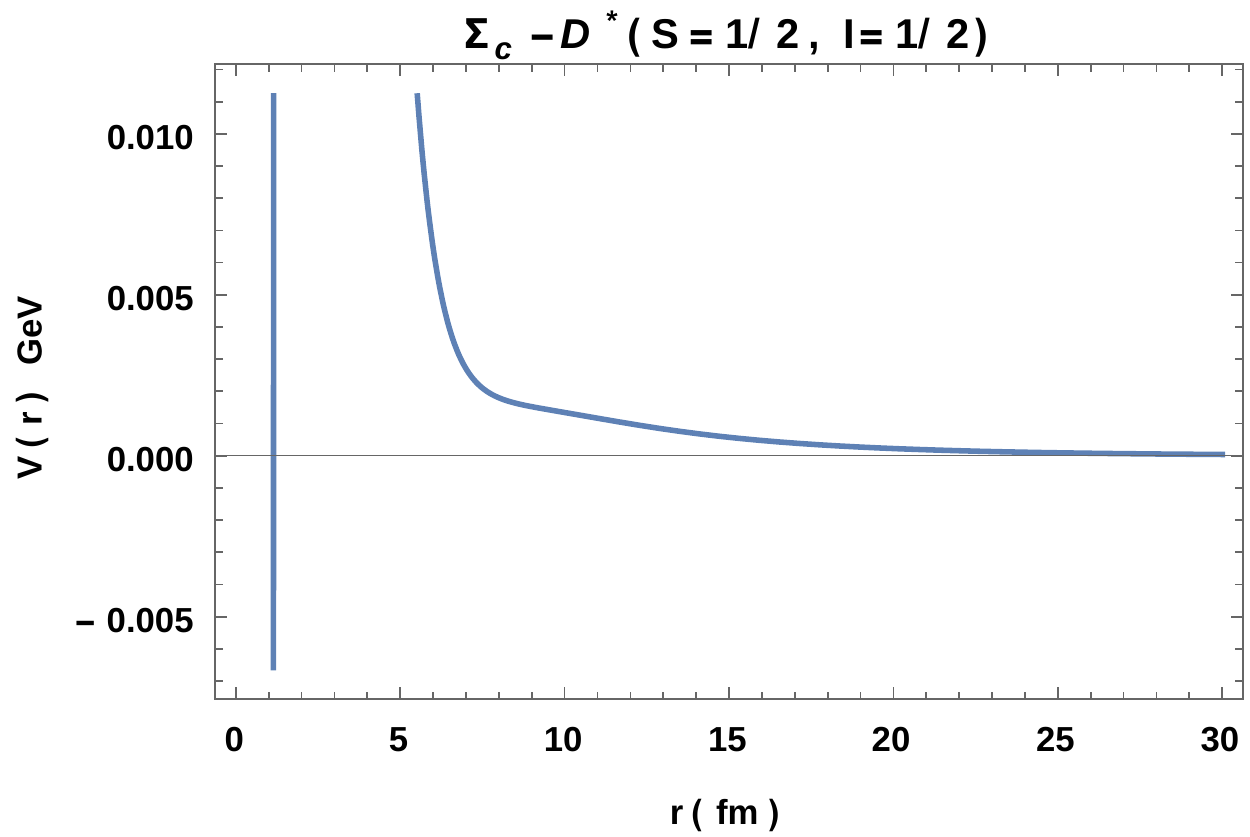}
\includegraphics[scale=0.335]{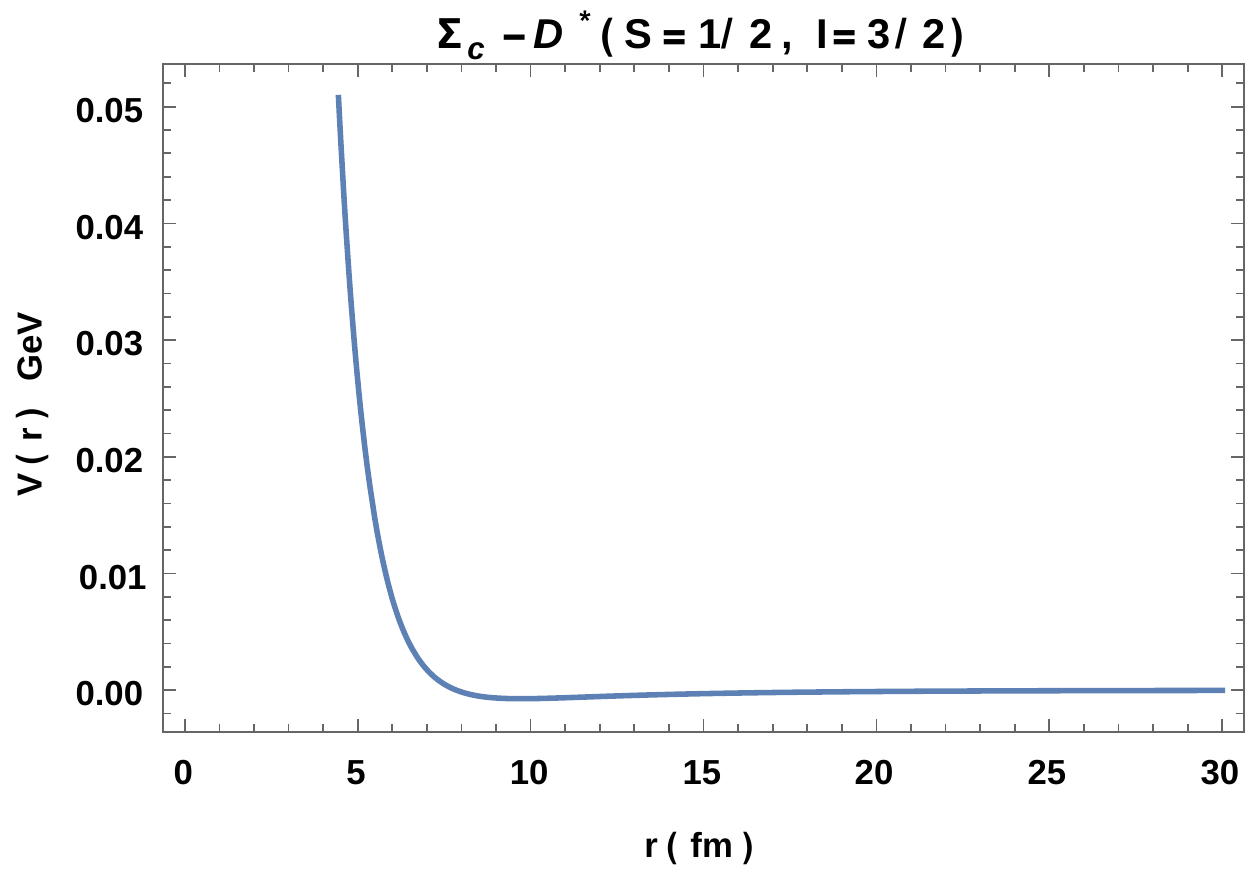}
\includegraphics[scale=0.335]{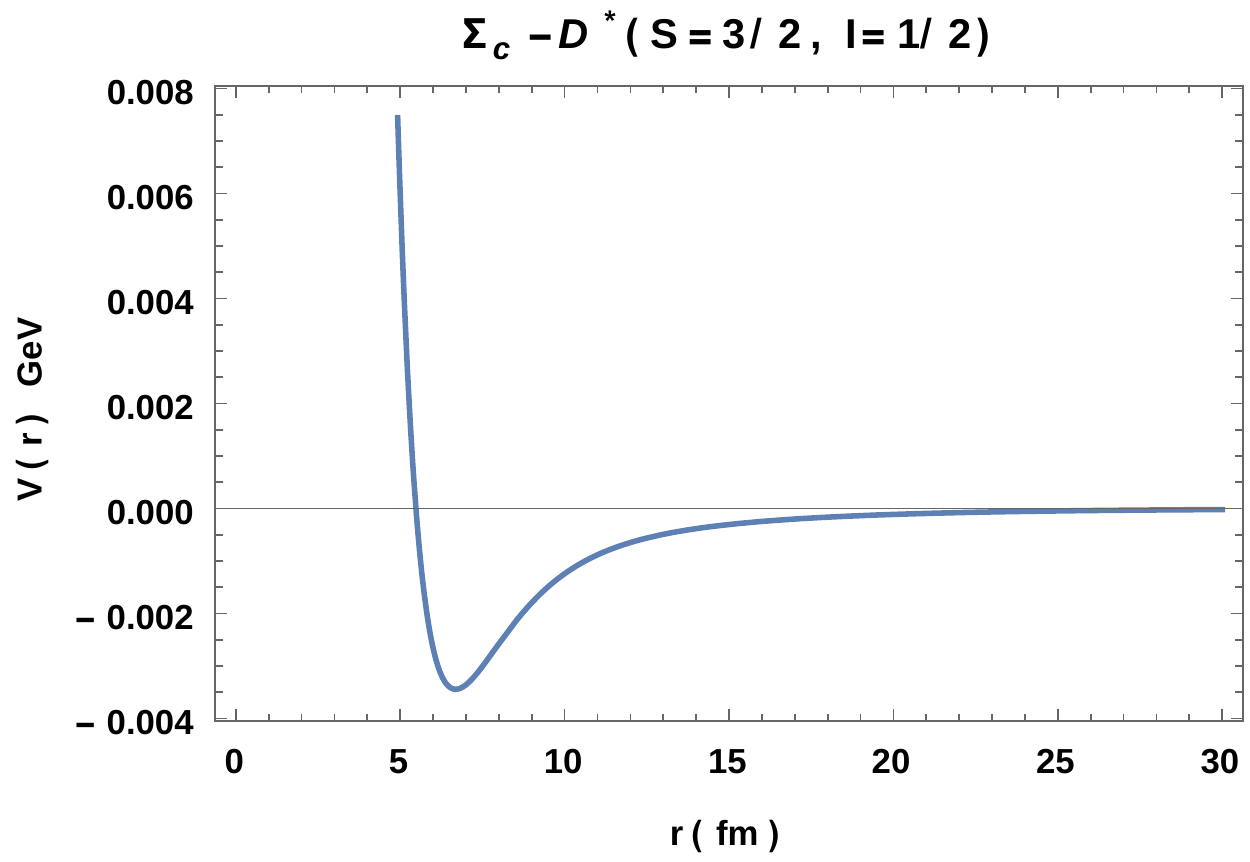}
\includegraphics[scale=0.335]{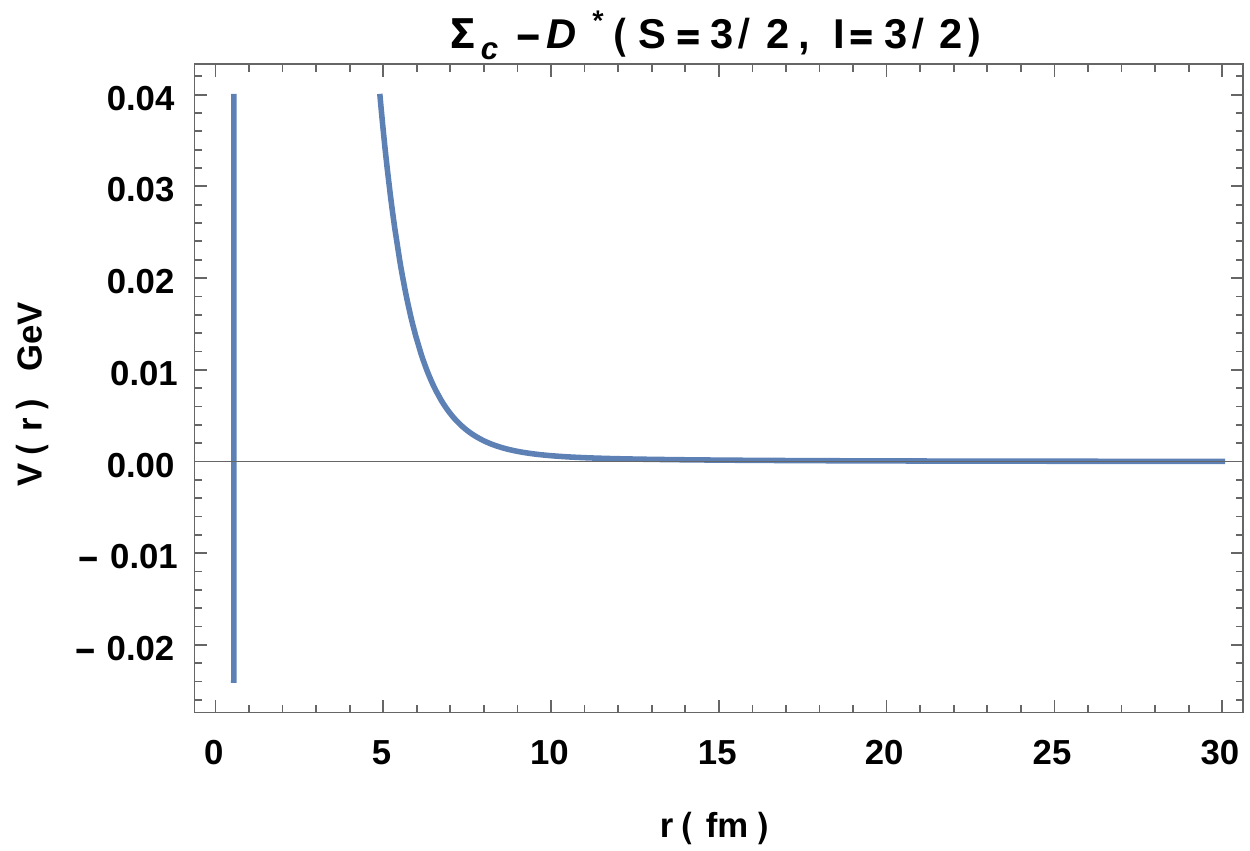}
\includegraphics[scale=0.335]{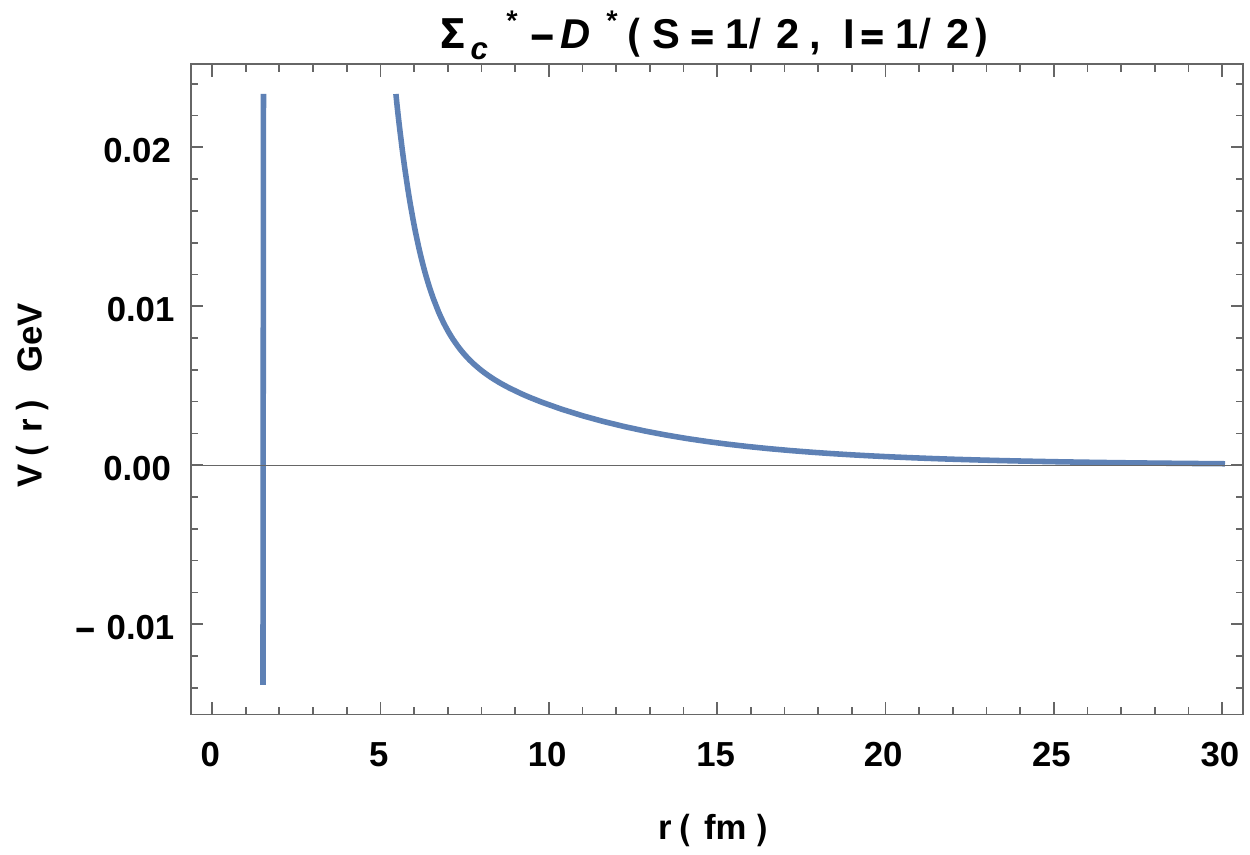}
\includegraphics[scale=0.335]{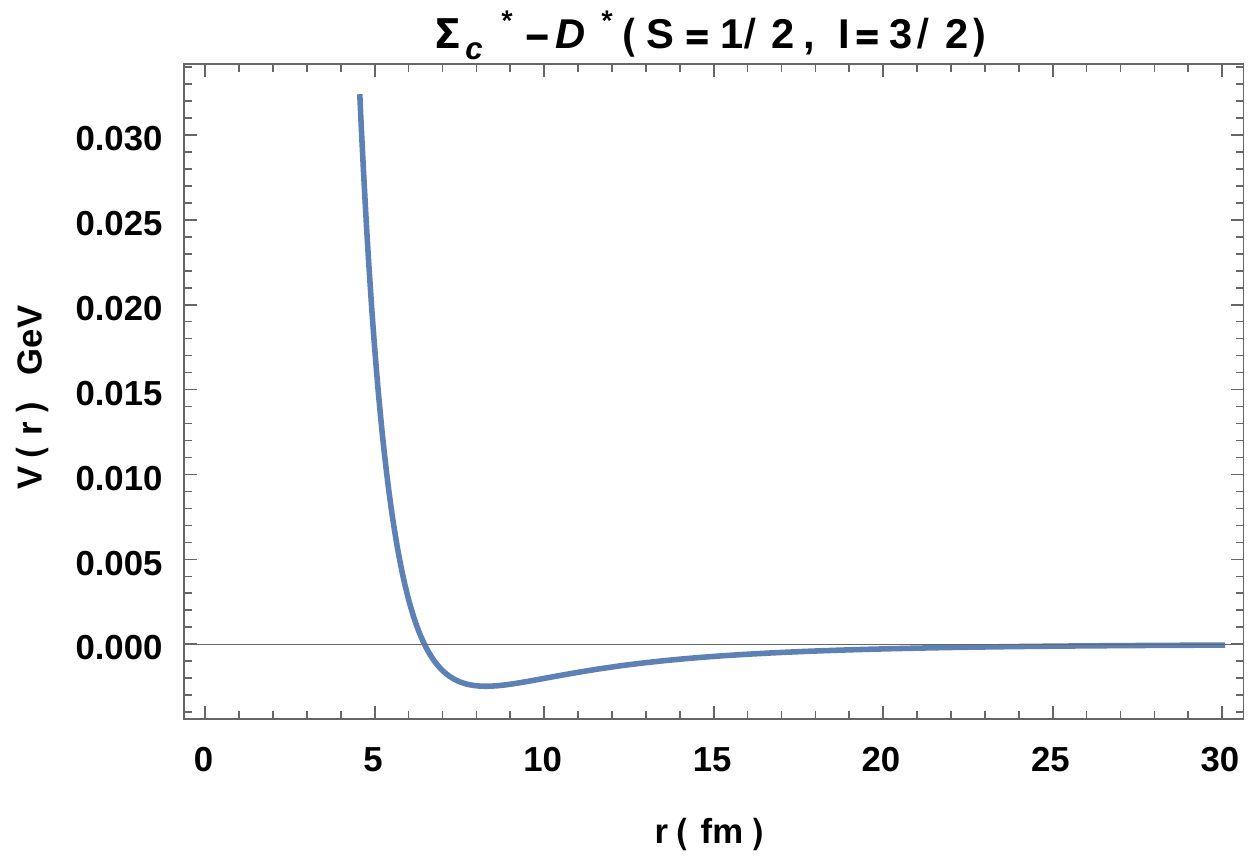}
\includegraphics[scale=0.335]{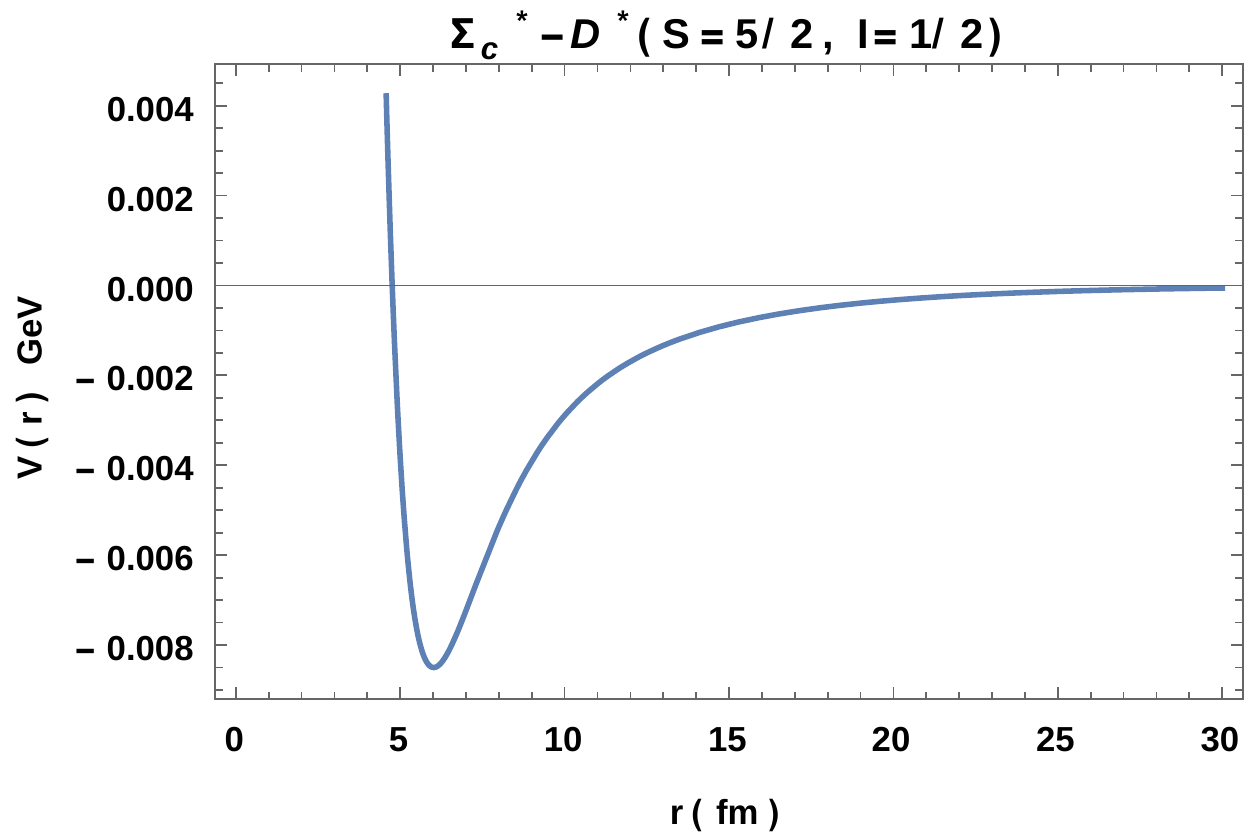}
\includegraphics[scale=0.335]{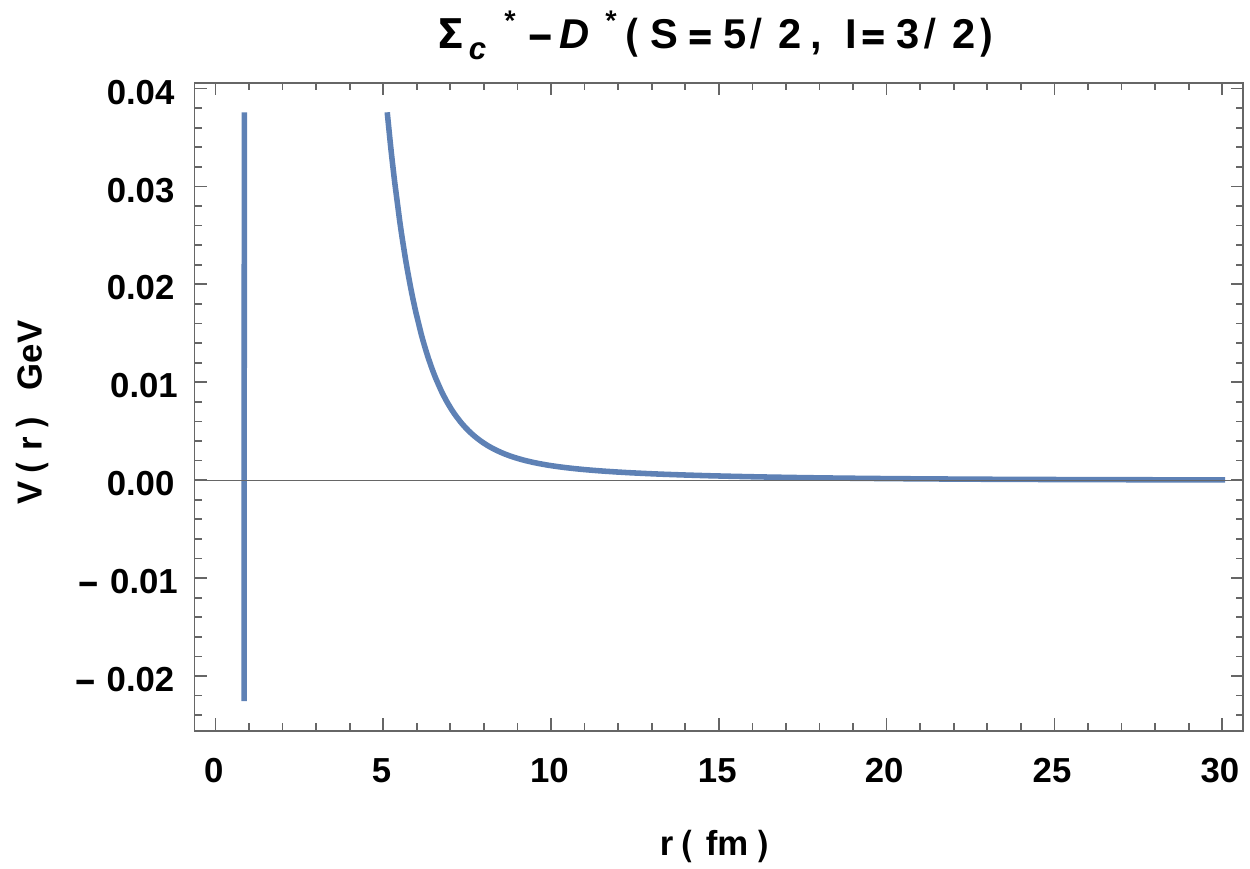}
\end{figure*}
\begin{figure*}
\caption{The characteristic nature of the s-wave OBE potential for the case of $\Xi_{c}D^{*}$ and $\Xi_{c}^{*}D^{*}$, for all possible spin-isospin channels}
\label{XiD* net OBE}
\includegraphics[scale=0.335]{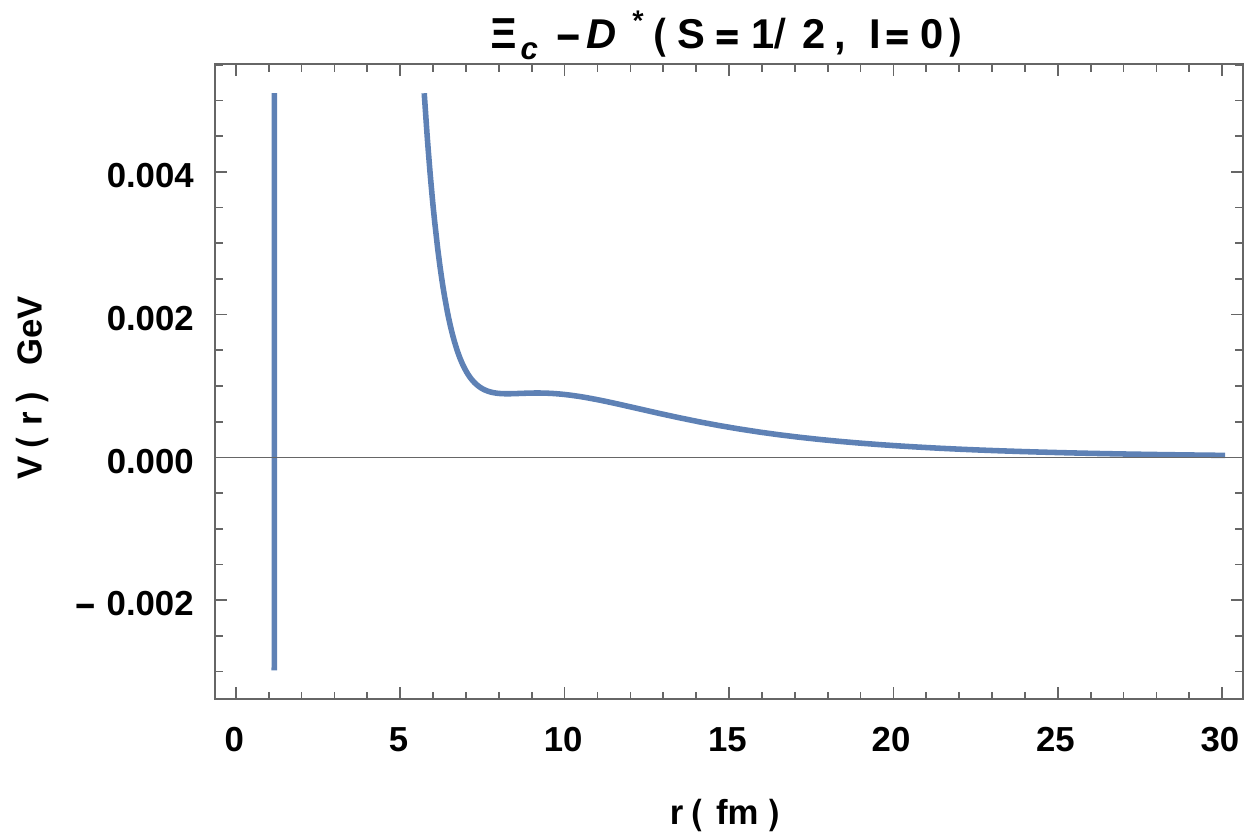}
\includegraphics[scale=0.335]{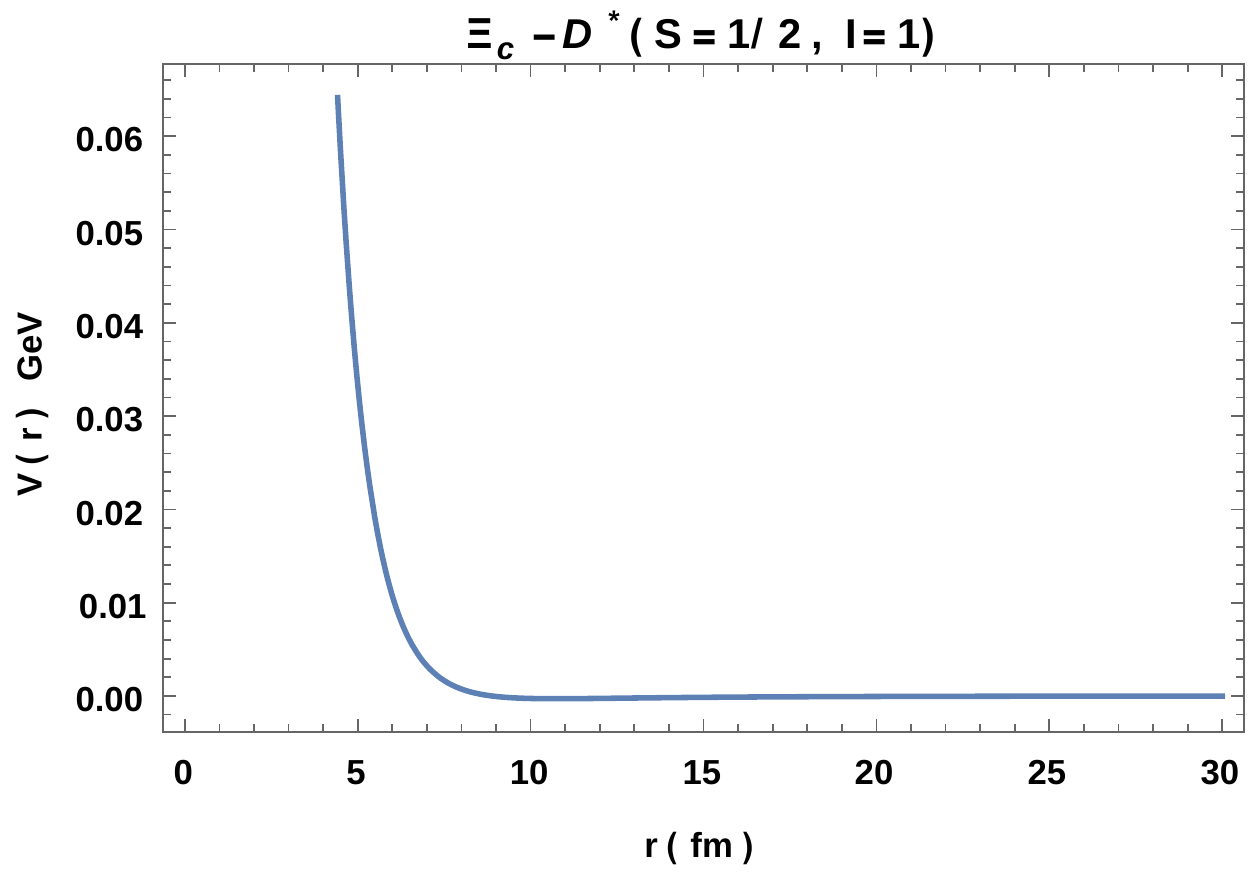}
\includegraphics[scale=0.335]{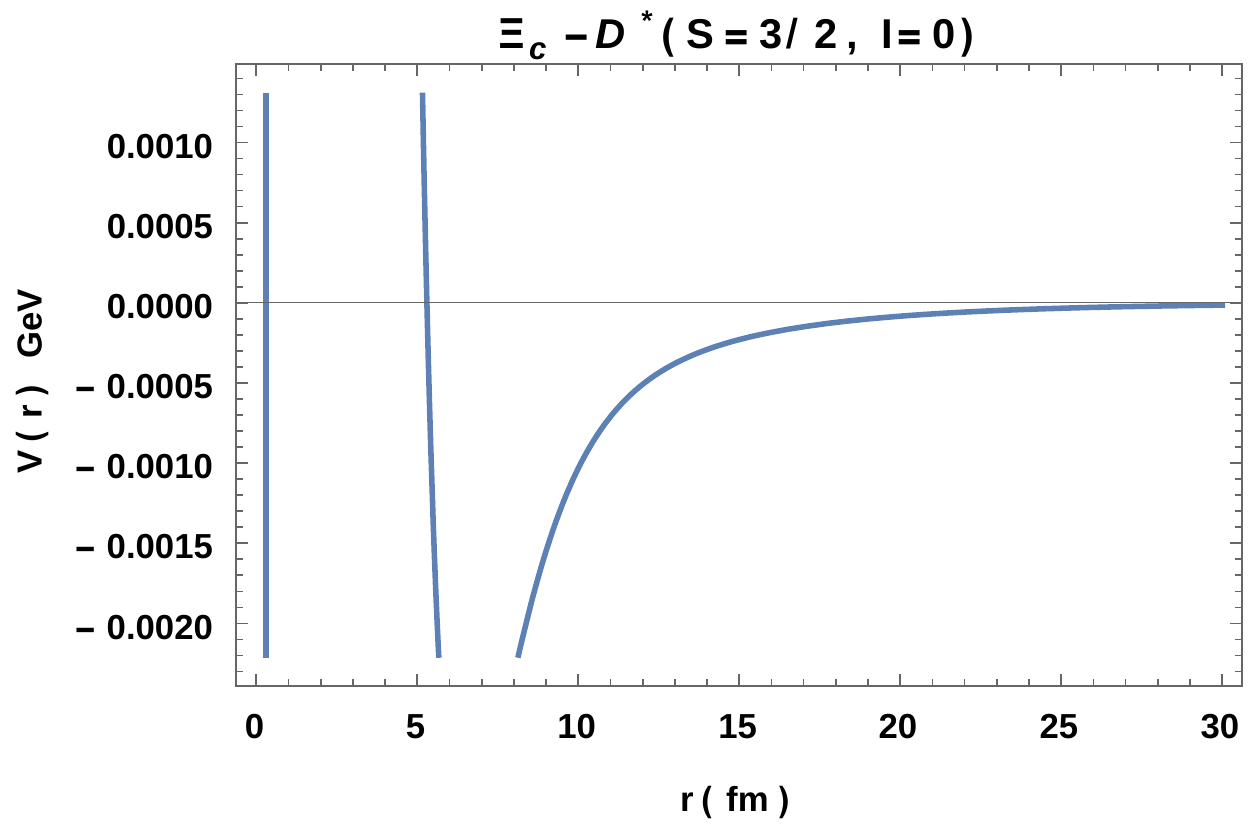}
\includegraphics[scale=0.335]{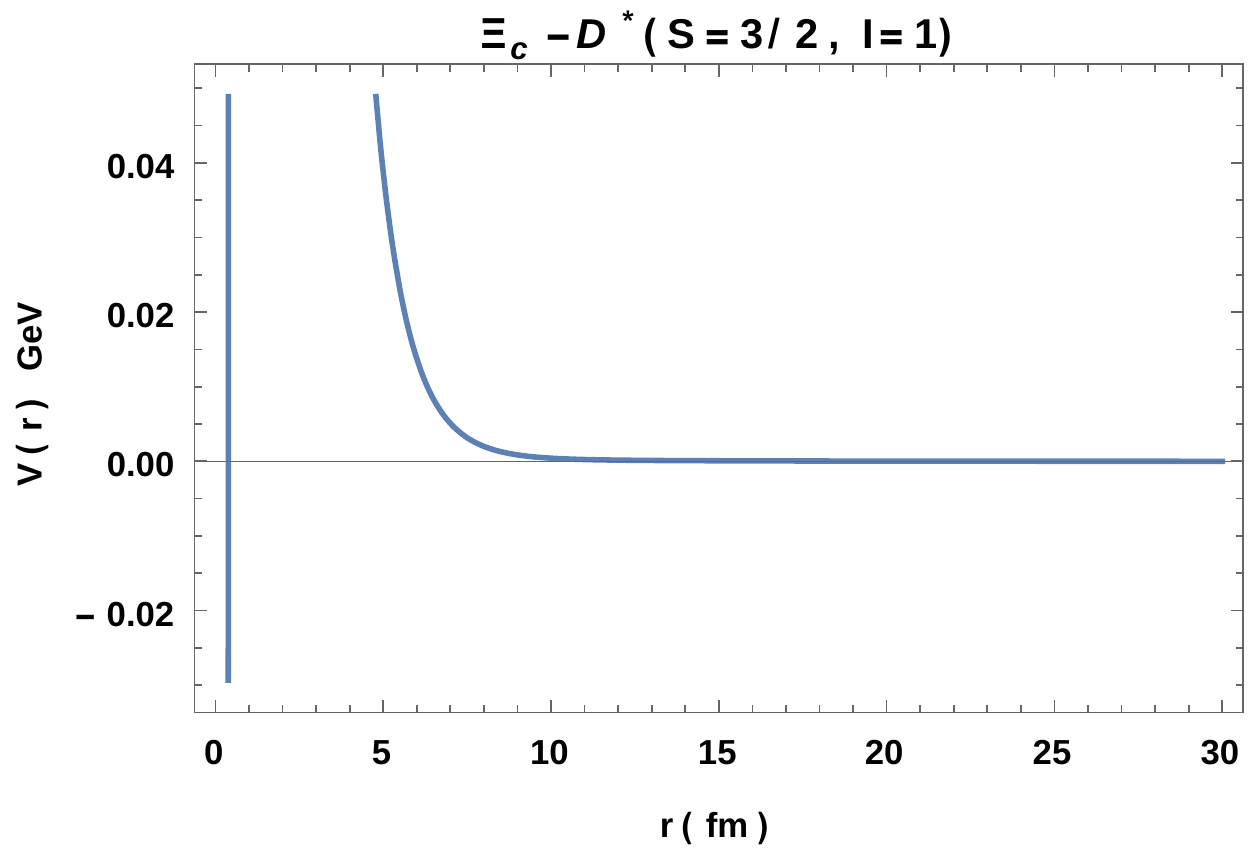}
\includegraphics[scale=0.335]{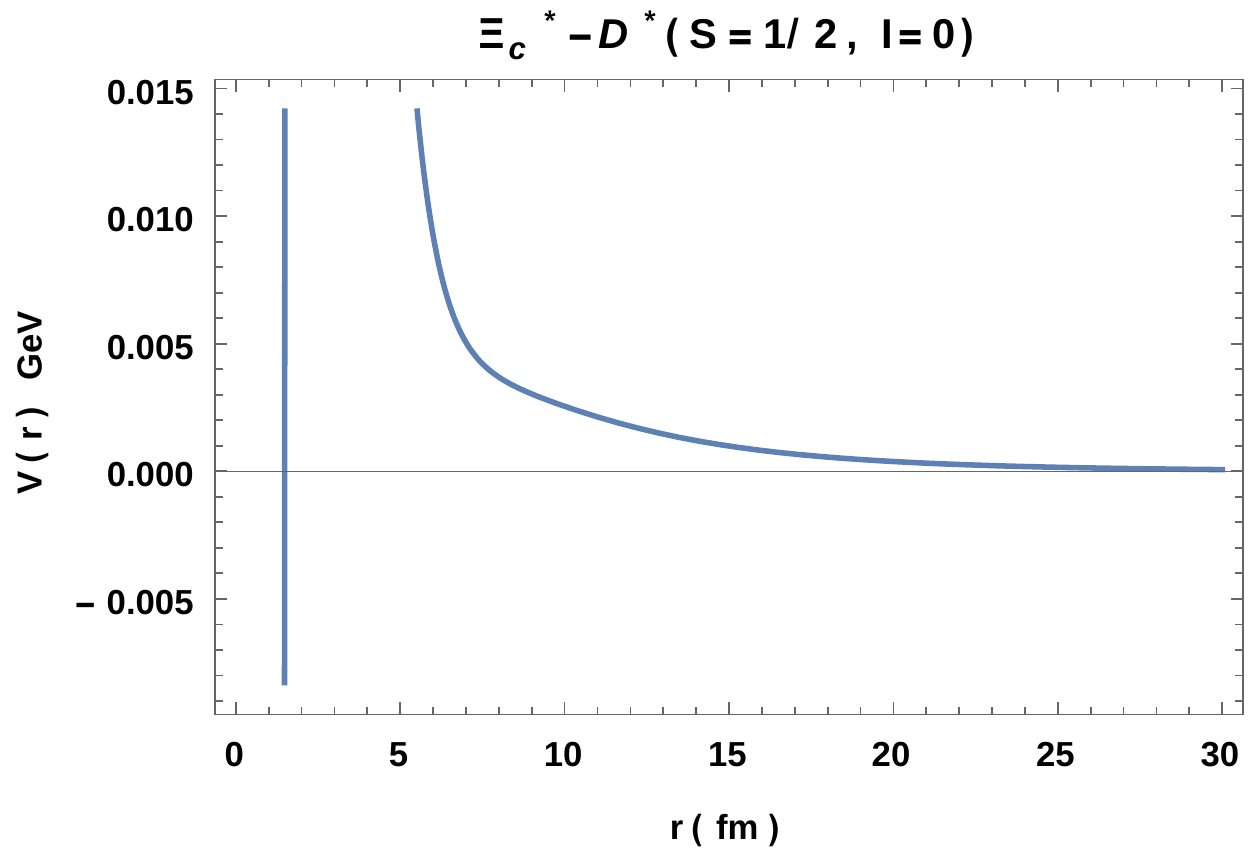}
\includegraphics[scale=0.335]{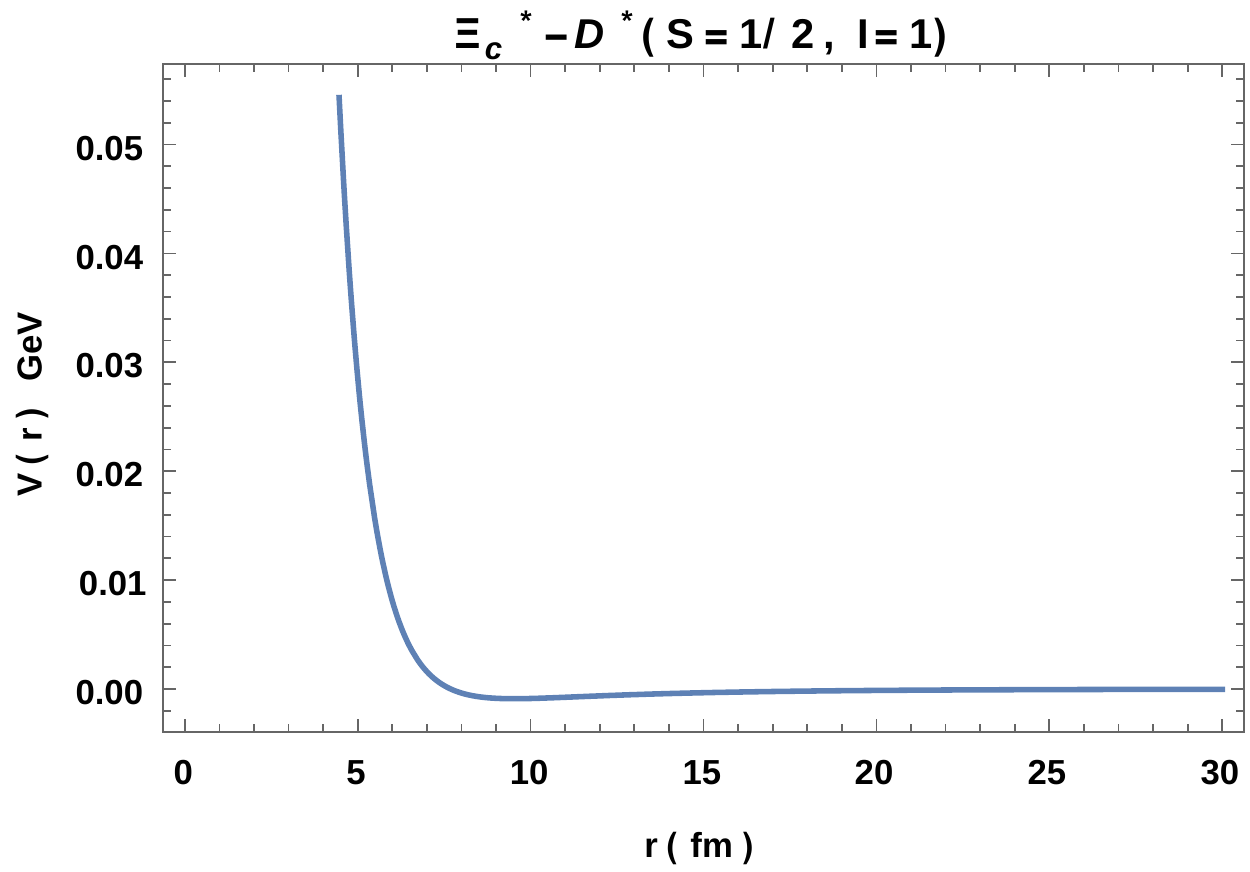}
\includegraphics[scale=0.335]{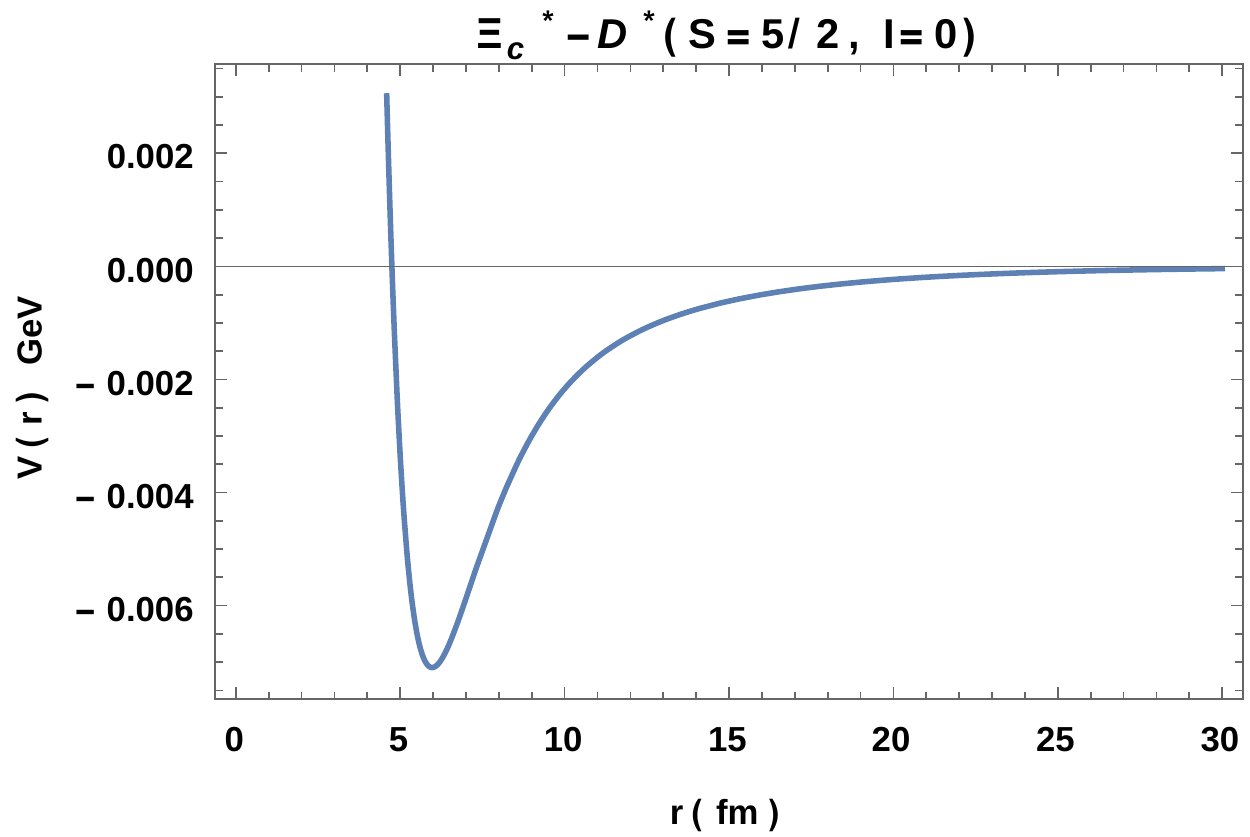}
\includegraphics[scale=0.335]{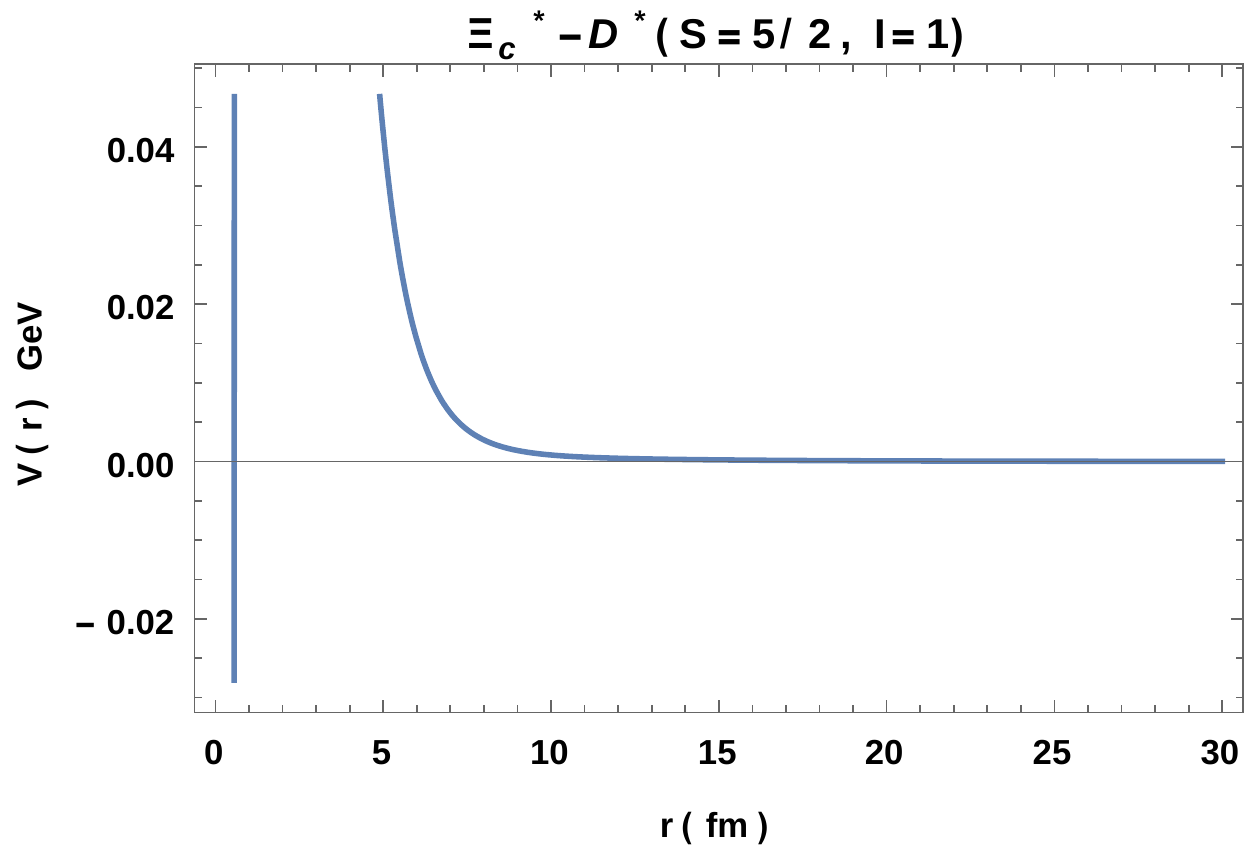}
\end{figure*}

%\subsection{Deuteron from the model}
In Fig.-\ref{p-n OBE and net plot}  the characteristic nature of the one meson exchange of OBE potential and in Fig-\ref{n-n OBE and net plot} net OBE, Yukawa screen like and effective potential are shown.  We can see from the Fig-\ref{p-n OBE and net plot}(a) $\&$(b) that the all individual meson exchange diminish exponentially at large distance, in which the pion exchange which is being the lightest meson exchange in OBE, contribute up to far distance while the sigma exchange contribute up to mid-range and other meson effectively contribute at short range.  In the case of deuteron, from Fig-\ref{p-n OBE and net plot}(a), we can see that the  pion, sigma and rho meson exchanges are attractive while eta, $a_{0}$ ($or$ $\delta$) and omega exchanges are repulsive. The net effective contribution of the s-wave OBE potential is shown in Fig-\ref{OBE net plot}. From Fig-\ref{OBE net plot}, we can see that in the case of p-n bound state the net effective s-wave OBE potential is very shallow attractive near 3 fm to 6 fm  while the potential is repulsive in case of neutron-neutron system. The overall contributory nature of s-wave OBE potential along with screen Yukawa-like and net effective potential can be seen in the Fig-\ref{n-n OBE and net plot}(a) (b). Fig-\ref{n-n OBE and net plot}(a) shows very shallow attractive  strength at large distance while the net effective potential shows attractive strength near 4 fm due to influence of the attractive Yukawa-like screen potential. Similarly,  Fig-\ref{p-n OBE and net plot}(b) shows the meson exchange behavior in the case of the n-n system. In general, the attractive and repulsiveness of the s-wave one meson exchange potentials are depends on their isospin-spin channels. We can see from Fig-\ref{n-n OBE and net plot}(b) that the net OBE potential is repulsive at short distance and exponentially diminish at long range.  Even, under the influence of the attractive Yukawa-like potential, the net effective potential get almost zero strength or just above the zero, while it is repulsive at short range. As a result, the Fig-\ref{n-n OBE and net plot}(b) shows that the resultant potential leads n-n systems unbound.

In the Table-\ref{deuteron-table from OBE}, the binding energy of the pn (deuteron) and nn bound states are tabulated by using c=0.0686 GeV with other fixed parameters which are already discussed in above section. The calculated binding energy is in agreement with experimental value 2.224 MeV whereas the obtained mean square root radius is nearly about 3 fm while the expected is near 2.1 fm.

%\begin{figure}[!h]
%\caption{The characteristic nature of the s-wave one meson exchange potential (a) in case of p-n (deuteron) system (b) in case of neutron-neutron system}
%\label{p-n OBE and net plot}
%\begin{center}
%\includegraphics[scale=0.2]{fig/PN-NNCoup-OBE.pdf}
%\caption*{(a)}
%\includegraphics[scale=0.2]{fig/NN-NNCoup-OBE.pdf}
%\caption*{(b)}
%\end{center}
%\end{figure}
%
%\begin{figure}[]
%\caption{The nature of the net s-wave OBE, Screen Yukawa-like and net effective potential (a) in case of p-n (deuteron) system (b)in case of neutron-neutron system}
%\label{n-n OBE and net plot}
%\includegraphics[scale=0.4]{fig/PN-NNCoup-NET.pdf}\caption*{(a)}
% \includegraphics[scale=0.4]{fig/NN-NNCoup-NET.pdf}\caption*{(b)}
%\end{figure}

\section{Mass spectra of meson-baryon systems}  
\label{Mass spectra of meson-baryon systems}
We have calculated the mass spectra of the $\Sigma_{s,c,b}K^{*}$, $\Sigma_{s,c,b}D^{*}$, $\Sigma_{s,c,b}B^{*}$, $\Xi_{s,c,b}K^{*}$, $\Xi_{s,c,b}D^{*}$ and $\Xi_{s,c,b}B^{*}$ systems. The graphs of potential strength verse range for individual meson exchange, net OBE potential, screen Yukawa-like potential and net effective potential are plotted, and are shown in Fig-\ref{SigmaD* net OBE} to \ref{pentaquark net-potential plots}. To understand the behavior of the potentials incorporated in this work with different spin-isospin channels for attempted di-hadronic systems, we present the analysis graphs for one example systems ($\Sigma_{c}D^{*}$ and $\Sigma_{c}^{*}D^{*}$). The plots for individual meson exchange and comparative plots of net s-wave OBE, Yukawa-like potential and net effective potentials are shown in Fig- \ref{pentaquark-OBE plots} and \ref{pentaquark net-potential plots}, and presented in appendix.  

Fig-\ref{SigmaD* net OBE} shows the contributory nature of the effective s-wave OBE potential in all possible spin-isospin channels for $\Sigma_{c}D^{*}$ and $\Sigma_{c}^{*}D^{*}$ systems, while Fig-\ref{XiD* net OBE} present the $\Xi_{c}D^{*}$ and $\Xi_{c}^{*}D^{*}$ systems. We can see from the Fig-\ref{SigmaD* net OBE}, the spin-isospin channel (S,I) = (5/2, 1/2), (3/2, 1/2) and (1/2, 3/2) gets attractive strength near 4-6 fm. Among these three channels, the channel (S,I) = (5/2, 1/2) seems strongest attractive channel. Whereas, the Fig-\ref{XiD* net OBE} shows that the spin-isospin channel (S,I) = (3/2, 0) and  (5/2, 0) channels are attractive. Indeed, with s-wave OBE interaction, all theses attractive channels provides the probabilities to get bound states with these specific spin-isospin combination, provided that these attractive strength should strong enough to overcome kinetic energy repulsion. 
In the present study, to get bound state of the proton-neutron (deuteron) system, the s-wave OBE potential do not get sufficient  attractive strength, hence, we need some additional attractive strength to get bound state. Thus, we have incorporated screen Yukawa-like potential along with s-wave OBE potential. All the potential parameters are tuned for deuteron experimental value of binding energy and taken same for all other calculations. The interesting graphs of one meson exchange potential and comparative  plots of screen Yukawa-like potential, net s-wave OBE potential and total effective potential are presented in the appendix.

\begin{table}[]
\caption{Mass spectra of meson-baryon $(\Sigma_{s,c,b} - K^{*})$ (molecular pentaquark) molecules. Masses of the meson and baryon are taken from PDG \cite{Patrignani-PDG2016} which are also listed in Table-\ref{HadronMaases-PDG}. Here, $\mu$ is variational parameter. I (isospin), G (G-parity), J (total angular momentum), Q (charge) and P (parity) are quantum numbers of the respective meson and baryon.}
\label{sigma-k* mass spectra}
\scalebox{0.8}{
\begin{tabular}{cccccccc}
\hline
\\
$[I_{1}(J_{1}^{P_{1}})]^{Q_{1}}$ - $[I_{2}(J_{2}^{P_{2}})]^{Q_{2}}$ & System & $I(J^{P})$ & $\mu$ & B.E. & Mass & $\sqrt{r^{2}}$ \\
& & & GeV & MeV & MeV & fm \\
\hline
\hline
\\
\multirow{4}{*}{$[1$($\frac{1}{2}^{+}$)$]^{0}$-$[\frac{1}{2}$($1^{-}$)$]^{0}$}
&\multirow{4}{*}{$\Sigma_{s} - K^{*}$}
& $\frac{1}{2}(\frac{1}{2}^{-})$ & 0.0071	& +0.0038 & 2.088 & 47.48\\
& & $\frac{3}{2}(\frac{1}{2}^{-})$ &0.0572	& -0.3704 & 2.088 & 05.97\\
& & $\frac{1}{2}(\frac{3}{2}^{-})$ &0.0767	& -0.9909 & 2.087 & 04.45\\
& & $\frac{3}{2}(\frac{3}{2}^{-})$ & 0.1077	& -1.5888 & 2.086 & 03.17\\
\\

\multirow{4}{*}{$[1$($\frac{3}{2}^{+}$)$]^{0}$-$[\frac{1}{2}$($1^{-}$)$]^{0}$}
& \multirow{4}{*}{$\Sigma_{s} - K^{*}$}
& $\frac{1}{2}(\frac{1}{2}^{-})$ &0.0066	& +0.0029 & 2.278 & 52.01 \\
& & $\frac{3}{2}(\frac{1}{2}^{-})$ &0.0466	& -0.2289 & 2.278 & 07.33 \\
& & $\frac{1}{2}(\frac{5}{2}^{-})$ &0.0555	& -0.4650 & 2.278 & 06.16 \\
& & $\frac{3}{2}(\frac{5}{2}^{-})$ &0.0072	& +0.0033 & 2.278 & 47.47 \\
\\
\hline
\\
\multirow{4}{*}{$[1$($\frac{1}{2}^{+}$)$]^{0}$-$[\frac{1}{2}$($1^{-}$)$]^{0}$}
& \multirow{4}{*}{$\Sigma_{c} - K^{*}$}
& $\frac{1}{2}(\frac{1}{2}^{-})$ &  0.0039	& +0.0007 & 3.349 & 86.62 \\
& & $\frac{3}{2}(\frac{1}{2}^{-})$ & 0.0781	& -1.0945 & 3.348 & 04.37 \\
& & $\frac{1}{2}(\frac{3}{2}^{-})$ & 0.1059	& -2.5205 & 3.347 & 03.22 \\
& & $\frac{3}{2}(\frac{3}{2}^{-})$ & 0.1003	& -1.8426 & 3.347 & 03.40 \\
\\
\multirow{4}{*}{$[1$($\frac{3}{2}^{+}$)$]^{0}$-$[\frac{1}{2}$($1^{-}$)$]^{0}$}
&\multirow{4}{*}{$\Sigma_{c} - K^{*}$}
&$\frac{1}{2}(\frac{1}{2}^{-})$  & 0.0039	& +0.0007 & 3.414 & 85.76 \\
&&$\frac{3}{2}(\frac{1}{2}^{-})$ & 0.0666	& -0.7637 & 3.413 & 05.13  \\
&&$\frac{1}{2}(\frac{5}{2}^{-})$ & 0.0819	& -1.4690 & 3.412 & 04.17  \\
&&$\frac{3}{2}(\frac{5}{2}^{-})$ & 0.1349	& -3.0589 & 3.411 & 02.53  \\
\\
\hline
\\

\multirow{4}{*}{$[1$($\frac{1}{2}^{+}$)$]^{0}$-$[\frac{1}{2}$($1^{-}$)$]^{0}$}
&\multirow{4}{*}{$\Sigma_{b} - K^{*}$}
&$\frac{1}{2}(\frac{1}{2}^{-})$ & 0.1666 & -6.2422 & 6.702 & 02.05  \\
&&$\frac{3}{2}(\frac{1}{2}^{-})$ & 0.0939 & -1.8751 & 6.707 & 03.64 \\
&&$\frac{1}{2}(\frac{3}{2}^{-})$ & 0.1309 & -4.2306 & 6.704 & 02.61 \\
&&$\frac{3}{2}(\frac{3}{2}^{-})$ & 0.1003 & -2.1359 & 6.706 & 03.40 \\
\\

\multirow{4}{*}{$[1$($\frac{3}{2}^{+}$)$]^{0}$-$[\frac{1}{2}$($1^{-}$)$]^{0}$}
&\multirow{4}{*}{$\Sigma_{b} - K^{*}$}
&$\frac{1}{2}(\frac{1}{2}^{-})$ & 0.0026 & +0.0002 & 6.727 & 130.6 \\
&&$\frac{3}{2}(\frac{1}{2}^{-})$ & 0.0886 & -1.6678 & 6.726 & 03.85 \\
&&$\frac{1}{2}(\frac{5}{2}^{-})$ & 0.1176 & -3.4806 & 6.724 & 02.90 \\
&&$\frac{3}{2}(\frac{5}{2}^{-})$ & 0.1054 & -2.3474 & 6.725 & 03.24 \\
\\
\hline
\hline
\end{tabular}
}
\end{table}

\begin{table}[!ht]
\caption{Mass spectra of meson-baryon (molecular pentaquark) $(\Sigma_{s,c,b} - D^{*})$  molecules. Masses of the meson and baryon are taken from PDG \cite{Patrignani-PDG2016} which are also listed in Table-\ref{HadronMaases-PDG}. Here, $\mu$ is variational parameter. I (isospin), G (G-parity), J (total angular momentum), Q (charge) and P (parity) are quantum numbers of the respective meson and baryon.}
\label{sigma-D* mass spectra}
\scalebox{0.8}{
\begin{tabular}{cccccccc}
\hline
\\
$[I_{1}(J_{1}^{P_{1}})]^{Q_{1}}$ - $[I_{2}(J_{2}^{P_{2}})]^{Q_{2}}$&System & $I(J^{P})$ & $\mu$ & B.E. & Mass&$\sqrt{r^{2}}$  \\
& & & GeV & MeV & GeV &fm \\
\hline
\hline
\\
\multirow{4}{*}{$[1$($\frac{1}{2}^{+}$)$]^{0}$-$[\frac{1}{2}$($1^{-}$)$]^{0}$}
&\multirow{4}{*}{$\Sigma_{s} - D^{*}$}
&$\frac{1}{2}(\frac{1}{2}^{-})$ &0.0031	& +0.0003 & 3.1996 & 110.1 \\
&&$\frac{3}{2}(\frac{1}{2}^{-})$ & 0.0779	& -1.1139 & 3.198 & 04.38 \\
&&$\frac{1}{2}(\frac{3}{2}^{-})$ & 0.1059	& -2.5417 & 3.197 & 03.22 \\
&&$\frac{3}{2}(\frac{3}{2}^{-})$ & 0.1034	& -1.9756 & 3.197 & 03.30\\
\\
\multirow{4}{*}{$[1$($\frac{3}{2}^{+}$)$]^{0}$-$[\frac{1}{2}$($1^{-}$)$]^{0}$}
&\multirow{4}{*}{$\Sigma_{s} - D^{*}$}
&$\frac{1}{2}(\frac{1}{2}^{-})$ & 0.0027	& +0.0002 & 3.389 & 126.2 \\
&&$\frac{3}{2}(\frac{1}{2}^{-})$ & 0.0669	& -0.7939 & 3.388 & 05.11 \\
&&$\frac{1}{2}(\frac{5}{2}^{-})$ & 0.0823	& -1.5068 & 3.388 & 04.15 \\
&&$\frac{3}{2}(\frac{5}{2}^{-})$ & 0.1434	& -3.3415 & 3.386 & 02.38 \\
\\
\hline
\\

\multirow{4}{*}{$[1$($\frac{1}{2}^{+}$)$]^{0}$-$[\frac{1}{2}$($1^{-}$)$]^{0}$}
&\multirow{4}{*}{$\Sigma_{c} - D^{*}$}
&$\frac{1}{2}(\frac{1}{2}^{-})$ & 0.0015	& +0.0000 & 4.460 & 221.1 \\
&&$\frac{3}{2}(\frac{1}{2}^{-})$ &0.0816	& -1.1805 & 4.459 & 04.18 \\
&&$\frac{1}{2}(\frac{3}{2}^{-})$ & 0.1149	& -2.8497 & 4.457 & 02.97 \\
&&$\frac{3}{2}(\frac{3}{2}^{-})$ & 0.0948	& -1.5949 & 4.459 & 03.60 \\
\\
\multirow{4}{*}{$[1$($\frac{3}{2}^{+}$)$]^{0}$-$[\frac{1}{2}$($1^{-}$)$]^{0}$}
&\multirow{4}{*}{$\Sigma_{c} - D^{*}$}
&$\frac{1}{2}(\frac{1}{2}^{-})$ & 0.0015	& +0.0000 & 4.525 & 221.4 \\
&&$\frac{3}{2}(\frac{1}{2}^{-})$ & 0.0729	& -0.9285 & 4.524 & 04.68 \\
&&$\frac{1}{2}(\frac{5}{2}^{-})$ & 0.0945	& -1.9401 & 4.523 & 03.61 \\
&&$\frac{3}{2}(\frac{5}{2}^{-})$ & 0.1078	& -2.0183 & 4.523 & 03.16 \\
\\
\hline
\\
\multirow{4}{*}{$[1$($\frac{1}{2}^{+}$)$]^{0}$-$[\frac{1}{2}$($1^{-}$)$]^{0}$}
&\multirow{4}{*}{$\Sigma_{b} - D^{*}$}
&$\frac{1}{2}(\frac{1}{2}^{-})$ & 0.1530 & -4.7770 & 7.8152 & 02.23 \\
&&$\frac{3}{2}(\frac{1}{2}^{-})$ & 0.1044 & -1.2479 & 7.8187 & 03.27 \\
&&$\frac{1}{2}(\frac{3}{2}^{-})$ & 0.1275 & -3.5372 & 7.8164 & 02.68 \\
&&$\frac{3}{2}(\frac{3}{2}^{-})$ & 0.0928 & -1.5886 & 7.8183 & 03.68 \\
\\
\multirow{4}{*}{$[1$($\frac{3}{2}^{+}$)$]^{0}$-$[\frac{1}{2}$($1^{-}$)$]^{0}$}
&\multirow{4}{*}{$\Sigma_{b} - D^{*}$}
&$\frac{1}{2}(\frac{1}{2}^{-})$ & 0.2285	& -7.7376 & 7.831 & 01.49 \\
&&$\frac{3}{2}(\frac{1}{2}^{-})$ & 0.0844	& -1.3127 & 7.837 & 04.05  \\
&&$\frac{1}{2}(\frac{5}{2}^{-})$ & 0.1167	& -3.0135 & 7.836 & 02.92 \\
&&$\frac{3}{2}(\frac{5}{2}^{-})$ & 0.0963	& -1.7036 & 7.837 & 03.54 \\
\\
\hline
\hline
\end{tabular}
}
\end{table}		        
\begin{table}[]
\begin{center}
\caption{Mass spectra of meson-baryon (molecular pentaquark) $(\Sigma_{s,c,b} - B^{*})$  molecules. Masses of the meson and baryon are taken from PDG \cite{Patrignani-PDG2016} which are also listed in Table-\ref{HadronMaases-PDG}. Here, $\mu$ is variational parameter. I (isospin), G (G-parity), J (total angular momentum), Q (charge) and P (parity) are quantum numbers of the respective meson and baryon.}
\label{sigma-B* mass spectra}
\scalebox{0.8}{
\begin{tabular}{cccccccc}
\hline
\\
$[I_{1}(J_{1}^{P_{1}})]^{Q_{1}}$ - $[I_{2}(J_{2}^{P_{2}})]^{Q_{2}}$ & System & $I(J^{P})$ & $\mu$ & B.E. & Mass & $\sqrt{r^{2}}$ \\
& & & GeV & MeV & MeV & fm \\
\hline
\hline
\\
\multirow{4}{*}{$[1$($\frac{1}{2}^{+}$)$]^{0}$-$[\frac{1}{2}$($1^{-}$)$]^{0}$}
&\multirow{4}{*}{$\Sigma_{s} - B^{*}$}
&$\frac{1}{2}(\frac{1}{2}^{-})$ & 0.1717	& -6.3149 & 6.511 & 01.99 \\
&&$\frac{3}{2}(\frac{1}{2}^{-})$ & 0.0927	& -1.7796 & 6.516 & 03.68 \\
&&$\frac{1}{2}(\frac{3}{2}^{-})$ &0.1306	& -4.0935 & 6.513 & 02.61 \\
&&$\frac{3}{2}(\frac{3}{2}^{-})$ & 0.0996	& -2.0467 & 6.515 & 03.43 \\
\\
\multirow{4}{*}{$[1$($\frac{3}{2}^{+}$)$]^{0}$-$[\frac{1}{2}$($1^{-}$)$]^{0}$}
&\multirow{4}{*}{$\Sigma_{s} - B^{*}$}
&$\frac{1}{2}(\frac{1}{2}^{-})$ & 0.0014	& +0.0000 & 6.707 & 242.4 \\
&&$\frac{3}{2}(\frac{1}{2}^{-})$ &0.0871	& -1.5525 & 6.706 & 03.92 \\
&&$\frac{1}{2}(\frac{5}{2}^{-})$ & 0.1169	& -3.3143 & 6.704 & 02.92\\
&&$\frac{3}{2}(\frac{5}{2}^{-})$ & 0.1043	& -2.2036 & 6.705 & 03.27 \\
\\
\hline
\\

\multirow{4}{*}{$[1$($\frac{1}{2}^{+}$)$]^{0}$-$[\frac{1}{2}$($1^{-}$)$]^{0}$}
&\multirow{4}{*}{$\Sigma_{c} - B^{*}$}
&$\frac{1}{2}(\frac{1}{2}^{-})$ & 0.1507 & -4.4956 & 7.774 & 02.26 \\
&&$\frac{3}{2}(\frac{1}{2}^{-})$ & 0.0863 & -1.3202 & 7.777 & 03.96  \\
&&$\frac{1}{2}(\frac{3}{2}^{-})$ & 0.1253 & -3.3102 & 7.775 & 02.72  \\
&&$\frac{3}{2}(\frac{3}{2}^{-})$ & 0.0908 & -1.4599 & 7.777 & 03.76 \\

\\

\multirow{4}{*}{$[1$($\frac{3}{2}^{+}$)$]^{0}$-$[\frac{1}{2}$($1^{-}$)$]^{0}$}
&\multirow{4}{*}{$\Sigma_{c} - B^{*}$}
&$\frac{1}{2}(\frac{1}{2}^{-})$ & 0.2219 & -7.1896 & 7.8368 & 01.53 \\
&&$\frac{3}{2}(\frac{1}{2}^{-})$ & 0.0822	& -1.1899 & 7.842 & 04.15\\
&&$\frac{1}{2}(\frac{5}{2}^{-})$ & 0.1144	& -2.7919 & 7.841 & 02.98\\
&&$\frac{3}{2}(\frac{5}{2}^{-})$ &0.0938	& -1.5453 & 7.842 & 03.64  \\
\\
\hline
\\

\multirow{4}{*}{$[1$($\frac{1}{2}^{+}$)$]^{0}$-$[\frac{1}{2}$($1^{-}$)$]^{0}$}
&\multirow{4}{*}{$\Sigma_{b} - B^{*}$}
&$\frac{1}{2}(\frac{1}{2}^{-})$  & 0.1279	& -2.9231 & 11.135 & 02.67 \\
&&$\frac{3}{2}(\frac{1}{2}^{-})$ & 0.0786	& -0.9082 & 11.137 & 04.34 \\
&&$\frac{1}{2}(\frac{3}{2}^{-})$ & 0.1177	& -2.5259 & 11.135 & 02.90  \\
&&$\frac{3}{2}(\frac{3}{2}^{-})$ & 0.0805	& -0.9545 & 11.137 & 04.24 \\
\\
\multirow{4}{*}{$[1$($\frac{3}{2}^{+}$)$]^{0}$-$[\frac{1}{2}$($1^{-}$)$]^{0}$}
&\multirow{4}{*}{$\Sigma_{b} - B^{*}$}
&$\frac{1}{2}(\frac{1}{2}^{-})$ & 0.1418	& -3.4602 & 11.153 & 02.41 \\
&&$\frac{3}{2}(\frac{1}{2}^{-})$ & 0.0768	& -0.8649 & 11.156 & 04.45 \\
&&$\frac{1}{2}(\frac{5}{2}^{-})$ & 0.1122	& -2.3129 & 11.154 & 03.04 \\
&&$\frac{3}{2}(\frac{5}{2}^{-})$ & 0.0819	& -0.9863 & 11.156 & 04.17 \\
\\
\hline
\hline
\end{tabular}
}
\end{center}
\end{table}
\subsubsection*{$\Sigma_{s,c,b}-K^{*}$, $\Sigma_{s,c,b}-D^{*}$ and $\Sigma_{s,c,b}-B^{*}$ :-}
The bound states of the di-hadronic systems with a baryon $\Sigma$ and a mesons $K^{*}$, $D^{*}$ and $B^{*}$  are calculated,  and the results are tabulated in the Table-\ref{sigma-k* mass spectra}, \ref{sigma-D* mass spectra} and \ref{sigma-B* mass spectra}. The strength and contribution of the s-wave OBE potential is the delicate cancellation of the individual meson exchange, hence, the Yukawa-like screen potential shows large impact on the net effective interaction potential, where this potential is sensitive to the parameter c (which is fixed at c=0.0686 GeV) and the value of residual coupling constant $k_{mol}$ (which is tabulated in Table-\ref{kmol and threshold mass}).  Thus, the results are found sensitive to the Yukawa-like screen potential.

The bound states of $\Sigma_{s,c,b}-K^{*}$ are found in (I,S) = (1/2, 3/2), (1/2, 5/2) and (3/2, 1/2)  channels. In the relatively heavier system $\Sigma_{c,b}-K^{*}$, the channels (3/2, 3/2) and (3/2, 5/2) are also found bound, even, in the $\Sigma_{b}-K^{*}$ system the channel (1/2, 1/2) is obtained bound state. The binding energy are appeared in the range from 0.2 MeV to 6 MeV.
In the $\Sigma_{s,c}-D^{*}$ systems all isospin-spin channels are emerged as bound state except (I,S)=(1/2, 1/2) channel, whereas in the $\Sigma_{b}-D^{*}$ systems all channel are found bound. In the case of the $\Sigma_{s,c,b}-D^{*}$  systems, the binding energy are appeared around  0.7 MeV to 7 MeV. On the other hand, the relatively heavier systems $\Sigma_{s,c,b}-B^{*}$,  all isospin-spin channels are appeared as a bound state and the binding energy are found nearly about 0.8 MeV to 7 MeV.

The meson-baryon molecular systems have also been studied  by others \cite{Wu-prc2011-84,Rui-Chen-npa2016-954-penta,Rui-Chen-cpc2017-41-penta,Azizi-prd2017-96,Azizi-prd2017-95,Shen-arxiv2017}. In Ref. \cite{Rui-Chen-npa2016-954-penta}, authors were predicted the bound states of molecular pentaquark within one pion exchange framework for possible ($I, J^{P}$) combination. For set of          values of regularization parameter, they were found binding energy between 1 MeV to 9 MeV. Our results are in agreement with results of Ref. \cite{Rui-Chen-npa2016-954-penta}. In Ref.\cite{Rui-Chen-cpc2017-41-penta}, within one boson exchange scheme, the strange hidden-charm pentaquarks was investigated and the binding energy of bound states were predicted about 0.5 MeV to 15 MeV. Jia-Jun Wu et. al. \cite{Wu-prc2011-84} predicted relatively narrow and dynamically generated meson-baryon resonance near 4.3 GeV within the coupled-channel unitary approach with the local hidden gauge formalism. Jia-Jun Wu et. al. \cite{Wu-prc2011-84} studied the coupled channel interaction of $\overline{D}\Sigma_{c}-\overline{D}\Lambda_{c}$, $\overline{D}^{*}\Sigma_{c}-\overline{D}^{*}\Lambda_{c}$ channels. We have not attempted the channel coupling in the present study. However, we obtained bound states near 4.46 GeV and 4.52 GeV in $\Sigma_{c}D^{*}$ and $\Sigma_{c}^{*}D^{*}$ system, respectively. Whereas, the bound states are appeared near 4.47 GeV and 4.66 GeV in        $\Xi_{c}D^{*}$ and $\Xi_{c}^{*}D^{*}$. Aziz et. al. \cite{Azizi-prd2017-95} studied  $P_{c}(4380)$ and $ P_{c}(4450)$ states as pentaquark states in the molecular picture with QCD sum rules,  moreover, they have predicted hidden bottom pentaquark states in \cite{Azizi-prd2017-96} as possible partner of these states.  These two new states as $P_{c}(4380)$ and $ P_{c}(4450)$ with the preferred $J^{P}$ assignments are of opposite parity, with one state having spin $\frac{3}{2}$ and the other $\frac{5}{2}$, were reported in 2015 by LHCb-collaboration \cite{Aaij-prl2015-115-pentaquark}.  In the present formalism  $ P_{c}(4450)$ is identified as $\Sigma_{c}-D^{*}$ molecular pentaquark with (I,S)=($\frac{1}{2}$,$\frac{3}{2}$) with negative parity whereas the  $P_{c}(4380)$ is not predicted within present study. The threshold of the $\Sigma_{c}-D^{*}$ is 4460.72 MeV, if   $P_{c}(4450)$ with mass $4449.8\pm1.7$ MeV and width $39\pm5\pm8$ MeV is a $\Sigma_{c}-D^{*}$ bound state molecule then it required binding energy approximately 10.92 MeV below the threshold. On the other hand, $P_{c}(4380)$ is required deep binding of approximately 80 MeV for molecular structure. Hence, $P_{c}(4450)$ is naturally look like loosely bound molecular candidate while $P_{c}(4380)$ with deep binding and large width is very unlikely to be molecular nature. The present formalism predict $ P_{c}(4450)$ as a $\Sigma_{c}-D^{*}$ bound state with $I(J^{P})=\frac{1}{2}(\frac{3}{2}^{-})$. 

\begin{table}[t]
\begin{center}
\caption{Mass spectra of meson-baryon (molecular pentaquark) $\Xi_{s,c,b}-K^{*}$ molecules. Masses of the meson and baryon are taken from PDG \cite{Patrignani-PDG2016} which are also listed in Table-\ref{HadronMaases-PDG}. Here, $\mu$ is variational parameter. I (isospin), G (G-parity), J (total angular momentum), Q (charge) and P (parity) are quantum numbers of the respective meson and baryon.}
\label{Xi-K* mass spectra}
\scalebox{0.8}{
\begin{tabular}{cccccccc}
\hline
\\
$[I_{1}(J_{1}^{P_{1}})]^{Q_{1}}$ - $[I_{2}(J_{2}^{P_{2}})]^{Q_{2}}$&System & $I(J^{P})$ & $\mu$ & B.E. & Mass & $\sqrt{r^{2}}$ \\
& & & GeV & MeV & MeV & fm \\
\hline
\hline
\\
\multirow{4}{*}{$[\frac{1}{2}$($\frac{1}{2}^{+}$)$]^{0}$-$[\frac{1}{2}$($1^{-}$)$]^{0}$}
&\multirow{4}{*}{$\Xi_{s} - K^{*}$}
&$0(\frac{1}{2}^{-})$ & 0.0063 &  +0.0027  &  2.210 &  54.54 \\
&&$1(\frac{1}{2}^{-})$ & 0.0596 &  -0.3899  &  2.210 &  05.73 \\
&&$0(\frac{3}{2}^{-})$ & 0.0994 &  -1.9759  &  2.208 &  03.43 \\
&&$1(\frac{3}{2}^{-})$ & 0.0764 &  -0.7226  &  2.210 &  04.47 \\
\\
\multirow{4}{*}{$[\frac{1}{2}$($\frac{3}{2}^{+}$)$]^{0}$-$[\frac{1}{2}$($1^{-}$)$]^{0}$}
&\multirow{4}{*}{$\Xi_{s} - K^{*}$}
&$0(\frac{1}{2}^{-})$ & 0.0057 &  +0.0019  &  2.427 &  59.95 \\
&&$1(\frac{1}{2}^{-})$ & 0.0535 &  -0.3139  &  2.427 &  06.39 \\
&&$0(\frac{5}{2}^{-})$ & 0.0732 &  -1.0271  &  2.426 &  04.66 \\
&&$1(\frac{5}{2}^{-})$ & 0.0947 &  -1.2009  &  2.426 &  03.60 \\
\\
\hline
\\

\multirow{4}{*}{$[\frac{1}{2}$($\frac{1}{2}^{+}$)$]^{0}$-$[\frac{1}{2}$($1^{-}$)$]^{0}$}
&\multirow{4}{*}{$\Xi_{c} - K^{*}$}
&$0(\frac{1}{2}^{-})$ & 0.0038 &  +0.0006 &  3.366 &  90.93 \\
&&$1(\frac{1}{2}^{-})$& 0.0737 &  -0.8884 &  3.365 &  04.63 \\
&&$0(\frac{3}{2}^{-})$ & 0.1306 &  -4.0298 &  3.362 &  02.61 \\
&&$1(\frac{3}{2}^{-})$ & 0.0824 &  -1.1341 &  3.365 &  04.14 \\
\\
\multirow{4}{*}{$[\frac{1}{2}$($\frac{3}{2}^{+}$)$]^{0}$-$[\frac{1}{2}$($1^{-}$)$]^{0}$}
&\multirow{4}{*}{$\Xi_{c} - K^{*}$}
&$0(\frac{1}{2}^{-})$ & 0.0037 &  +0.0006 &  3.550 &  92.80 \\
&&$1(\frac{1}{2}^{-})$ & 0.0683 &  -0.7507 &  3.550 &  05.00 \\
&&$0(\frac{5}{2}^{-})$ & 0.1027 &  -2.5459 &  3.548 &  03.32 \\
&&$1(\frac{5}{2}^{-})$ & 0.0891 &  -1.3392 &  3.549 &  03.83 \\
\\
\hline
\\
\multirow{4}{*}{$[\frac{1}{2}$($\frac{1}{2}^{+}$)$]^{0}$-$[\frac{1}{2}$($1^{-}$)$]^{0}$}
&\multirow{4}{*}{$\Xi_{b} - K^{*}$}
&$0(\frac{1}{2}^{-})$ & 0.2133 &  -9.5451 &  6.674 &  01.60 \\
&&$1(\frac{1}{2}^{-})$ & 0.0843 &  -1.3986 &  6.682 &  04.05\\
&&$0(\frac{3}{2}^{-})$ & 0.1608 &  -6.4365 &  6.677 &  02.12 \\
&&$1(\frac{3}{2}^{-})$ & 0.0868 &  -1.4863 &  6.682 &  03.93 \\
\\
\hline
\hline
\end{tabular}
}
\end{center}
\end{table} 
      
\begin{table}[t]
\begin{center}
\caption{Mass spectra of meson-baryon (molecular pentaquark) $\Xi_{s,c,b}-D^{*}$ molecules. Masses of the meson and baryon are taken from PDG \cite{Patrignani-PDG2016} which are also listed in Table-\ref{HadronMaases-PDG}. Here, $\mu$ is variational parameter. I (isospin), G (G-parity), J (total angular momentum), Q (charge) and P (parity) are quantum numbers of the respective meson and baryon.}
\label{Xi-D* mass spectra}
\scalebox{0.8}{
\begin{tabular}{cccccccc}
\hline
\\
$[I_{1}(J_{1}^{P_{1}})]^{Q_{1}}$ - $[I_{2}(J_{2}^{P_{2}})]^{Q_{2}}$&System & $I(J^{P})$ & $\mu$ & B.E. & Mass & $\sqrt{r^{2}}$ \\
& & & GeV & MeV & MeV & fm \\
\hline
\hline
\\
\multirow{4}{*}{$[\frac{1}{2}$($\frac{1}{2}^{+}$)$]^{0}$-$[\frac{1}{2}$($1^{-}$)$]^{0}$}
&\multirow{4}{*}{$\Xi_{s} - D^{*}$}
&$0(\frac{1}{2}^{-})$ & 0.0027 & +0.0002 & 3.321 & 128.7 \\
&&$1(\frac{1}{2}^{-})$ & 0.0744 & -0.9313 & 3.320 & 04.59  \\
&&$0(\frac{3}{2}^{-})$ &0.1333 & -4.1971 & 3.317 & 02.56  \\
&&$1(\frac{3}{2}^{-})$ & 0.0837 & -1.1954 & 3.320 & 04.08 \\
\\
\multirow{4}{*}{$[\frac{1}{2}$($\frac{3}{2}^{+}$)$]^{0}$-$[\frac{1}{2}$($1^{-}$)$]^{0}$}
&\multirow{4}{*}{$\Xi_{s} - D^{*}$}
&$0(\frac{1}{2}^{-})$ & 0.0023 & +0.0001 & 3.538 & 148.4 \\
&&$1(\frac{1}{2}^{-})$ & 0.0685 & -0.7754 & 3.537 & 04.98 \\
&&$0(\frac{5}{2}^{-})$ & 0.1038 & -2.6032 & 3.536 & 03.29 \\
&&$1(\frac{5}{2}^{-})$ & 0.0903 & -1.3857 & 3.537 & 03.78  \\
\\
\hline
\\
\multirow{4}{*}{$[\frac{1}{2}$($\frac{1}{2}^{+}$)$]^{0}$-$[\frac{1}{2}$($1^{-}$)$]^{0}$}
&\multirow{4}{*}{$\Xi_{c} - D^{*}$}
&$0(\frac{1}{2}^{-})$ & 0.0015 & +0.0000 & 4.477 & 232.8 \\
&&$1(\frac{1}{2}^{-})$ & 0.0744 & -0.8932 & 4.476 & 04.59 \\
&&$0(\frac{3}{2}^{-})$ & 0.1438 & -4.5849 & 4.473 & 02.37 \\
&&$1(\frac{3}{2}^{-})$ & 0.0795 & -1.0254 & 4.476 & 04.30 \\
\\
\multirow{4}{*}{$[\frac{1}{2}$($\frac{3}{2}^{+}$)$]^{0}$-$[\frac{1}{2}$($1^{-}$)$]^{0}$}
&\multirow{4}{*}{$\Xi_{c} - D^{*}$}
&$0(\frac{1}{2}^{-})$ & 0.0014 & +0.0000 & 4.661 & 242.9 \\
&&$1(\frac{1}{2}^{-})$ & 0.0703 & -0.7859 & 4.661 & 04.85 \\
&&$0(\frac{5}{2}^{-})$ & 0.1182 & -3.2173 & 4.658 & 02.89 \\
&&$1(\frac{5}{2}^{-})$ & 0.0826 & -1.0998 & 4.660 & 04.13 \\
\\
\hline
\\
	
\multirow{4}{*}{$[\frac{1}{2}$($\frac{1}{2}^{+}$)$]^{0}$-$[\frac{1}{2}$($1^{-}$)$]^{0}$}
&\multirow{4}{*}{$\Xi_{b} - D^{*}$}
&$0(\frac{1}{2}^{-})$ & 0.1989 & -7.6632 & 7.787 & 01.71 \\
&&$1(\frac{1}{2}^{-})$ & 0.0781 & -1.0319 & 7.793 & 04.37 \\
&&$0(\frac{3}{2}^{-})$ & 0.1599 & -5.6126 & 7.789 & 02.13 \\
&&$1(\frac{3}{2}^{-})$ & 0.0798 & -1.0799 & 7.793 & 04.28 \\
\\
\hline
\hline
\end{tabular}
}
\end{center}
\end{table}				
\begin{table}[]
\caption{Mass spectra of meson-baryon (molecular pentaquark) $\Xi_{s,c,b}-B^{*}$ molecules. Masses of the meson and baryon are taken from PDG \cite{Patrignani-PDG2016} which are also listed in Table-\ref{HadronMaases-PDG}. Here, $\mu$ is variational parameter. I (isospin), G (G-parity), J (total angular momentum), Q (charge) and P (parity) are quantum numbers of the respective meson and baryon.}
\label{Xi-B* mass spectra}
\scalebox{0.8}{
\begin{tabular}{cccccccc}
\hline
\\
$[I_{1}(J_{1}^{P_{1}})]^{Q_{1}}$ - $[I_{2}(J_{2}^{P_{2}})]^{Q_{2}}$&System & $I(J^{P})$ & $\mu$ & B.E. & Mass & $\sqrt{r^{2}}$ \\
& & & GeV & MeV & MeV & fm \\
\hline
\hline
\\
\multirow{4}{*}{$[\frac{1}{2}$($\frac{1}{2}^{+}$)$]^{0}$-$[\frac{1}{2}$($1^{-}$)$]^{0}$}
&\multirow{4}{*}{$\Xi_{s} - B^{*}$}
&$0(\frac{1}{2}^{-})$ & 0.2238 & -9.6429 & 6.630 & 01.52 \\
&&$1(\frac{1}{2}^{-})$ & 0.0824 & -1.2782 & 6.638 & 04.14 \\
&&$0(\frac{3}{2}^{-})$ & 0.1619 & -6.2356 & 6.633 & 02.10  \\
&&$1(\frac{3}{2}^{-})$ & 0.0849 & -1.3602 & 6.638 & 04.02  \\
\\
\multirow{4}{*}{$[\frac{1}{2}$($\frac{3}{2}^{+}$)$]^{0}$-$[\frac{1}{2}$($1^{-}$)$]^{0}$}
&\multirow{4}{*}{$\Xi_{s} - B^{*}$}
&$0(\frac{1}{2}^{-})$ & 0.0012 & +0.0000 & 6.857 & 287.6 \\
&&$1(\frac{1}{2}^{-})$ & 0.0789 & -1.1323 & 6.855 & 04.33 \\
&&$0(\frac{5}{2}^{-})$ & 0.1434 & -4.9749 & 6.852 & 02.38 \\
&&$1(\frac{5}{2}^{-})$ & 0.0851 & -1.3221 & 6.855 & 04.01 \\
\\
\hline
\\
\multirow{4}{*}{$[\frac{1}{2}$($\frac{1}{2}^{+}$)$]^{0}$-$[\frac{1}{2}$($1^{-}$)$]^{0}$}
&\multirow{4}{*}{$\Xi_{c} - B^{*}$}
&$0(\frac{1}{2}^{-})$ & 0.1959 & -7.2349 & 7.788 & 01.74 \\
&&$1(\frac{1}{2}^{-})$ & 0.0759 & -0.9353 & 7.795 & 04.49\\
&&$0(\frac{3}{2}^{-})$ & 0.1574 & -5.2851 & 7.790 & 02.17\\
&&$1(\frac{3}{2}^{-})$ & 0.0777 & -0.9789 & 7.795 & 04.39 \\
\\
\multirow{4}{*}{$[\frac{1}{2}$($\frac{3}{2}^{+}$)$]^{0}$-$[\frac{1}{2}$($1^{-}$)$]^{0}$}
&\multirow{4}{*}{$\Xi_{c} - B^{*}$}
&$0(\frac{1}{2}^{-})$ & 0.0006 & $+2.3\times10^{-6}$ & 7.980 & 569.9 \\
&&$1(\frac{1}{2}^{-})$ & 0.0738 & -0.8626 & 7.979 & 04.63 \\
&&$0(\frac{5}{2}^{-})$ & 0.1425 & -4.4301 & 7.975 & 02.39 \\
&&$1(\frac{5}{2}^{-})$ & 0.0779 & -0.9682 & 7.979 & 04.38 \\
\\
\hline
\\     
\multirow{4}{*}{$[\frac{1}{2}$($\frac{1}{2}^{+}$)$]^{0}$-$[\frac{1}{2}$($1^{-}$)$]^{0}$}
&\multirow{4}{*}{$\Xi_{b} - B^{*}$}
&$0(\frac{1}{2}^{-})$ & 0.1641 & -4.8455 & 11.108 & 02.08 \\
&&$1(\frac{1}{2}^{-})$ & 0.0681 & -0.6105 & 11.112 & 05.02\\
&&$0(\frac{3}{2}^{-})$& 0.1492 & -4.1776 & 11.109 & 02.29 \\
&&$1(\frac{3}{2}^{-})$& 0.0688 & -0.6244 & 11.112 & 04.96 \\
\\
\hline
\hline
\end{tabular}
}
\end{table}

\subsubsection*{$\Xi_{s,c,b}-K^{*}$, $\Xi_{s,c,b}-D^{*}$ and $\Xi_{s,c,b}-B^{*}$ :-}
The bound states of the di-hadronic systems with a Baryon $\Xi$ and a mesons $K^{*}$, $D^{*}$ and $B^{*}$  are calculated,  and the results are tabulated in the Table-\ref{Xi-K* mass spectra},\ref{Xi-D* mass spectra} and \ref{Xi-B* mass spectra}. The bound states of $\Xi_{s,c,b}-K^{*}$ are  appeared in (I,S) = (0, 3/2), (0, 5/2), (1, 5/2), (1, 3/2), (1, 1/2) channels.  While, the (0, 1/2) channel obtained as a bound state only for  $\Xi_{b}-K^{*}$ system. The binding energies are appeared around 0.3 MeV to 9 MeV. Similarly, for the $\Xi_{s,c,b}-D^{*}$ systems, the (0, 1/2) channel is bound in only for  $\Xi_{b}-D^{*}$ case, whereas all other channels are found bound state in all cases. The binding energy are found about 0.7 MeV to 7 MeV.  In the case of relatively heavier system $\Xi_{s,c,b}-B^{*}$, the (I,S)=(0, 1/2) is unbound for $\Xi_{s,c}^{*}-B^{*}$ systems, whereas, all other isospin-spin channels are found bound states in the $\Xi_{s,c,b}-B^{*}$ systems. The binding energies are appeared around 0.6 MeV to 9 MeV. Rui-Chen et. al.\cite{Rui-Chen-cpc2017-41-penta} have investigated strange hidden-charm pentaquarks states within one boson exchange scheme, and predicted $\Xi_{c}^{'}\overline{D}^{*}$ state with I($J^{P}$) = 0($\frac{1}{2}^{-}$) and $\Xi_{c}^{*}\overline{D}^{*}$ state with 0($\frac{1}{2}^{-}$) and 0($\frac{3}{2}^{-}$). They have predicted binding energies in the range from 0.5 to 15 MeV. Our results of binding energies for these isospin-spin channels are in agreement with reported in Ref. \cite{Rui-Chen-cpc2017-41-penta}. 
These results are shown very interesting near threshold bound states possibilities.
\section{Results of $a_{s}$ and $r_{e}$ from compositeness theorem}
\label{Results of as and re from compositeness theorem}
In the Sixties, Weinberg \cite{Weinberg-RhysRev1965} suggested in a sophisticated way that the deuteron were a composite particle. In his novel work, he tried to show an elegant  model-independent  way to identify whether a particle is in a bare elementary state or in a composite state. The conclusion was based on a generalization of Levinson's theorem which gives the formulas for scattering length $a_{s}$ and effective range $r_{e}$ in terms of Z, where Z is the ``field renormalization" constant \cite{Weinberg-RhysRev1965}, 
\begin{eqnarray}
\label{as and re equation}
&&a_{s}=\left[2(1-Z)/(2-Z)\right]R+{\cal{O}}(1/\beta) \nonumber \\
&&r_{e}=\left[-Z/(1-Z)\right]R+{\cal{O}}(1/\beta)
\end{eqnarray}
where R is the size of the molecular or composite state and is determined by 
$R\equiv \frac{1}{\sqrt{2\mu \epsilon}}$, here, $\epsilon$ is the binding energy and $\mu$ is the reduced mass of the composite system (note that we chose binding energy $\epsilon$ positive for Eq.(\ref{as and re equation}), the bound state is located at $\epsilon$ = -$\epsilon$).  The ${\cal{O}}(1/\beta)$ is the range correction and $\beta$ is the inverse range of the force and could be calculated if one know the information of the interaction and it is expected to be of the order of magnitude of the inverse of the mass of the exchange particle, in some extent, it is expected to be $m_{\pi}^{-1}$ $\simeq$ 1.41 fm. 
In order to determine the state of the particle as in a bare elementary or in a composite state, Weinberg argued that the renormalization constant Z take the value  $0 \leq Z \leq 1$. If Z=0 then the particle is in a pure composite state while for Z=1 it becomes a purely elementary.

for Z=0  (deuteron as a composite particle), Eq(1) becomes $a_{s}=R$ and $r_{e}={\cal{O}}(1/\beta)$ which is in agreement with the experimental vales : $a_{s}=+5.41$ fm, $r_{e}=+1.75$ fm.  
for deuteron binding energy $\epsilon = 2.22457 MeV \Rightarrow \sqrt{2\mu \epsilon} = 45.7 MeV = 0.23$ $fm^{-1}$.
In contrast, if the deuteron has a significant probability Z ($>0.2$) of being found in an elementary (confined) state then $a_{s}$ would be less than R, and $r_{e}$ would be large and negative which would be in contradict with experimental values.

The results of the scattering length ($a_{s}$) and effective range ($r_{e}$) are obtained by using Eq(\ref{as and re equation}).The results of $a_{s}$ and $r_{e}$ for meson-baryon  systems, by using the calculated binding energy  are shown in the Table-\ref{as and re for meson-baryon states}. 

The state $P_{c}(4450)$ for which the calculated binding energy is underestimated about few MeV, while with the expected binding energy (10.92 MeV) the effective range gain negative value from Z=0.6. The expected binding energy is almost close to its natural energy scale which is about 9 MeV if it has $\Sigma_{c}D^{*}$ in its substructure. The large scattering length and positive $r_{e}$ for Z$\rightarrow$0 and binding energy near to expected natural energy scale for $P_{c}(4450)$ are indicating molecular structure of the state.

\begin{table*}[]
\begin{center}
\caption{\small The scattering length $a_{s}$ and effective range $r_{e}$ are calculated for meson-baryon systems by using Eq.(\ref{as and re equation}) for different values of renormalization constant Z.  The range correction ${\cal{O}}(1/\beta)$ is considered as $m_{\pi}^{-1}$. The values of $a_{s}$ and $r_{e}$ are in fm. Binding energies are taken from the results of chapter-\ref{Meson-Baryon and Di-baryonic molecules}}
\label{as and re for meson-baryon states}
\scalebox{1}{
\begin{tabular}{cccccccccccccccc}
\hline
I$(J^{P})$ & State   &\multicolumn{2}{c}{Z=0} &\multicolumn{2}{c}{Z=0.2} &\multicolumn{2}{c}{Z=0.4} &\multicolumn{2}{c}{Z=0.5} &\multicolumn{2}{c}{Z=0.6} &\multicolumn{2}{c}{Z=0.9} &\multicolumn{2}{c}{Z=1} \\
\cline{3-16}
 & &$a_{s}$ &$r_{e}$ &$a_{s}$ &$r_{e}$& $a_{s}$ &$r_{e}$ &$a_{s}$ &$r_{e}$ & $a_{s}$ &$r_{e}$ & $a_{s}$ &$r_{e}$ & $a_{s}$ &$r_{e}$\\
\hline
\hline
\\
$\frac{3}{2}(\frac{1}{2}^{-})$  & $\Sigma_{s}K^{*}$   & 11.6  &  1.46  &  10.47  &  -1.07  &  9.06  &  -5.3  &  8.22  &  -8.67  &  7.25  &  -13.74  &  3.3  &  -89.76  &  1.46  &   - \\
$\frac{1}{2}(\frac{3}{2}^{-})$ & $\Sigma_{s}K^{*}$  &  7.66  &  1.46  &  6.97  &  -0.09  &  6.11  &  -2.67  &  5.59  &  -4.74  &  5  &  -7.83  &  2.59  &  -54.31  &  1.46  &   - \\
$\frac{3}{2}(\frac{3}{2}^{-})$ & $\Sigma_{s}K^{*}$ &   6.36  &  1.46  &  5.81  &  0.24  &  5.13  &  -1.8  &  4.72  &  -3.43  &  4.26  &  -5.88  &  2.35  &  -42.59  &  1.46  & -   \\
$\frac{1}{2}(\frac{5}{2}^{-}) $  & $\Sigma_{s}^{*}K^{*}$  &  10.24  &  1.46  &  9.26  &  -0.73  &  8.04  &  -4.39  &  7.31  &  -7.31  &  6.48  &  -11.7  &  3.06  &  -77.52  &  1.46  &  -  \\
  \\  
$\frac{3}{2}(\frac{1}{2}^{-})$  &  $\Sigma_{s}D^{*}$ &   6.3  &  1.46  &  5.76  &  0.25  &  5.09  &  -1.76  &  4.68  &  -3.37  &  4.22  &  -5.79  &  2.34  &  -42.04  &  1.46  &   - \\
$\frac{1}{2}(\frac{3}{2}^{-}) $  & $\Sigma_{s}D^{*}$  &  4.66  &  1.46  &  4.31  &  0.66  &  3.86  &  -0.67  &  3.6  &  -1.74  &  3.29  &  -3.34  &  2.04  &  -27.34  &  1.46  &  -  \\
$\frac{3}{2}(\frac{3}{2}^{-}) $  & $\Sigma_{s}D^{*}$  &  5.09  &  1.46  &  4.69  &  0.55  &  4.18  &  -0.96  &  3.88  &  -2.17  &  3.54  &  -3.98  &  2.12  &  -31.2  &  1.46  &   - \\
$\frac{1}{2}(\frac{5}{2}^{-}) $  & $\Sigma_{s}^{*}D^{*}$  &  5.43  &  1.46  &  4.99  &  0.47  &  4.44  &  -1.19  &  4.11  &  -2.51  &  3.73  &  -4.5  &  2.18  &  -34.29  &  1.46  & -   \\
$\frac{3}{2}(\frac{5}{2}^{-})$  &  $\Sigma_{s}^{*}D^{*}$ &   4.13  &  1.46  &  3.83  &  0.8  &  3.46  &  -0.32  &  3.24  &  -1.21  &  2.99  &  -2.54  &  1.95  &  -22.55  &  1.46  &  -  \\
\\
$\frac{1}{2}(\frac{1}{2}^{-}) $  & $\Sigma_{s}B^{*}$  &  3.24  &  1.46  &  3.04  &  1.02  &  2.8  &  0.28  &  2.65  &  -0.32  &  2.48  &  -1.21  &  1.79  &  -14.55  &  1.46  & -   \\
$\frac{3}{2}(\frac{1}{2}^{-}) $  & $\Sigma_{s}B^{*}$ &   4.81  &  1.46  &  4.44  &  0.62  &  3.98  &  -0.77  &  3.7  &  -1.89  &  3.38  &  -3.56  &  2.07  &  -28.7  &  1.46  &  -  \\
$\frac{1}{2}(\frac{3}{2}^{-}) $  & $\Sigma_{s}B^{*}$ &   3.67  &  1.46  &  3.43  &  0.91  &  3.12  &  -0.01  &  2.93  &  -0.75  &  2.72  &  -1.85  &  1.86  &  -18.42  &  1.46  &   - \\
$\frac{3}{2}(\frac{3}{2}^{-})$  &  $\Sigma_{s}B^{*}$  &  4.59  &  1.46  &  4.24  &  0.68  &  3.81  &  -0.62  &  3.54  &  -1.66  &  3.25  &  -3.22  &  2.03  &  -26.66  &  1.46  &  -  \\
$\frac{1}{2}(\frac{5}{2}^{-})$  &  $\Sigma_{s}^{*}B^{*}$ &   3.78  &  1.46  &  3.52  &  0.88  &  3.2  &  -0.08  &  3  &  -0.85  &  2.78  &  -2.01  &  1.88  &  -19.36  &  1.46  &   - \\
$\frac{3}{2}(\frac{5}{2}^{-})$  & $\Sigma_{s}^{*}B^{*}$ & 4.3  &  1.46  &  3.98  &  0.75  &  3.59  &  -0.43  &  3.35  &  -1.38  &  3.08  &  -2.79  &  1.98  &  -24.07  &  1.46  & -   \\

\\
\hline
\\
$\frac{3}{2}(\frac{1}{2}^{-})$  &  $\Sigma_{c}K^{*}$  &  6.67  &  1.46  &  6.09  &  0.16  &  5.37  &  -2.01  &  4.93  &  -3.74  &  4.44  &  -6.35  &  2.41  &  -45.39  &  1.46  &  -  \\
$\frac{1}{2}(\frac{3}{2}^{-})$  &  $\Sigma_{c}K^{*}$  &  4.89  &  1.46  &  4.51  &  0.6  &  4.03  &  -0.83  &  3.75  &  -1.97  &  3.42  &  -3.68  &  2.09  &  -29.41  &  1.46  &  -  \\
$\frac{3}{2}(\frac{3}{2}^{-})$  &  $\Sigma_{c}K^{*}$  &  5.47  &  1.46  &  5.03  &  0.46  &  4.47  &  -1.21  &  4.14  &  -2.55  &  3.75  &  -4.56  &  2.19  &  -34.65  &  1.46  & -   \\
$\frac{1}{2}(\frac{5}{2}^{-})$  &  $\Sigma_{c}^{*}K^{*}$  &  5.94  &  1.46  &  5.44  &  0.34  &  4.82  &  -1.52  &  4.45  &  -3.02  &  4.02  &  -5.26  &  2.28  &  -38.85  &  1.46  & -   \\
$\frac{3}{2}(\frac{5}{2}^{-})$  &  $\Sigma_{c}^{*}K^{*}$  &  4.57  &  1.46  &  4.22  &  0.69  &  3.79  &  -0.61  &  3.53  &  -1.64  &  3.24  &  -3.19  &  2.03  &  -26.47  &  1.46  &  -  \\
       \\                                                       
$\frac{3}{2}(\frac{1}{2}^{-})$  &  $\Sigma_{c}D^{*}$  &  5.33  &  1.46  &  4.9  &  0.5  &  4.36  &  -1.11  &  4.04  &  -2.4  &  3.67  &  -4.34  &  2.16  &  -33.32  &  1.46  &  -  \\
$\frac{1}{2}(\frac{3}{2}^{-})$  &  $\Sigma_{c}D^{*}$  &  3.95  &  1.46  &  3.67  &  0.84  &  3.33  &  -0.2  &  3.12  &  -1.03  &  2.88  &  -2.27  &  1.91  &  -20.93  &  1.46  &  -  \\
$\frac{3}{2}(\frac{3}{2}^{-})$  &  $\Sigma_{c}D^{*}$  &  4.79  &  1.46  &  4.42  &  0.63  &  3.96  &  -0.75  &  3.68  &  -1.86  &  3.36  &  -3.53  &  2.07  &  -28.46  &  1.46  &  -  \\
$\frac{1}{2}(\frac{5}{2}^{-})$  &  $\Sigma_{c}^{*}D^{*}$  &  4.46  &  1.46  &  4.13  &  0.71  &  3.71  &  -0.54  &  3.46  &  -1.54  &  3.17  &  -3.03  &  2.01  &  -25.51  &  1.46  &  -  \\
$\frac{3}{2}(\frac{5}{2}^{-})$  &  $\Sigma_{c}^{*}D^{*}$  &  4.4  &  1.46  &  4.07  &  0.73  &  3.67  &  -0.5  &  3.42  &  -1.48  &  3.14  &  -2.95  &  2  &  -24.99  &  1.46  &  -  \\
          \\                                                    
$\frac{1}{2}(\frac{1}{2}^{-})$  &  $\Sigma_{c}B^{*}$  &  3.07  &  1.46  &  2.89  &  1.06  &  2.67  &  0.39  &  2.53  &  -0.14  &  2.38  &  -0.95  &  1.75  &  -12.99  &  1.46  & -   \\
$\frac{3}{2}(\frac{1}{2}^{-})$  &  $\Sigma_{c}B^{*}$  &  4.42  &  1.46  &  4.1  &  0.72  &  3.68  &  -0.51  &  3.44  &  -1.5  &  3.16  &  -2.98  &  2  &  -25.21  &  1.46  &  -  \\
$\frac{1}{2}(\frac{3}{2}^{-})$  &  $\Sigma_{c}B^{*}$  &  3.33  &  1.46  &  3.13  &  0.99  &  2.87  &  0.21  &  2.71  &  -0.41  &  2.53  &  -1.34  &  1.8  &  -15.38  &  1.46  & -   \\
$\frac{3}{2}(\frac{3}{2}^{-})$  &  $\Sigma_{c}B^{*}$  &  4.28  &  1.46  &  3.97  &  0.76  &  3.58  &  -0.42  &  3.34  &  -1.36  &  3.07  &  -2.76  &  1.97  &  -23.9  &  1.46  &  -  \\
$\frac{1}{2}(\frac{5}{2}^{-})$  &  $\Sigma_{c}^{*}B^{*}$  &  3.48  &  1.46  &  3.26  &  0.96  &  2.98  &  0.12  &  2.81  &  -0.56  &  2.62  &  -1.57  &  1.83  &  -16.71  &  1.46  &   - \\
$\frac{3}{2}(\frac{5}{2}^{-})$  &  $\Sigma_{c}^{*}B^{*}$  &  4.18  &  1.46  &  3.87  &  0.78  &  3.5  &  -0.35  &  3.27  &  -1.25  &  3.01  &  -2.61  &  1.96  &  -22.97  &  1.46  &   - \\
\\
\hline
\\
$\frac{1}{2}(\frac{1}{2}^{-})$  &  $\Sigma_{b}K^{*}$  &  3.47  &  1.46  &  3.24  &  0.96  &  2.97  &  0.13  &  2.8  &  -0.54  &  2.61  &  -1.54  &  1.83  &  -16.58  &  1.46  &  -  \\
$\frac{3}{2}(\frac{1}{2}^{-})$  &  $\Sigma_{b}K^{*}$  &  5.12  &  1.46  &  4.71  &  0.55  &  4.2  &  -0.98  &  3.9  &  -2.2  &  3.55  &  -4.02  &  2.13  &  -31.45  &  1.46  &   - \\
$\frac{1}{2}(\frac{3}{2}^{-})$  &  $\Sigma_{b}K^{*}$  &  3.9  &  1.46  &  3.63  &  0.85  &  3.29  &  -0.16  &  3.09  &  -0.97  &  2.85  &  -2.19  &  1.9  &  -20.45  &  1.46  &   - \\
$\frac{3}{2}(\frac{3}{2}^{-})$  &  $\Sigma_{b}K^{*}$  &  4.89  &  1.46  &  4.51  &  0.61  &  4.03  &  -0.82  &  3.75  &  -1.96  &  3.42  &  -3.68  &  2.08  &  -29.38  &  1.46  &  -  \\
$\frac{1}{2}(\frac{5}{2}^{-})$  &  $\Sigma_{b}^{*}K^{*}$  &  4.15  &  1.46  &  3.85  &  0.79  &  3.47  &  -0.33  &  3.25  &  -1.22  &  3  &  -2.56  &  1.95  &  -22.69  &  1.46  &  -  \\
$\frac{3}{2}(\frac{5}{2}^{-})$  &  $\Sigma_{b}^{*}K^{*}$  &  4.73  &  1.46  &  4.37  &  0.64  &  3.91  &  -0.72  &  3.64  &  -1.81  &  3.33  &  -3.44  &  2.06  &  -27.95  &  1.46  &  -  \\
    \\                                                          
$\frac{1}{2}(\frac{1}{2}^{-})$  &  $\Sigma_{b}D^{*}$  &  3.11  &  1.46  &  2.93  &  1.05  &  2.7  &  0.36  &  2.56  &  -0.19  &  2.41  &  -1.02  &  1.76  &  -13.41  &  1.46  &  -  \\
$\frac{3}{2}(\frac{1}{2}^{-})$  &  $\Sigma_{b}D^{*}$  &  4.7  &  1.46  &  4.34  &  0.65  &  3.89  &  -0.69  &  3.62  &  -1.77  &  3.31  &  -3.39  &  2.05  &  -27.64  &  1.46  &  -  \\
$\frac{1}{2}(\frac{3}{2}^{-})$  &  $\Sigma_{b}D^{*}$  &  3.38  &  1.46  &  3.17  &  0.98  &  2.9  &  0.18  &  2.74  &  -0.46  &  2.56  &  -1.42  &  1.81  &  -15.83  &  1.46  &  -  \\
$\frac{3}{2}(\frac{3}{2}^{-})$  &  $\Sigma_{b}D^{*}$  &  4.33  &  1.46  &  4.01  &  0.75  &  3.61  &  -0.45  &  3.37  &  -1.4  &  3.1  &  -2.84  &  1.98  &  -24.33  &  1.46  &   - \\
$\frac{1}{2}(\frac{5}{2}^{-})$  &  $\Sigma_{b}^{*}D^{*}$  &  3.54  &  1.46  &  3.31  &  0.94  &  3.02  &  0.08  &  2.85  &  -0.62  &  2.65  &  -1.66  &  1.84  &  -17.26  &  1.46  &  -  \\
$\frac{3}{2}(\frac{5}{2}^{-})$  &  $\Sigma_{b}^{*}D^{*}$  &  4.23  &  1.46  &  3.92  &  0.77  &  3.54  &  -0.38  &  3.31  &  -1.3  &  3.04  &  -2.69  &  1.96  &  -23.44  &  1.46  &   - \\
   \hline  
   \hline
\end{tabular}
}
\end{center}
\end{table*}

\begin{table*}[]
\begin{center}
\addtocounter{table}{-1}
\caption{to be continued.. }
\label{as and re for meson-baryon states}
\scalebox{1}{
\begin{tabular}{cccccccccccccccc}
\hline
I$(J^{P})$ & State   &\multicolumn{2}{c}{Z=0} &\multicolumn{2}{c}{Z=0.2} &\multicolumn{2}{c}{Z=0.4} &\multicolumn{2}{c}{Z=0.5} &\multicolumn{2}{c}{Z=0.6} &\multicolumn{2}{c}{Z=0.9} &\multicolumn{2}{c}{Z=1} \\
\cline{3-16}
 & &$a_{s}$ &$r_{e}$ &$a_{s}$ &$r_{e}$& $a_{s}$ &$r_{e}$ &$a_{s}$ &$r_{e}$ & $a_{s}$ &$r_{e}$ & $a_{s}$ &$r_{e}$ & $a_{s}$ &$r_{e}$\\
\hline
\hline
\\                                                       
$\frac{1}{2}(\frac{1}{2}^{-})$  &  $\Sigma_{b}B^{*}$  &  3.01  &  1.46  &  2.84  &  1.07  &  2.62  &  0.43  &  2.49  &  -0.09  &  2.35  &  -0.86  &  1.74  &  -12.47  &  1.46  &   - \\
$\frac{3}{2}(\frac{1}{2}^{-})$  &  $\Sigma_{b}B^{*}$  &  4.24  &  1.46  &  3.93  &  0.77  &  3.54  &  -0.39  &  3.31  &  -1.32  &  3.05  &  -2.7  &  1.97  &  -23.53  &  1.46  &  -  \\
$\frac{1}{2}(\frac{3}{2}^{-})$  &  $\Sigma_{b}B^{*}$  &  3.13  &  1.46  &  2.94  &  1.05  &  2.71  &  0.35  &  2.57  &  -0.2  &  2.41  &  -1.04  &  1.76  &  -13.53  &  1.46  & -   \\
$\frac{3}{2}(\frac{3}{2}^{-})$  &  $\Sigma_{b}B^{*}$  &  4.17  &  1.46  &  3.87  &  0.78  &  3.49  &  -0.34  &  3.27  &  -1.25  &  3.01  &  -2.6  &  1.95  &  -22.92  &  1.46  &  -  \\
$\frac{1}{2}(\frac{5}{2}^{-})$  &  $\Sigma_{b}^{*}B^{*}$  &  3.2  &  1.46  &  3.01  &  1.03  &  2.77  &  0.3  &  2.62  &  -0.28  &  2.46  &  -1.15  &  1.78  &  -14.19  &  1.46  &  -  \\
$\frac{3}{2}(\frac{5}{2}^{-})$  &  $\Sigma_{b}^{*}B^{*}$  &  4.13  &  1.46  &  3.83  &  0.8  &  3.46  &  -0.31  &  3.24  &  -1.2  &  2.98  &  -2.53  &  1.95  &  -22.51  &  1.46  &   - \\
\\
\hline
\\
$1(\frac{1}{2}^{-})$  &  $\Xi_{s}K^{*}$  &  11.14  &  1.46  &  10.07  &  -0.96  &  8.72  &  -4.99  &  7.92  &  -8.22  &  6.99  &  -13.06  &  3.22  &  -85.67  &  1.46  &   - \\
$0(\frac{3}{2}^{-})$  &  $\Xi_{s}K^{*}$  &  5.76  &  1.46  &  5.28  &  0.39  &  4.69  &  -1.4  &  4.33  &  -2.84  &  3.92  &  -4.99  &  2.24  &  -37.24  &  1.46  &  -  \\
$1(\frac{3}{2}^{-})$  &  $\Xi_{s}K^{*}$  &  8.57  &  1.46  &  7.78  &  -0.32  &  6.8  &  -3.28  &  6.2  &  -5.65  &  5.53  &  -9.2  &  2.75  &  -62.54  &  1.46  &  -  \\
$0(\frac{5}{2}^{-})$  &  $\Xi_{s}K^{*}$  &  7.25  &  1.46  &  6.61  &  0.01  &  5.81  &  -2.4  &  5.32  &  -4.33  &  4.77  &  -7.22  &  2.51  &  -50.66  &  1.46  &  -  \\
$1(\frac{5}{2}^{-})$  &  $\Xi_{s}K^{*}$  &  6.82  &  1.46  &  6.22  &  0.12  &  5.48  &  -2.11  &  5.03  &  -3.89  &  4.52  &  -6.57  &  2.44  &  -46.74  &  1.46  &   - \\
\\                                                              
$1(\frac{1}{2}^{-})$  &  $\Xi_{s}D^{*}$  &  6.59  &  1.46  &  6.02  &  0.18  &  5.31  &  -1.96  &  4.88  &  -3.67  &  4.39  &  -6.23  &  2.39  &  -44.71  &  1.46  & -   \\
$0(\frac{3}{2}^{-})$  &  $\Xi_{s}D^{*}$  &  3.88  &  1.46  &  3.61  &  0.86  &  3.27  &  -0.15  &  3.07  &  -0.95  &  2.84  &  -2.16  &  1.9  &  -20.29  &  1.46  &  -  \\
$1(\frac{3}{2}^{-})$  &  $\Xi_{s}D^{*}$  &  5.99  &  1.46  &  5.49  &  0.33  &  4.86  &  -1.56  &  4.48  &  -3.07  &  4.05  &  -5.33  &  2.29  &  -39.29  &  1.46  &  -  \\
$0(\frac{5}{2}^{-})$  &  $\Xi_{s}D^{*}$  &  4.4  &  1.46  &  4.07  &  0.73  &  3.66  &  -0.49  &  3.42  &  -1.47  &  3.14  &  -2.94  &  2  &  -24.95  &  1.46  &  -  \\
$1(\frac{5}{2}^{-})$  &  $\Xi_{s}D^{*}$  &  5.48  &  1.46  &  5.04  &  0.46  &  4.48  &  -1.22  &  4.14  &  -2.56  &  3.76  &  -4.57  &  2.19  &  -34.73  &  1.46  & -   \\
        \\                                                      
$0(\frac{1}{2}^{-})$  &  $\Xi_{s}B^{*}$  &  2.85  &  1.46  &  2.69  &  1.12  &  2.5  &  0.54  &  2.38  &  0.08  &  2.25  &  -0.61  &  1.71  &  -10.99  &  1.46  &  -  \\
$1(\frac{1}{2}^{-})$  &  $\Xi_{s}B^{*}$  &  5.26  &  1.46  &  4.84  &  0.51  &  4.31  &  -1.07  &  4  &  -2.34  &  3.63  &  -4.24  &  2.15  &  -32.74  &  1.46  &    -\\
$0(\frac{3}{2}^{-})$  &  $\Xi_{s}B^{*}$  &  3.18  &  1.46  &  2.99  &  1.03  &  2.75  &  0.31  &  2.61  &  -0.26  &  2.45  &  -1.12  &  1.77  &  -14.02  &  1.46  &  -  \\
$1(\frac{3}{2}^{-})$  &  $\Xi_{s}B^{*}$  &  5.15  &  1.46  &  4.74  &  0.54  &  4.23  &  -0.99  &  3.92  &  -2.22  &  3.57  &  -4.06  &  2.13  &  -31.7  &  1.46  &  -  \\
$0(\frac{5}{2}^{-})$  &  $\Xi_{s}B^{*}$  &  3.28  &  1.46  &  3.07  &  1.01  &  2.82  &  0.25  &  2.67  &  -0.35  &  2.5  &  -1.26  &  1.79  &  -14.86  &  1.46  &   - \\
$1(\frac{5}{2}^{-})$  &  $\Xi_{s}B^{*}$  &  4.98  &  1.46  &  4.59  &  0.58  &  4.1  &  -0.88  &  3.81  &  -2.06  &  3.47  &  -3.82  &  2.1  &  -30.2  &  1.46  &  -  \\
\\
\hline
\\
$1(\frac{1}{2}^{-})$  &  $\Xi_{c}K^{*}$  &  7.24  &  1.46  &  6.59  &  0.02  &  5.79  &  -2.39  &  5.31  &  -4.31  &  4.76  &  -7.2  &  2.51  &  -50.5  &  1.46  &  - \\
$0(\frac{3}{2}^{-})$  &  $\Xi_{c}K^{*}$  &  4.17  &  1.46  &  3.87  &  0.78  &  3.49  &  -0.35  &  3.27  &  -1.25  &  3.01  &  -2.6  &  1.95  &  -22.93  &  1.46  & - \\
$1(\frac{3}{2}^{-})$  &  $\Xi_{c}K^{*}$  &  6.57  &  1.46  &  6  &  0.18  &  5.29  &  -1.94  &  4.87  &  -3.65  &  4.38  &  -6.2  &  2.39  &  -44.53  &  1.46  &  - \\
$0(\frac{5}{2}^{-})$  &  $\Xi_{c}^{*}K^{*}$  &  4.84  &  1.46  &  4.47  &  0.62  &  4  &  -0.79  &  3.71  &  -1.92  &  3.39  &  -3.61  &  2.08  &  -28.95  &  1.46  &  - \\
$1(\frac{5}{2}^{-})$  &  $\Xi_{c}^{*}K^{*}$  &  6.12  &  1.46  &  5.6  &  0.3  &  4.96  &  -1.64  &  4.57  &  -3.2  &  4.12  &  -5.53  &  2.31  &  -40.47  &  1.46  &  - \\
  \\                                                            
$1(\frac{1}{2}^{-})$  &  $\Xi_{c}D^{*}$  &  5.9  &  1.46  &  5.41  &  0.35  &  4.79  &  -1.5  &  4.42  &  -2.97  &  4  &  -5.19  &  2.27  &  -38.47  &  1.46  &  - \\
$0(\frac{3}{2}^{-})$  &  $\Xi_{c}D^{*}$  &  3.42  &  1.46  &  3.2  &  0.97  &  2.93  &  0.16  &  2.77  &  -0.5  &  2.58  &  -1.48  &  1.82  &  -16.16  &  1.46  &  - \\
$1(\frac{3}{2}^{-})$  &  $\Xi_{c}D^{*}$  &  5.6  &  1.46  &  5.14  &  0.43  &  4.57  &  -1.3  &  4.22  &  -2.68  &  3.83  &  -4.75  &  2.21  &  -35.8  &  1.46  &  - \\
$0(\frac{5}{2}^{-})$  &  $\Xi_{c}^{*}D^{*}$  &  3.76  &  1.46  &  3.51  &  0.89  &  3.19  &  -0.07  &  3  &  -0.84  &  2.78  &  -1.99  &  1.88  &  -19.25  &  1.46  & - \\
$1(\frac{5}{2}^{-})$  &  $\Xi_{c}^{*}D^{*}$  &  5.4  &  1.46  &  4.96  &  0.48  &  4.41  &  -1.16  &  4.09  &  -2.47  &  3.71  &  -4.44  &  2.18  &  -33.96  &  1.46  &  - \\
  \\                                                            
$0(\frac{1}{2}^{-})$  &  $\Xi_{c}B^{*}$  &  2.72  &  1.46  &  2.58  &  1.15  &  2.41  &  0.62  &  2.3  &  0.2  &  2.18  &  -0.43  &  1.69  &  -9.9  &  1.46  &  - \\
$1(\frac{1}{2}^{-})$  &  $\Xi_{c}B^{*}$  &  4.97  &  1.46  &  4.58  &  0.58  &  4.1  &  -0.88  &  3.8  &  -2.05  &  3.47  &  -3.81  &  2.1  &  -30.14  &  1.46  &  - \\
$0(\frac{3}{2}^{-})$  &  $\Xi_{c}B^{*}$  &  2.94  &  1.46  &  2.78  &  1.09  &  2.57  &  0.48  &  2.45  &  -0.02  &  2.31  &  -0.75  &  1.73  &  -11.83  &  1.46  &  - \\
$1(\frac{3}{2}^{-})$  &  $\Xi_{c}B^{*}$  &  4.89  &  1.46  &  4.51  &  0.6  &  4.04  &  -0.83  &  3.75  &  -1.97  &  3.42  &  -3.69  &  2.09  &  -29.43  &  1.46  & - \\
$0(\frac{5}{2}^{-})$  &  $\Xi_{c}^{*}B^{*}$  &  3.04  &  1.46  &  2.86  &  1.07  &  2.64  &  0.41  &  2.51  &  -0.11  &  2.36  &  -0.9  &  1.75  &  -12.71  &  1.46  &  - \\
$1(\frac{5}{2}^{-})$  &  $\Xi_{c}^{*}B^{*}$  &  4.83  &  1.46  &  4.46  &  0.62  &  3.99  &  -0.78  &  3.71  &  -1.91  &  3.39  &  -3.59  &  2.07  &  -28.86  &  1.46  & - \\
\\
\hline
\end{tabular}
}
\end{center}
\end{table*}

\begin{table*}[]
\begin{center}
\addtocounter{table}{-1}
\caption{\small to be continued.. }
\label{as and re for meson-baryon states}
\scalebox{1}{
\begin{tabular}{cccccccccccccccc}
\hline
I$(J^{P})$ & State   &\multicolumn{2}{c}{Z=0} &\multicolumn{2}{c}{Z=0.2} &\multicolumn{2}{c}{Z=0.4} &\multicolumn{2}{c}{Z=0.5} &\multicolumn{2}{c}{Z=0.6} &\multicolumn{2}{c}{Z=0.9} &\multicolumn{2}{c}{Z=1} \\
\cline{3-16}
 & &$a_{s}$ &$r_{e}$ &$a_{s}$ &$r_{e}$& $a_{s}$ &$r_{e}$ &$a_{s}$ &$r_{e}$ & $a_{s}$ &$r_{e}$ & $a_{s}$ &$r_{e}$ & $a_{s}$ &$r_{e}$\\
\hline
\hline
\\               
$0(\frac{1}{2}^{-})$  &  $\Xi_{b}K^{*}$  &  3.08  &  1.46  &  2.9  &  1.06  &  2.68  &  0.38  &  2.54  &  -0.16  &  2.39  &  -0.97  &  1.76  &  -13.13  &  1.46  &  -  \\
$1(\frac{1}{2}^{-})$  &  $\Xi_{b}K^{*}$  &  5.7  &  1.46  &  5.23  &  0.4  &  4.64  &  -1.36  &  4.29  &  -2.77  &  3.88  &  -4.89  &  2.23  &  -36.66  &  1.46  &  -  \\
$0(\frac{3}{2}^{-})$  &  $\Xi_{b}K^{*}$  &  3.44  &  1.46  &  3.22  &  0.97  &  2.94  &  0.15  &  2.78  &  -0.51  &  2.59  &  -1.5  &  1.82  &  -16.31  &  1.46  &  -  \\
$1(\frac{3}{2}^{-})$  &  $\Xi_{b}K^{*}$  &  5.57  &  1.46  &  5.11  &  0.43  &  4.54  &  -1.28  &  4.2  &  -2.65  &  3.81  &  -4.7  &  2.21  &  -35.52  &  1.46  &  -  \\
     \\                                                         
$0(\frac{1}{2}^{-})$  &  $\Xi_{b}D^{*}$  &  2.77  &  1.46  &  2.62  &  1.14  &  2.44  &  0.59  &  2.33  &  0.16  &  2.21  &  -0.5  &  1.7  &  -10.29  &  1.46  &  -  \\
$1(\frac{1}{2}^{-})$  &  $\Xi_{b}D^{*}$  &  5.02  &  1.46  &  4.62  &  0.57  &  4.13  &  -0.91  &  3.83  &  -2.1  &  3.5  &  -3.88  &  2.11  &  -30.56  &  1.46  &  -  \\
$0(\frac{3}{2}^{-})$  &  $\Xi_{b}D^{*}$  &  2.99  &  1.46  &  2.82  &  1.08  &  2.61  &  0.44  &  2.48  &  -0.06  &  2.33  &  -0.83  &  1.74  &  -12.27  &  1.46  & -  \\
$1(\frac{3}{2}^{-})$  &  $\Xi_{b}D^{*}$  &  4.94  &  1.46  &  4.55  &  0.59  &  4.07  &  -0.86  &  3.78  &  -2.02  &  3.45  &  -3.76  &  2.09  &  -29.84  &  1.46  &  -  \\
    \\                                                          
$0(\frac{1}{2}^{-})$  &  $\Xi_{b}B^{*}$  &  2.67  &  1.46  &  2.53  &  1.16  &  2.36  &  0.66  &  2.26  &  0.26  &  2.15  &  -0.34  &  1.68  &  -9.37  &  1.46  &  -  \\
$1(\frac{1}{2}^{-})$  &  $\Xi_{b}B^{*}$  &  4.85  &  1.46  &  4.48  &  0.61  &  4.01  &  -0.8  &  3.72  &  -1.93  &  3.4  &  -3.62  &  2.08  &  -29.06  &  1.46  &  -  \\
$0(\frac{3}{2}^{-})$  &  $\Xi_{b}B^{*}$  &  2.76  &  1.46  &  2.61  &  1.14  &  2.43  &  0.6  &  2.33  &  0.17  &  2.2  &  -0.48  &  1.7  &  -10.2  &  1.46  &   - \\
$1(\frac{3}{2}^{-})$  &  $\Xi_{b}B^{*}$  &  4.82  &  1.46  &  4.44  &  0.62  &  3.98  &  -0.77  &  3.7  &  -1.89  &  3.38  &  -3.57  &  2.07  &  -28.72  &  1.46  &  -  \\
\\
\hline
\hline
\end{tabular}
}
\end{center}
\end{table*}

\section{Discussion and summary}
\label{Discussion and summary}
In this chapter, we have used the s-wave One Boson Exchange (OBE) potential. We have discussed the characteristic contribution of the individual s-wave one meson exchange potential to the net s-wave OBE potential in respective isospin-spin channels. The strength and contribution of the s-wave OBE potential is the delicate cancellation of the individual meson exchange, hence, the Yukawa-like screen potential shows large impact on the net effective interaction potential, where this potential is sensitive to the parameter c (which is fixed at c=0.0686 GeV) and the value of residual coupling constant $k_{mol}$. Thus, the results are found sensitive to the Yukawa-like screen potential. We have proposed this additional Yukawa-like screen potential to get additional attractive strength with approximation that the two hadrons are experienced the dipole-like interaction.

With the effective interaction potential, we are able to calculate the mass spectra of meson-baryon and di-baryon (antibaryon) molecular states. The obtained results have predicted some interesting molecular pentaquark and hexaquark states with open as well as hidden flavour (strange, charm, bottom), such as $\Sigma_{s,c,b}K^{*}$, $\Sigma_{s,c,b}D^{*}$, $\Sigma_{s,c,b}B^{*}$, $\Xi_{s,c,b}K^{*}$, $\Xi_{s,c,b}D^{*}$, $\Xi_{s,c,b}B^{*}$, $\Sigma_{s,c,b}\Sigma_{s,c,b}$, $\Sigma_{s,c,b}\overline{\Sigma_{s,c,b}}$, $\Xi_{s,c,b}\Xi_{s,c,b}$ and $\Xi_{s,c,b}\overline{\Xi_{s,c,b}}$. The calculated results have compared with work of others. 
We have predicted some interesting near threshold and shallow bound states. The recently observed state $P_{c}(4450)$ which is supposed to have minimum five quarks in its internal structure is identified  as  a bound $\Sigma_{c}D^{*}$ state with I($J^{P}$) = $\frac{1}{2}(\frac{3}{2}^{-})$. This state $P_{c}(4450)$ together with $P_{c}(4380)$ were reported by LHCb collaboration in 2015 \cite{Aaij-prl2015-115-pentaquark}. The entire mass spectra presented in this paper have provided the possibilities of resonance like structure, either just above the threshold or bound state (just below the threshold) and provides the reference for searches of such molecular structures in future experiments.    
%%%%%%%%%%%%%%%%%%%%%%%%%%% Figure %%%%%%%%%%%%%%%%%%%%%%%%%%%%%%%%%
\pagebreak
\appendix

\section{s-wave OBE potential and net effective potential} 
It is interesting to see the s-wave one meson exchange contribution to the OBE interaction potential, which is shown in the Fig-\ref{pentaquark-OBE plots} when effective s-wave OBE, Yukawa-like screen and net effective potential shown in Fig-\ref{pentaquark net-potential plots}. The graphs are plotted for all possible isospin-spin channels. 

We can see from Fig-\ref{pentaquark-OBE plots} that in all possible isospin-spin channels, the s-wave $\sigma$-exchange potential is an attractive in nature while the s-wave $\omega$-exchange is repulsive.

The behavior of s-wave $\pi$-exchange potential is repulsive in (I,S)=(1/2, 1/2) and (3/2, 5/2) channels when it is an attractive in other channels, moreover it seems relatively strong attractive in (I,S)=(1/2, 5/2) channel.

We can see from the figure that the strength of the s-wave $\eta$ and $\rho$ meson exchange are relatively week. However, the $\eta$-exchange is an attractive in (I,S)=(1/2, 1/2), (3/2, 1/2) and (1/2, 3/2) channels while $\rho$-exchange is an attractive in (3/2, 1/2),(1/2, 3/2) and (1/2, 5/2) channels.

The scalar s-wave $\delta$-exchange (which is also known as $a_{0}$-exchange) is attractive in (I,S)=(3/2, 1/2), (3/2, 3/2) and (3/2, 5/2) channels while its strength is seems relatively weak compared to s-wave $\sigma$-exchange. 

From Fig-\ref{pentaquark net-potential plots}, we can see the behavior of the net s-wave OBE potential, Yukawa-like screen potential and net effective potential in possible isospin-spin channels.

It can be seen from Fig-\ref{pentaquark net-potential plots} that the strength of the net effective potential highly influenced by attractive Yukawa-like potential.

The effective s-wave OBE gets attractive in (I,S)=(1/2, 5/2), (1/2, 3/2) and (3/2, 1/2) channels, in which (1/2, 5/2) channel get more attractive depth than (1/2, 3/2) channel while channel (3/2, 1/2) being the weakest attractive channel.

However, these attractive channels of s-wave OBE get depth around 4-6 fm.

Thus, even pure s-wave OBE have a possibilities of shallow bound states in these particular channels.

Moreover, the attractive Yukawa-like potential have shown high impact on net effective interaction potential which reflect in the calculated results.         
\begin{figure*}[b]
\caption{The characteristic nature of the individual s-wave meson exchange potential,  in a respective isospin-spin channels. The graphs are plotted for the $\Sigma_{s,c,b}-D^{*}$ systems. These graphs can be consider as generalize plots to understand the behavior of the potentials in respective isospin-spin channels for other systems, as the similar nature have been found in different meson-baryon systems with small scaling.}
\label{pentaquark-OBE plots}
\includegraphics[scale=0.48]{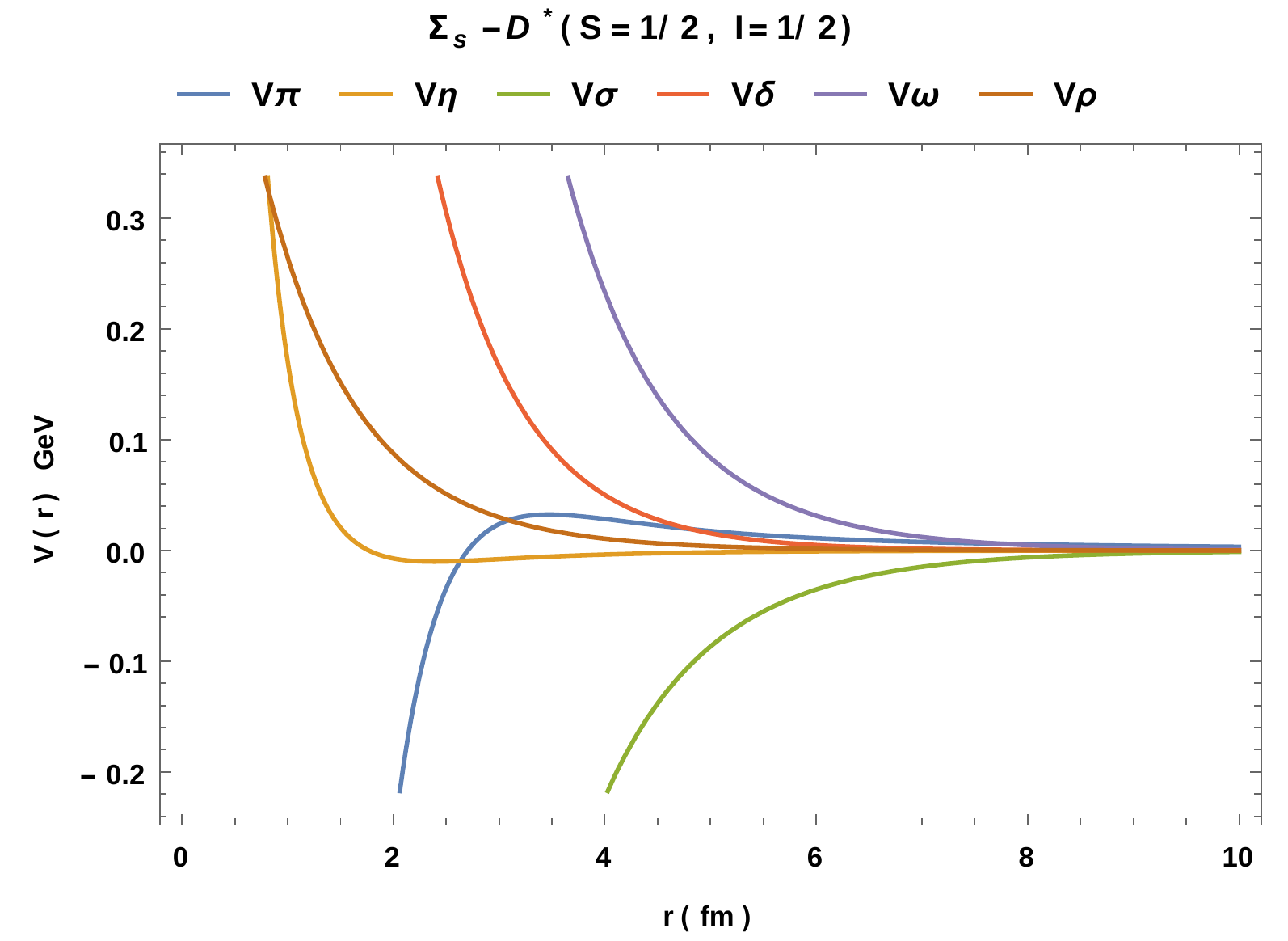}
\includegraphics[scale=0.48]{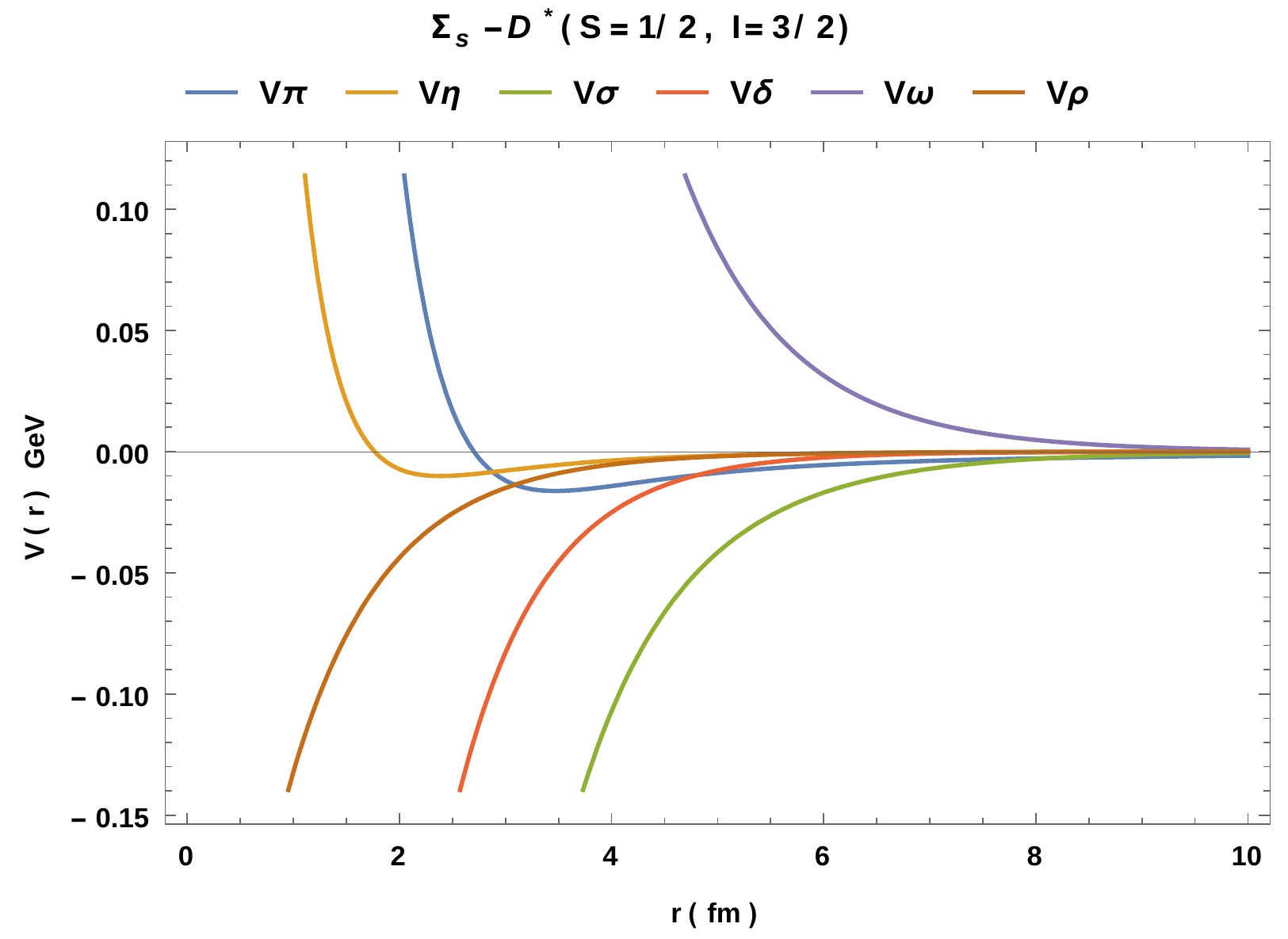}
\includegraphics[scale=0.48]{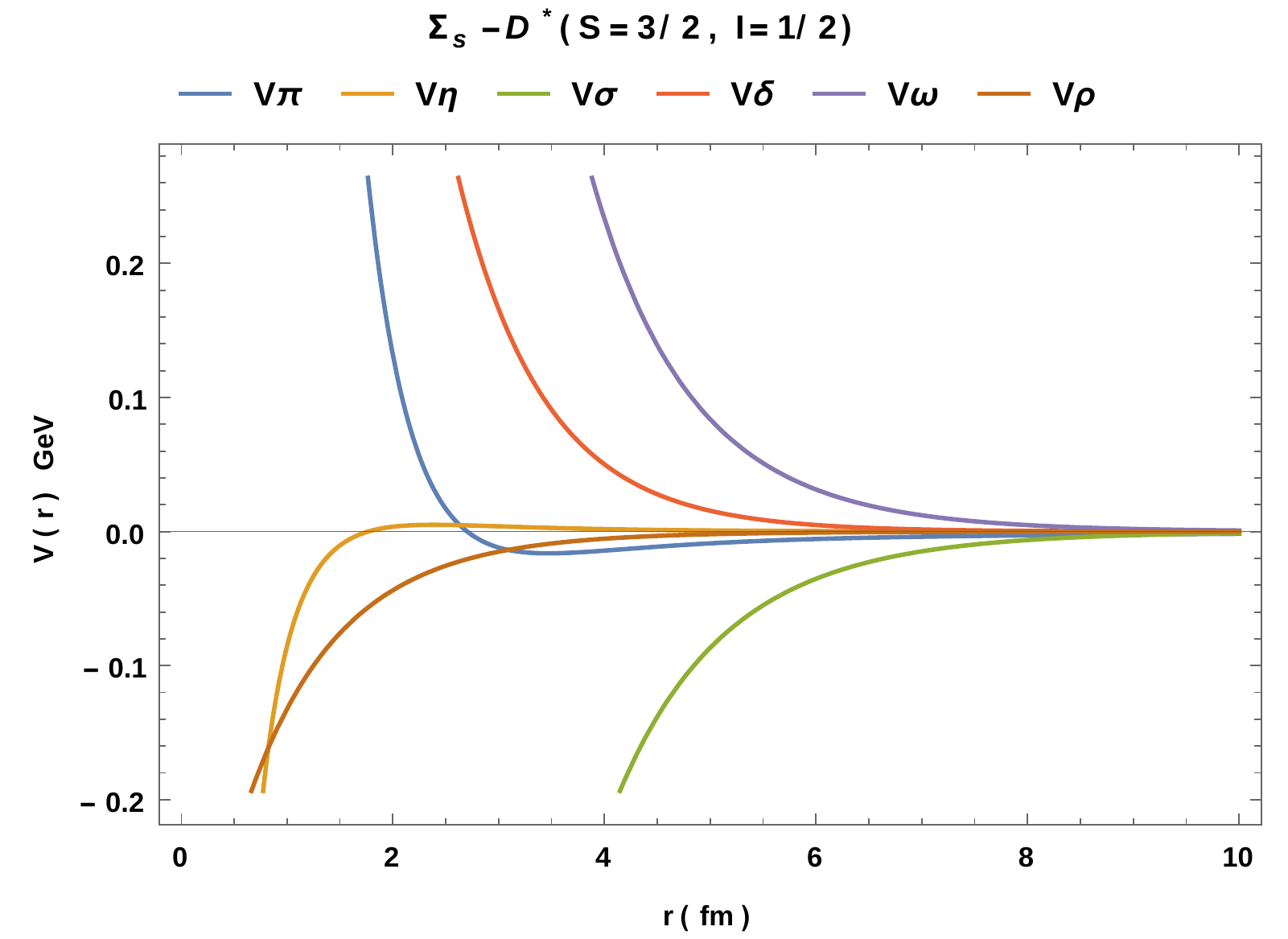}
\includegraphics[scale=0.48]{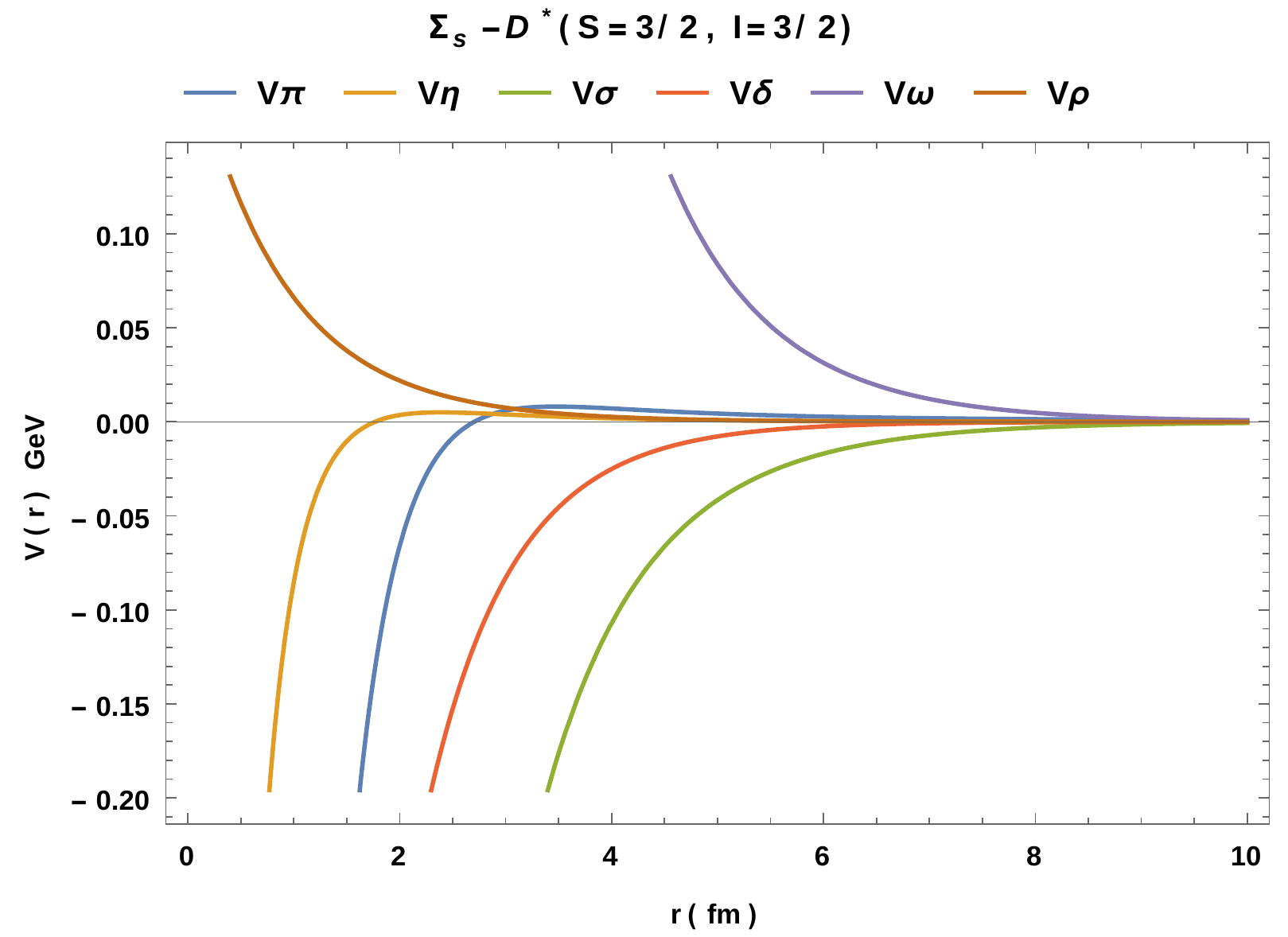}
\includegraphics[scale=0.48]{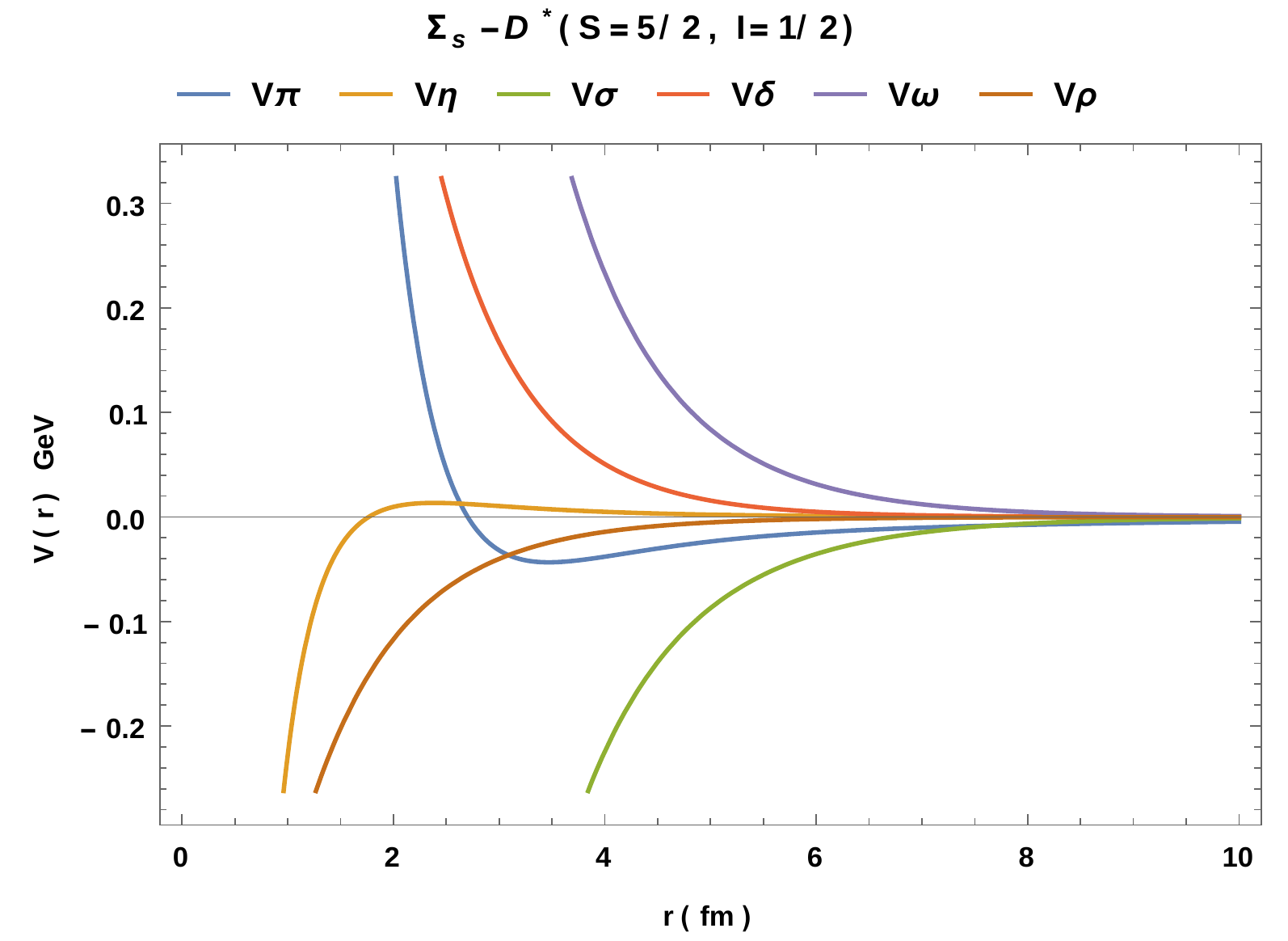}
\includegraphics[scale=0.48]{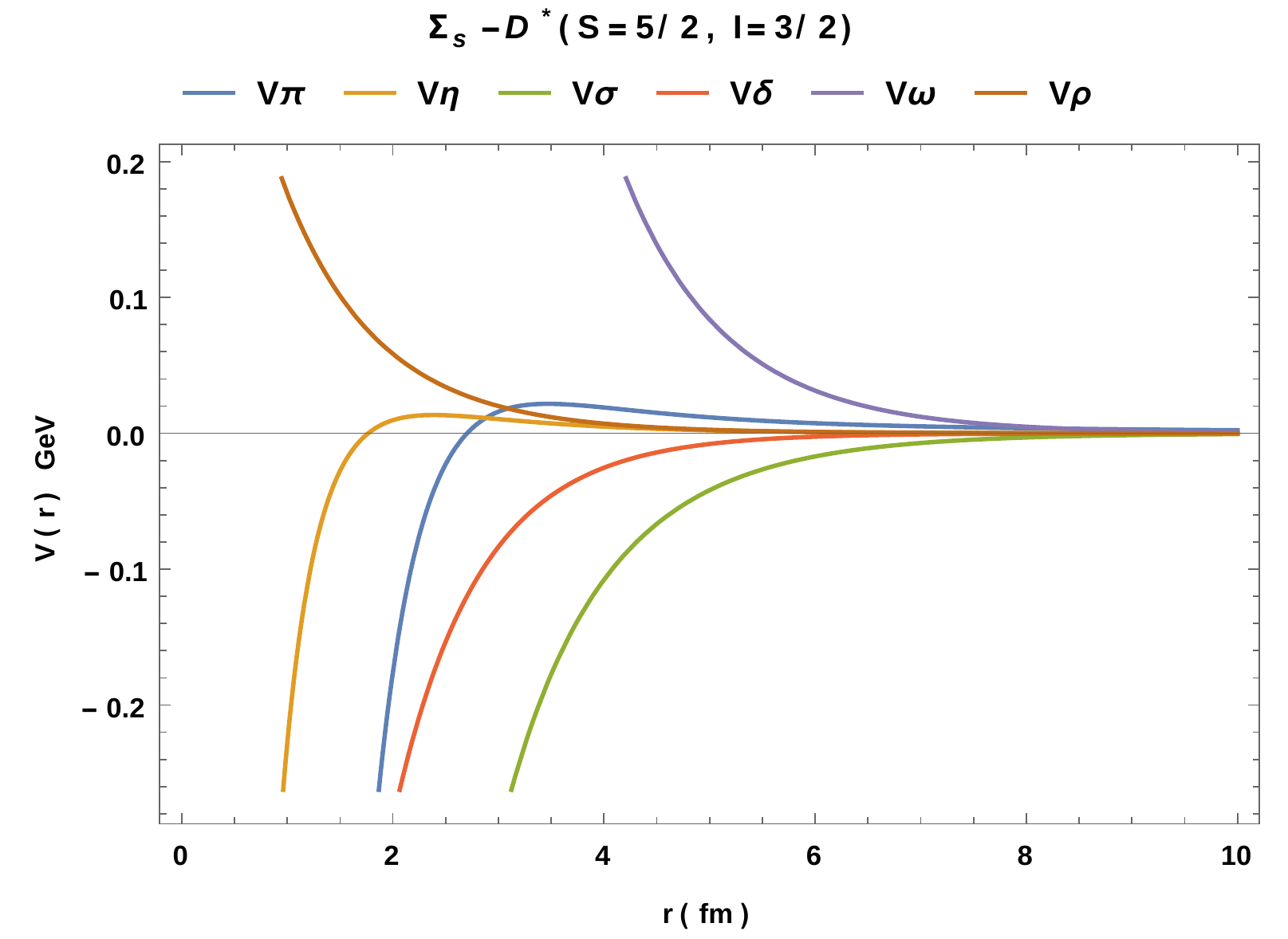}
\end{figure*}

\begin{figure*}[b]
\addtocounter{figure}{-1}
\caption{to be continued..}

\includegraphics[scale=0.52]{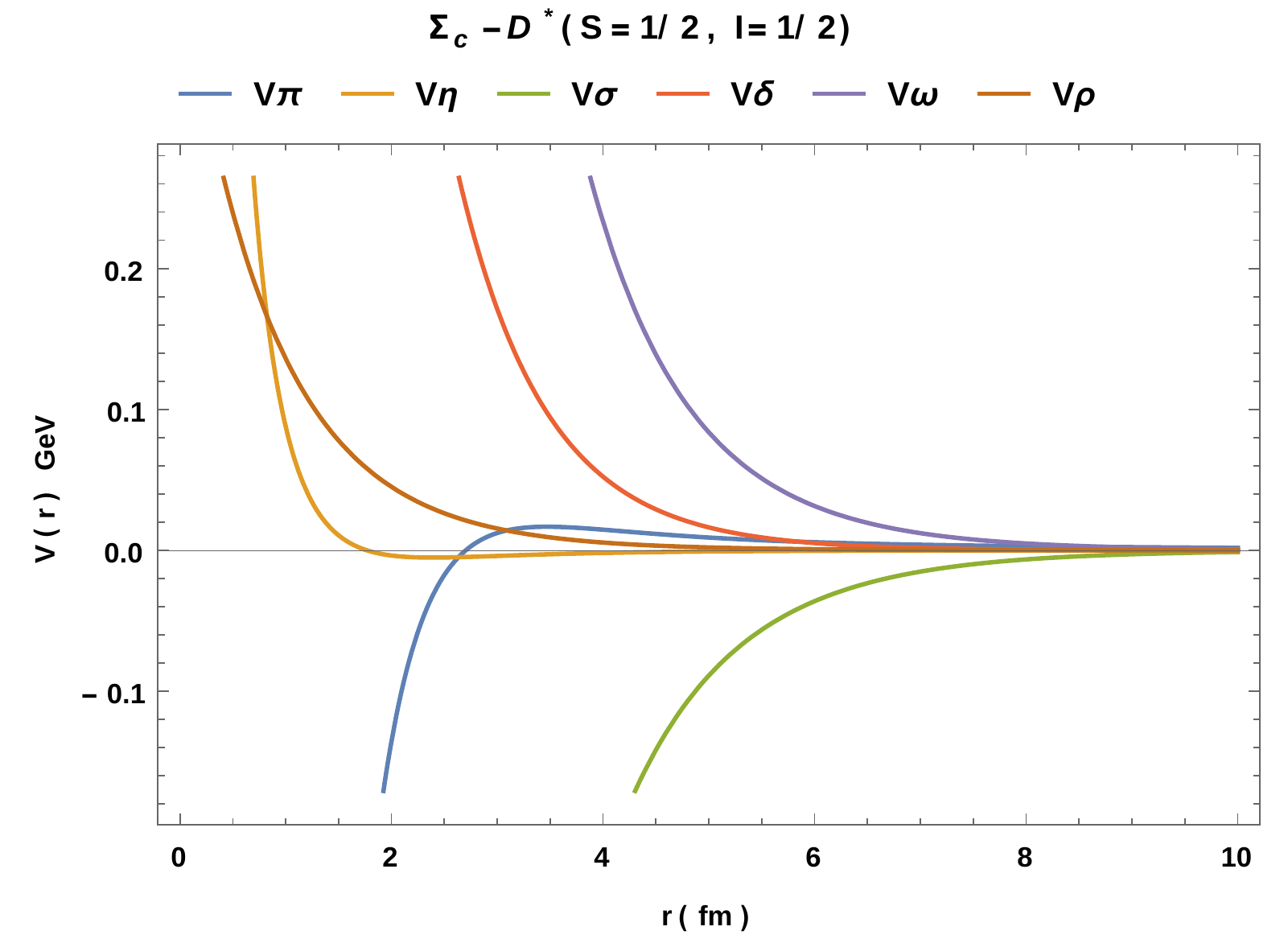}
\includegraphics[scale=0.52]{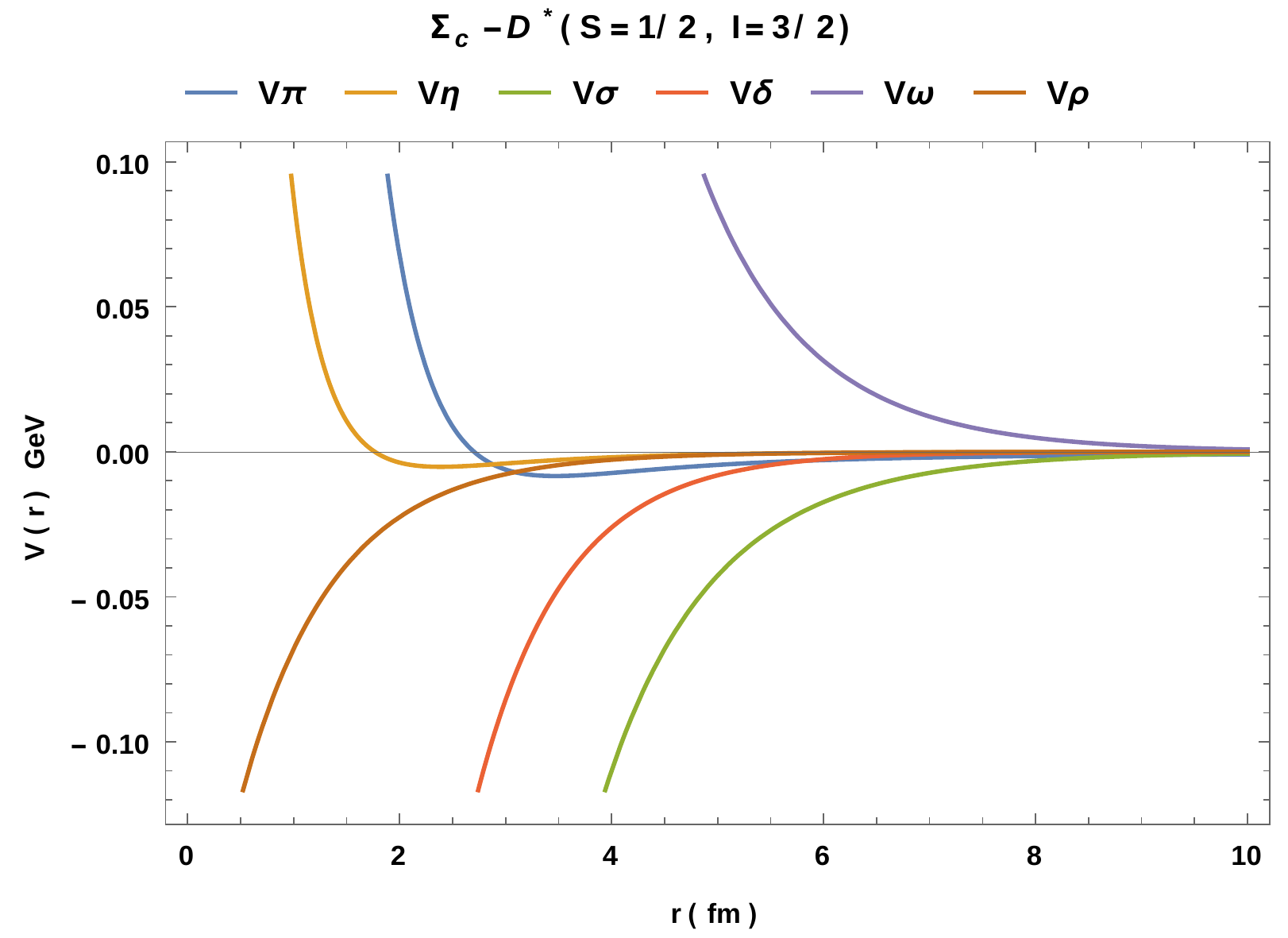}
\includegraphics[scale=0.52]{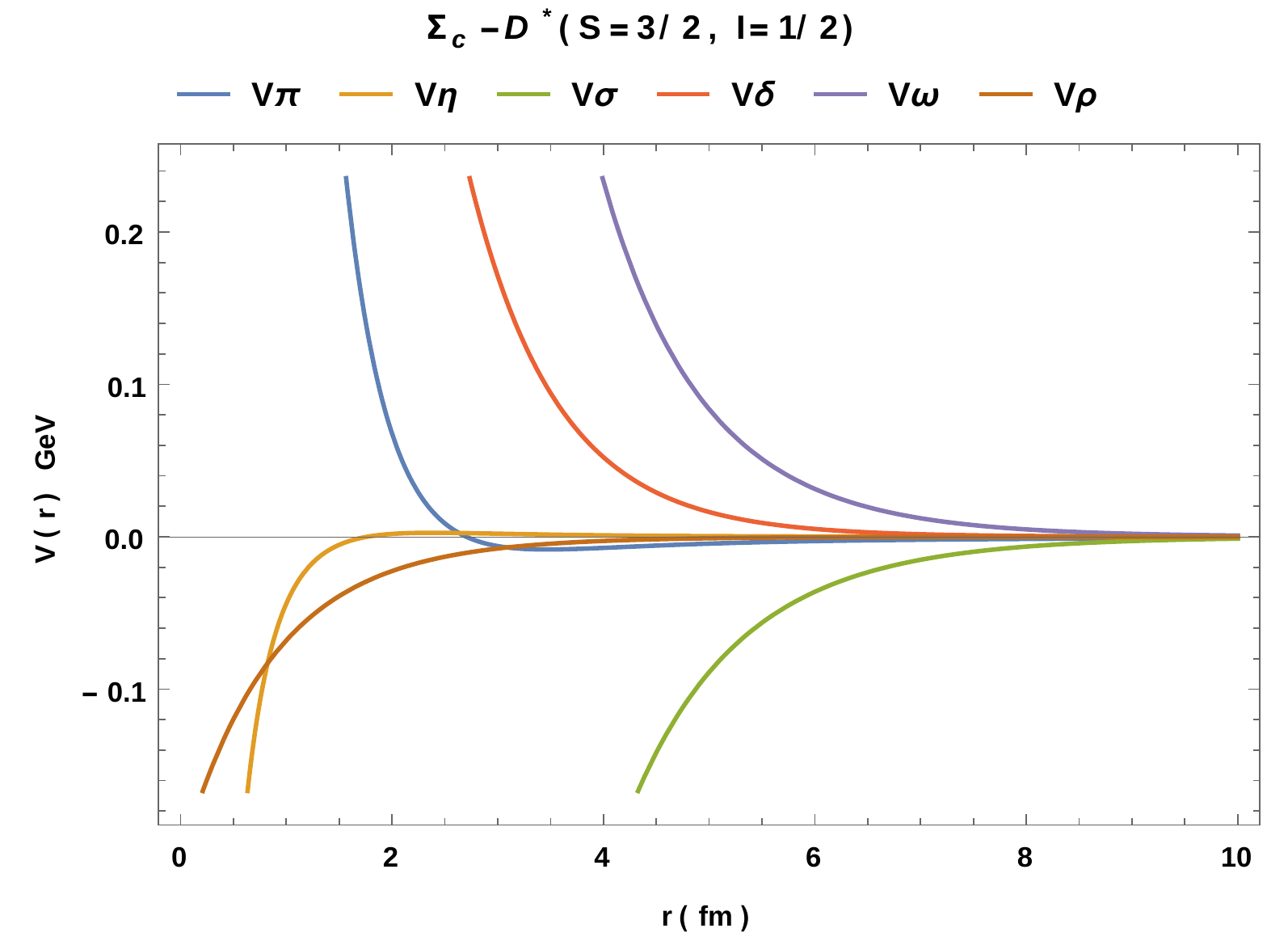}
\includegraphics[scale=0.52]{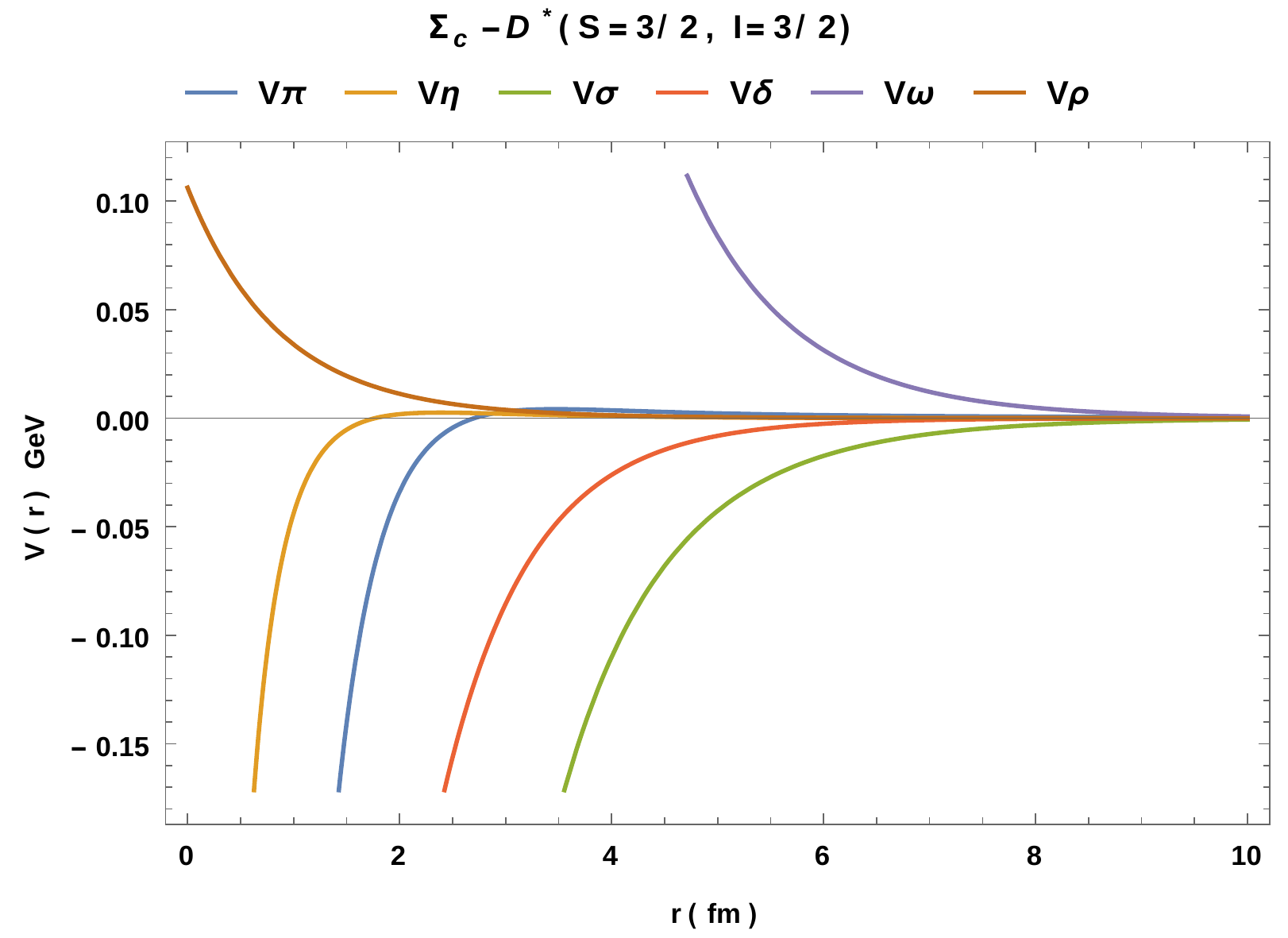}
\includegraphics[scale=0.52]{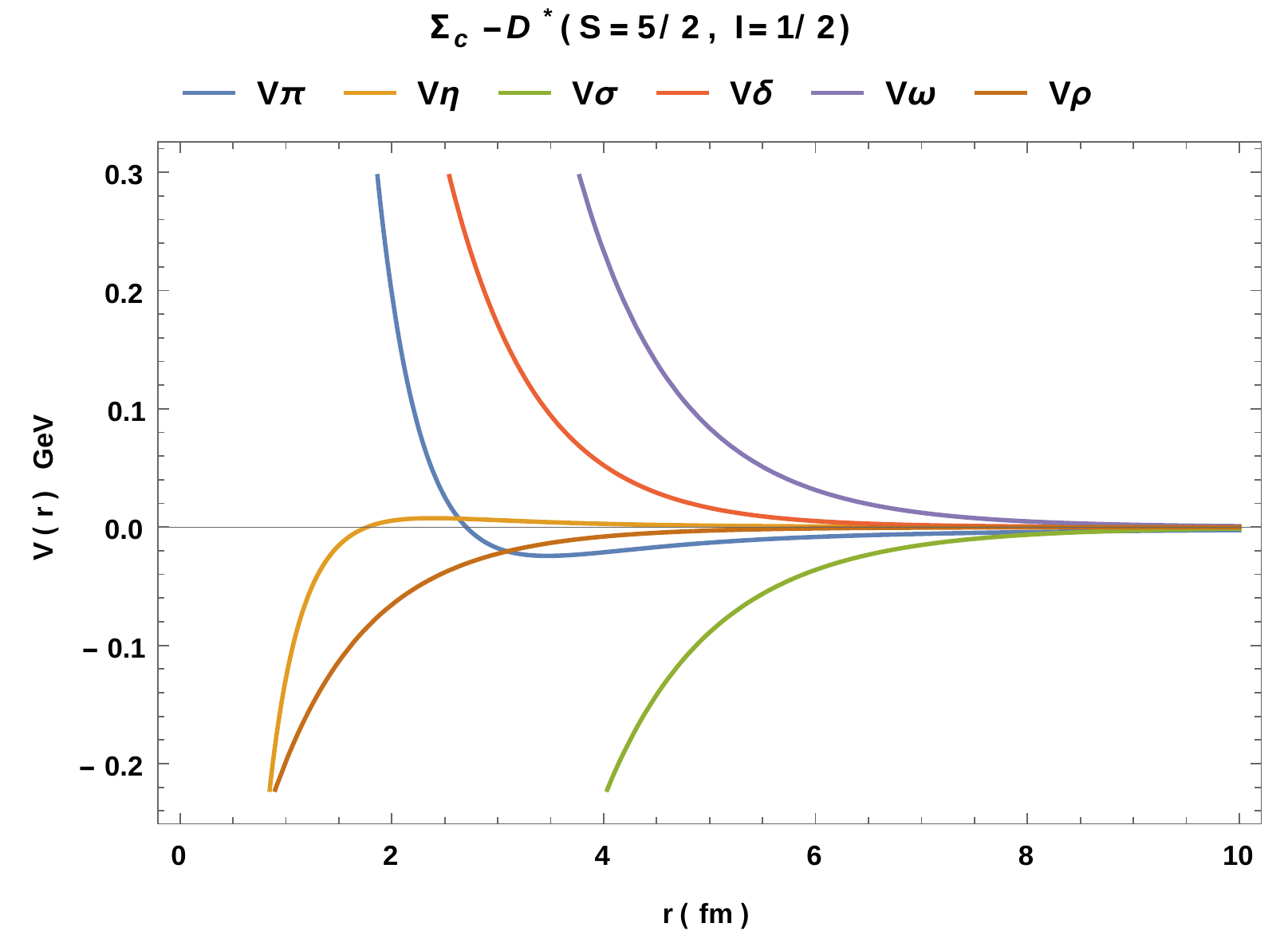}
\includegraphics[scale=0.52]{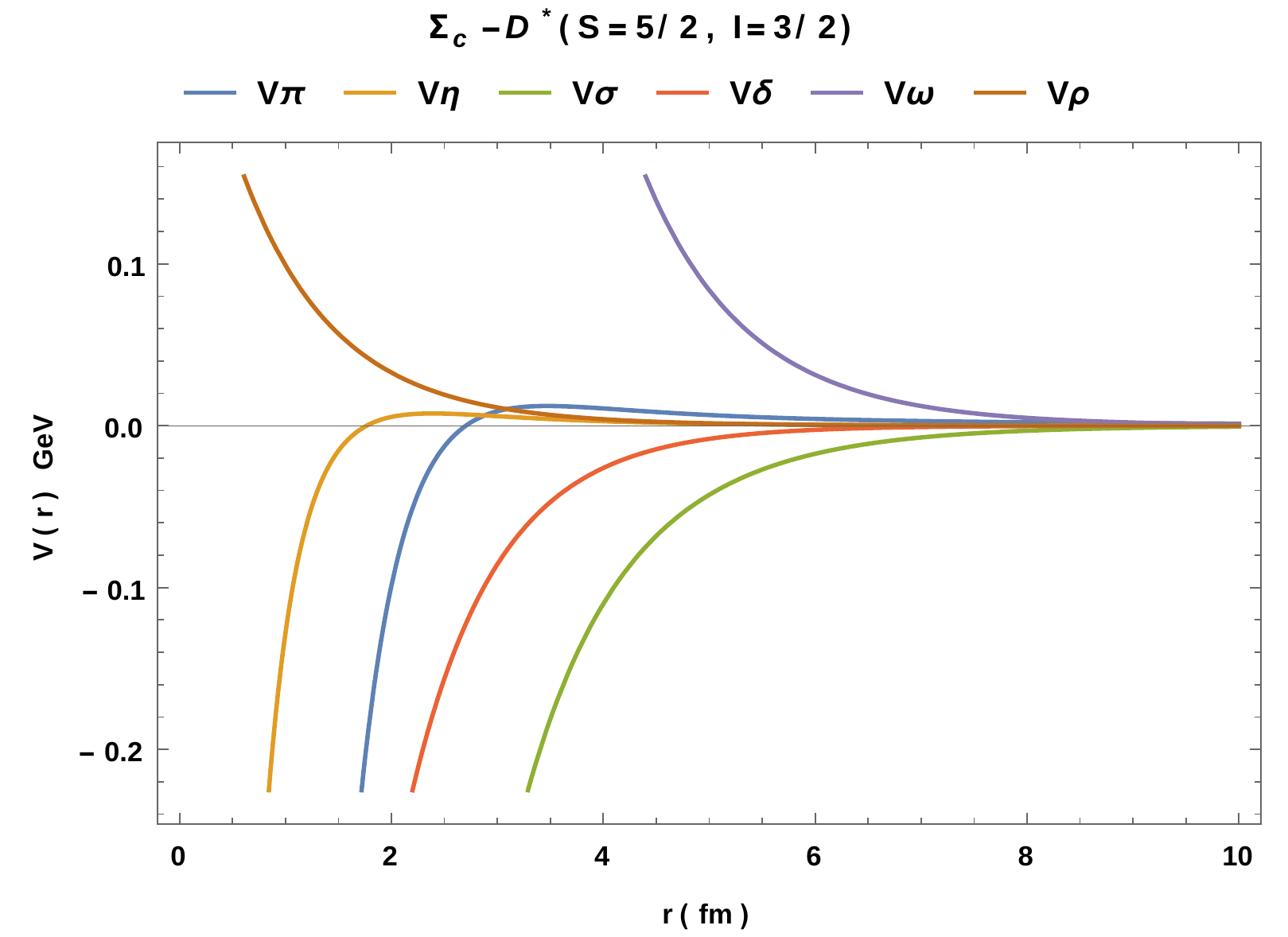}
\end{figure*}

\begin{figure*}[b]
\addtocounter{figure}{-1}
\caption{to be continued..}
\includegraphics[scale=0.52]{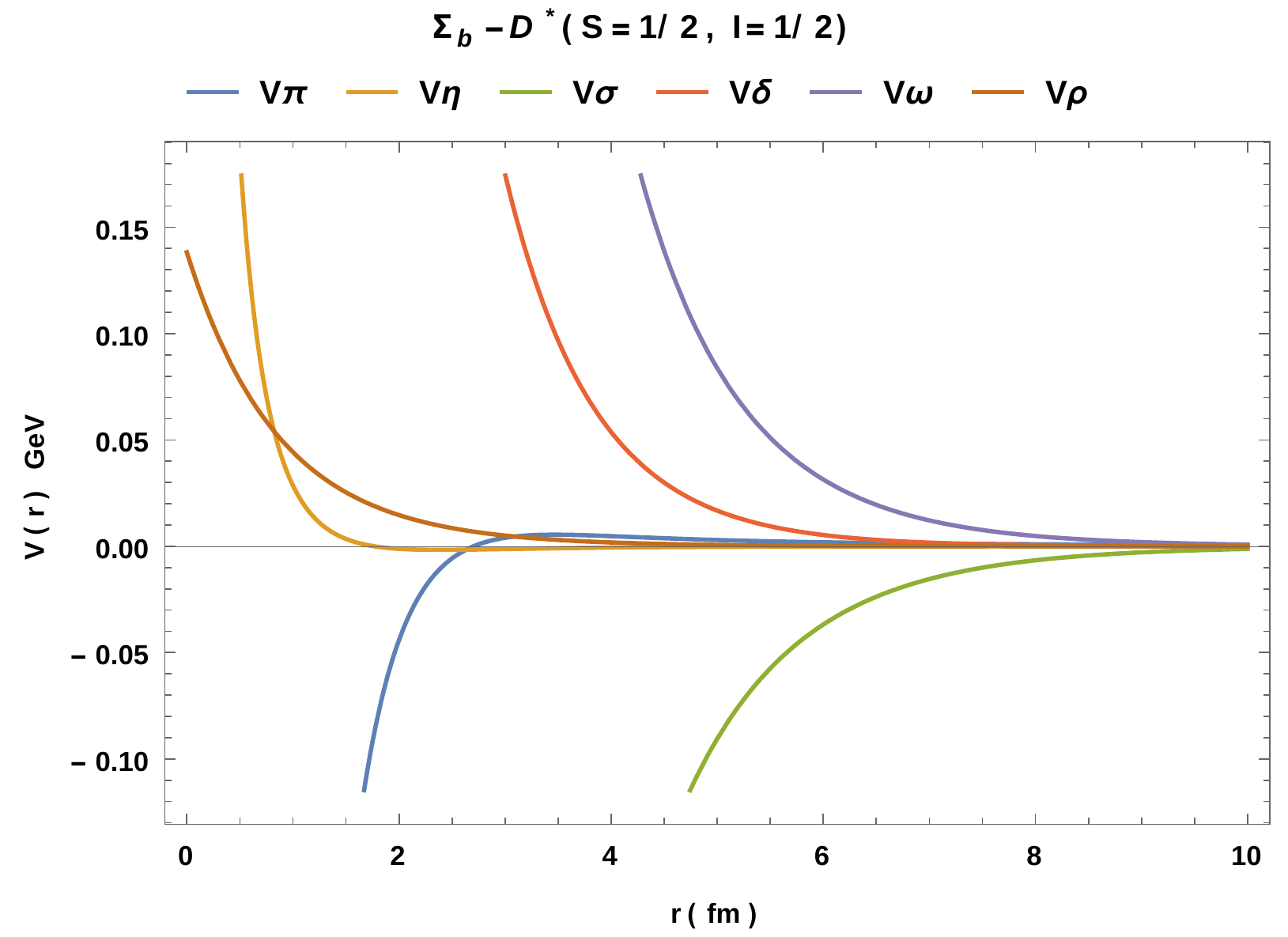}
\includegraphics[scale=0.52]{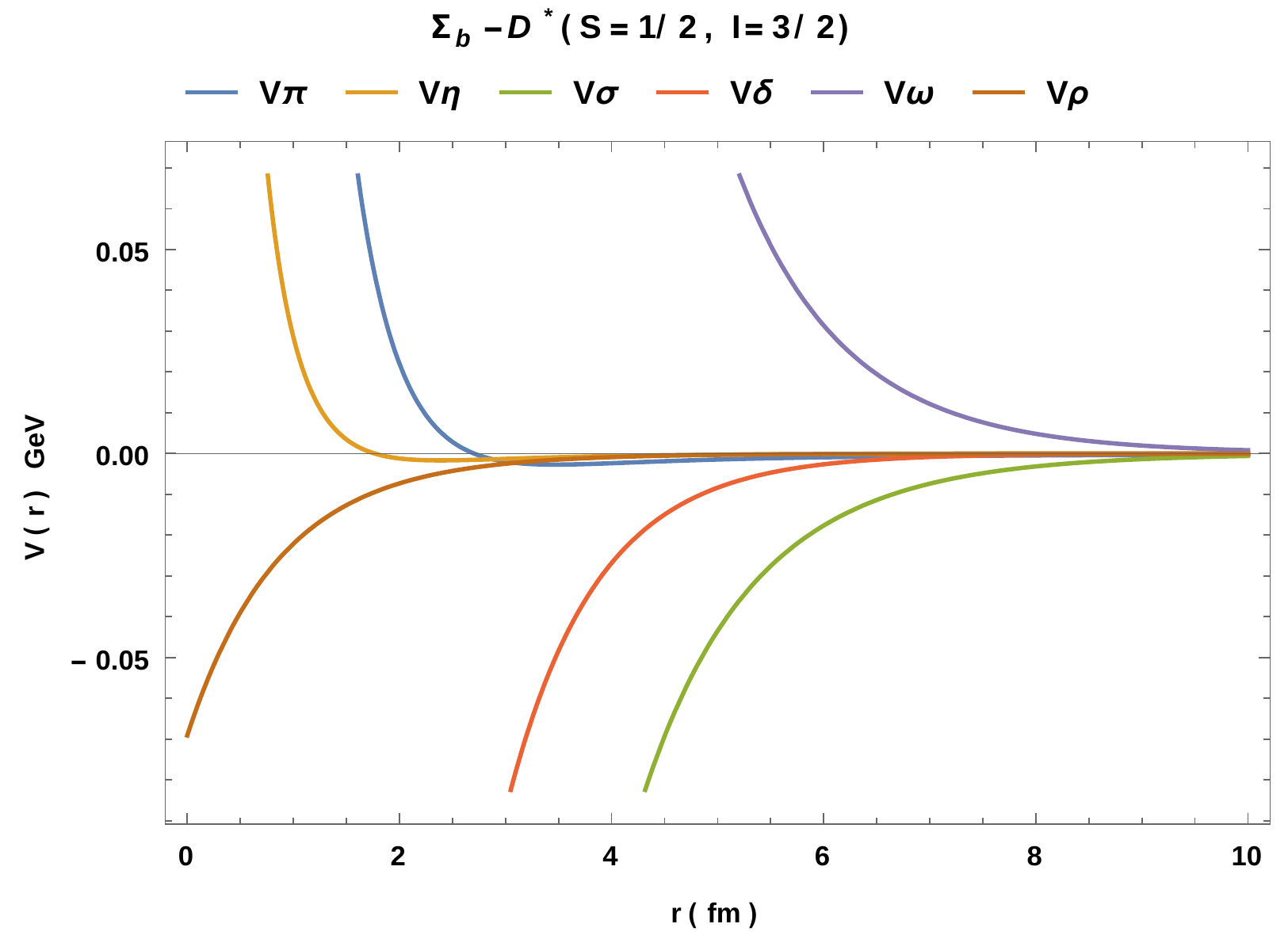}
\includegraphics[scale=0.52]{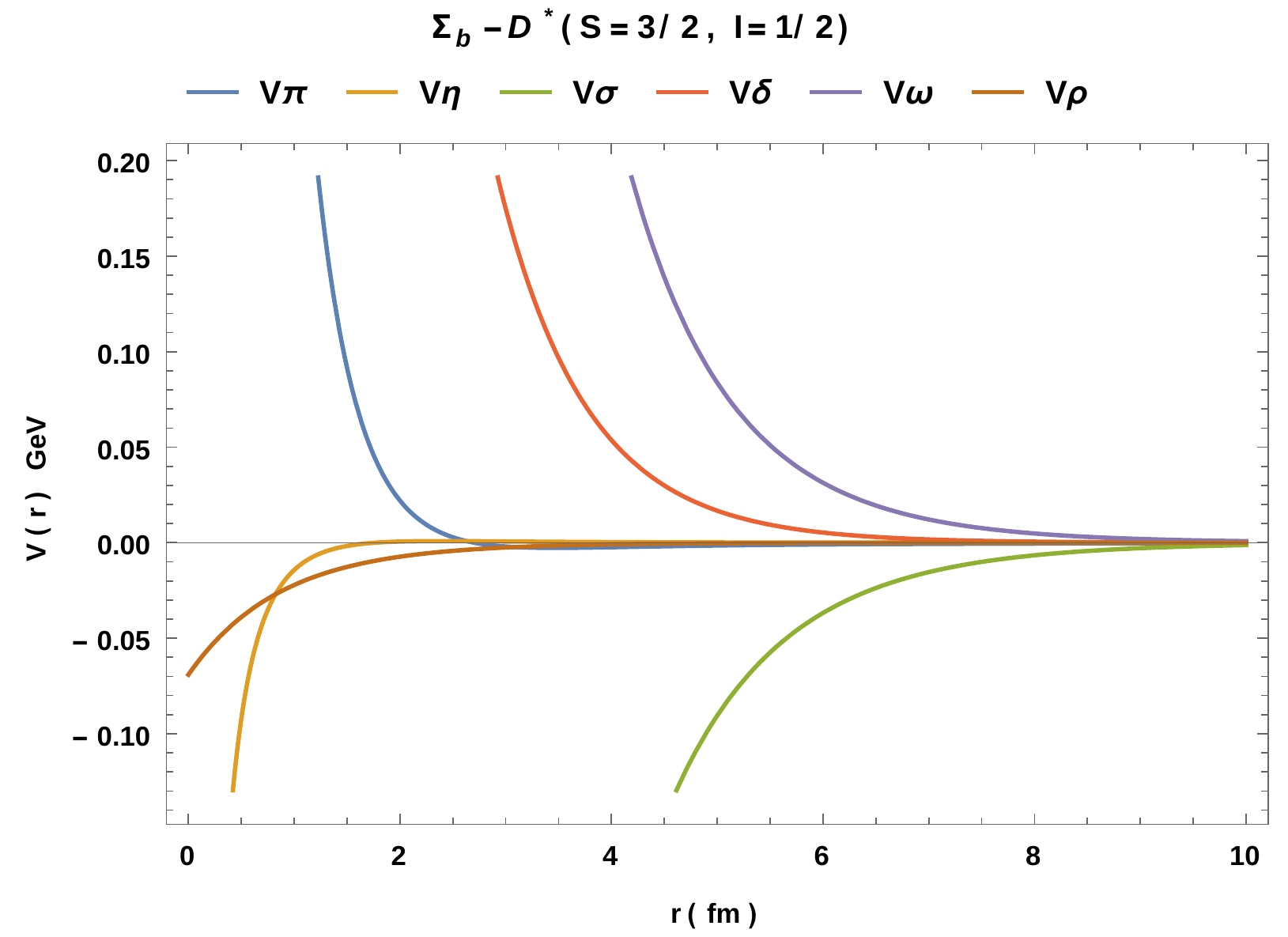}
\includegraphics[scale=0.52]{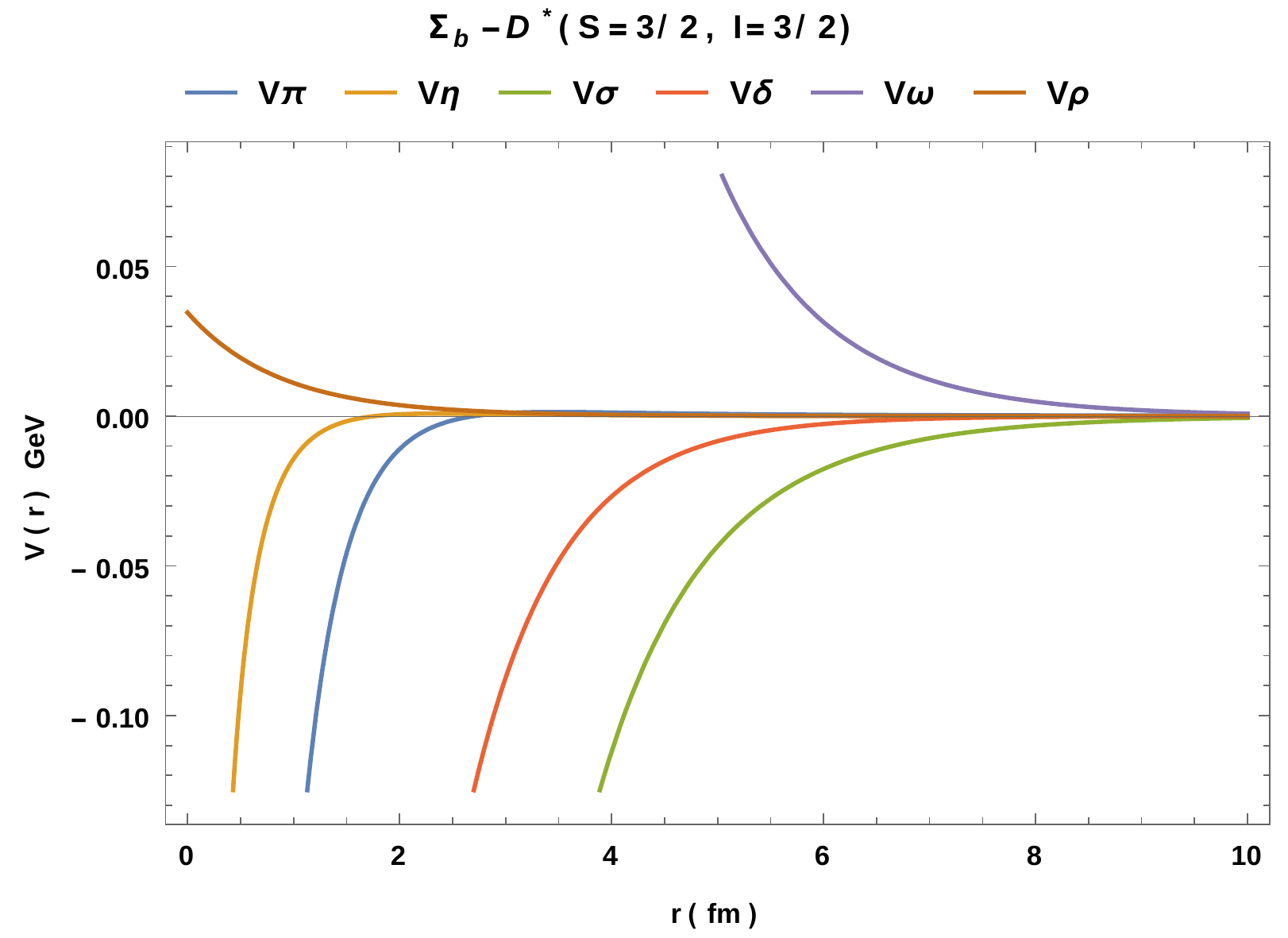}
\includegraphics[scale=0.52]{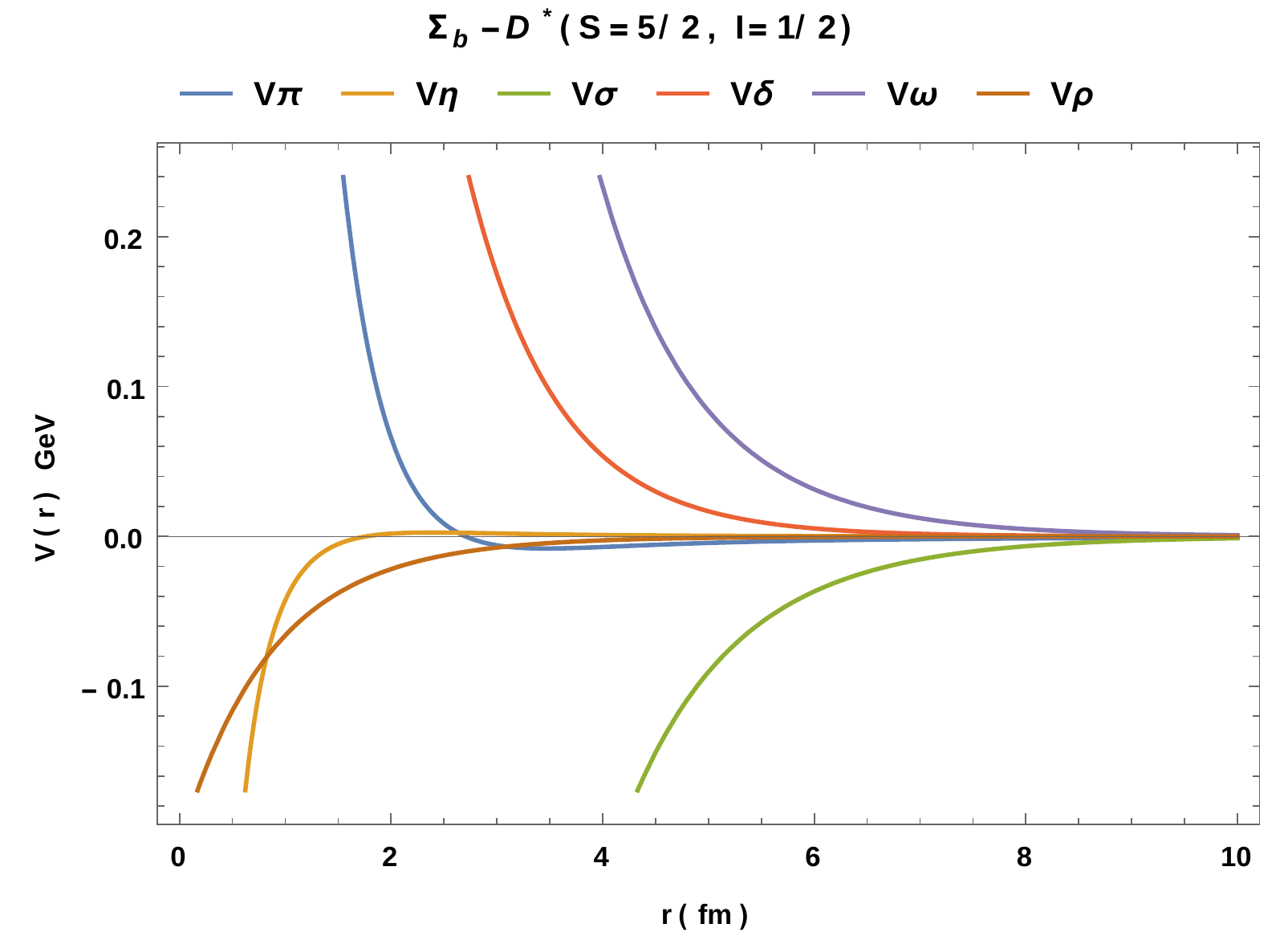}
\includegraphics[scale=0.52]{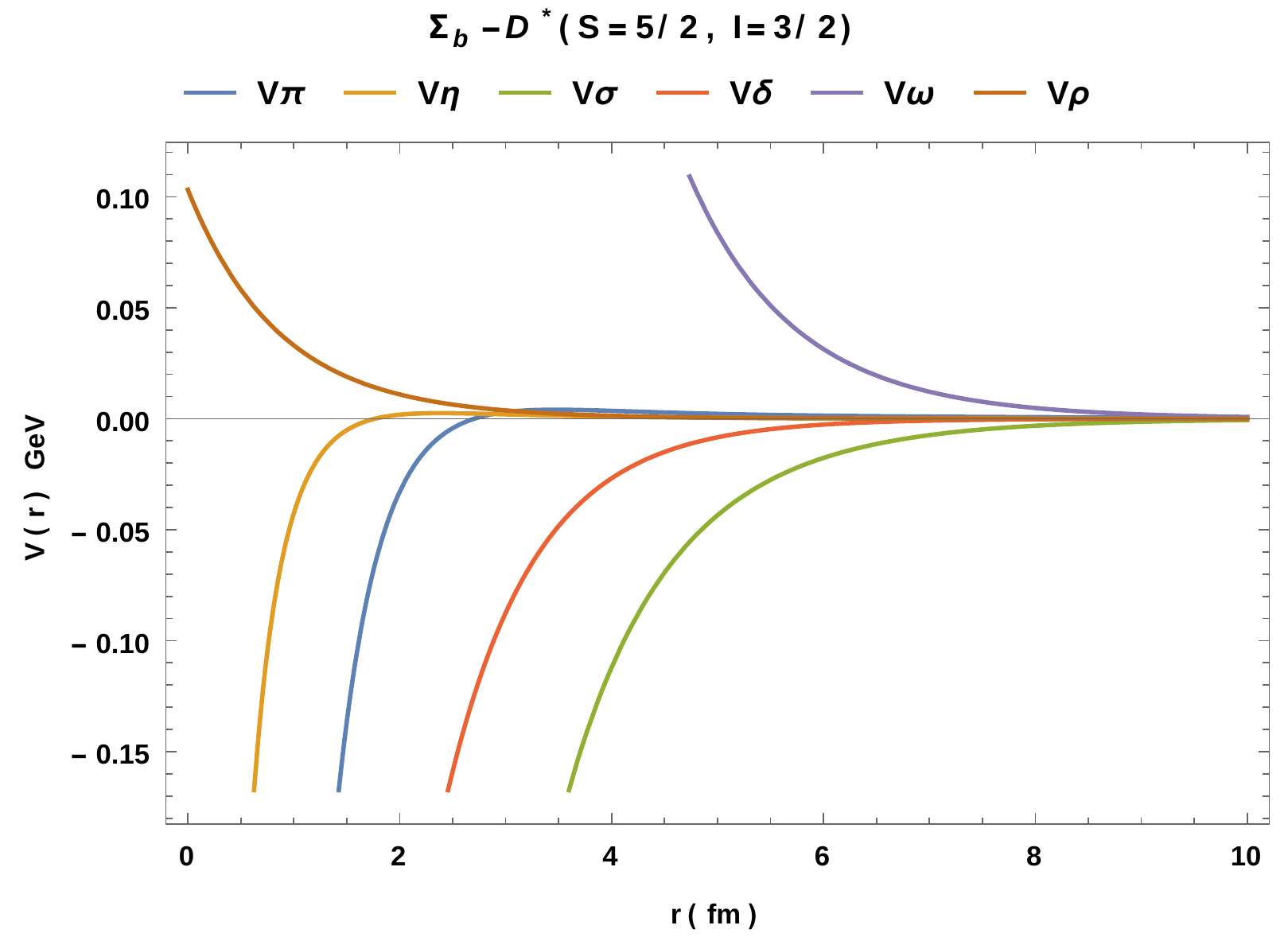}
\end{figure*}

\begin{figure*}[b]
\caption{The contributory nature of the s-wave OBE potential (Vobe), Yukawa-like screen potential (Vy) and net effective potential(Veff = Vobe + Vy) in a respective isospin-spin channels are shown. The graphs are plotted for the $\Sigma_{s,c,b}-D^{*}$ systems. These graphs can be consider as generalize plots to understand the behavior of the potentials in respective isospin-spin channels for other systems, as the similar nature have been found in different meson-baryon systems with small scaling.}
\label{pentaquark net-potential plots}
\includegraphics[scale=0.67]{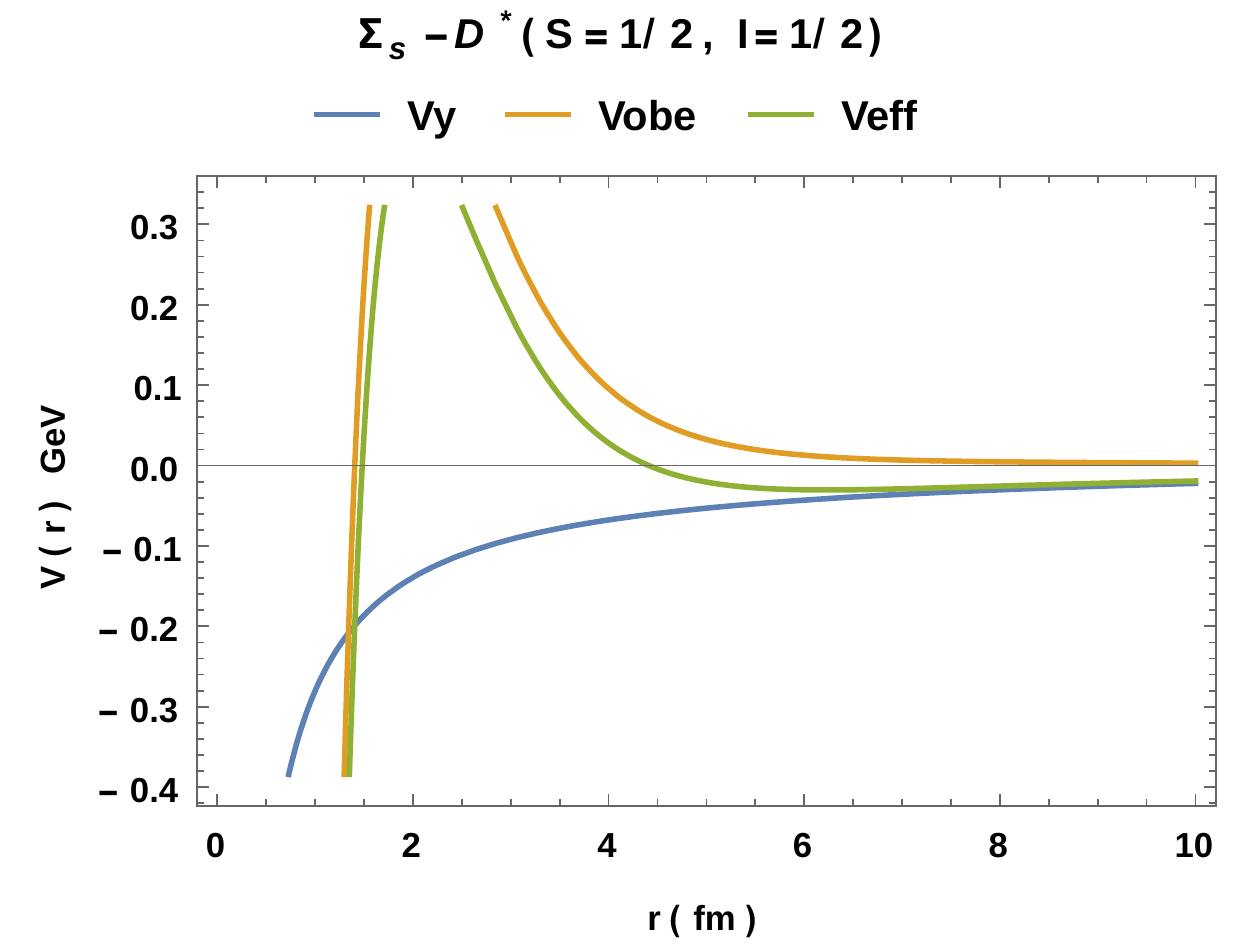}
\includegraphics[scale=0.67]{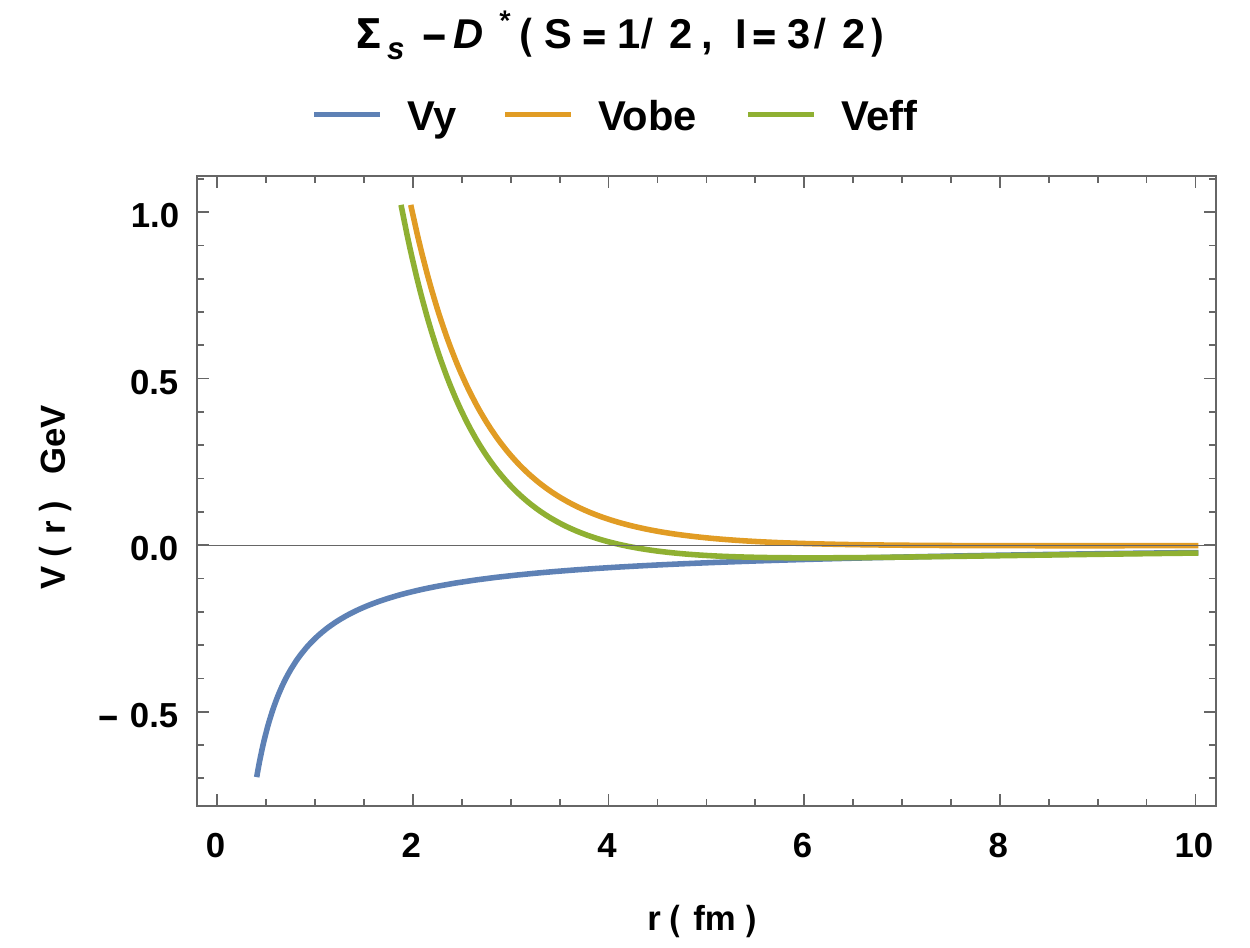}
\includegraphics[scale=0.67]{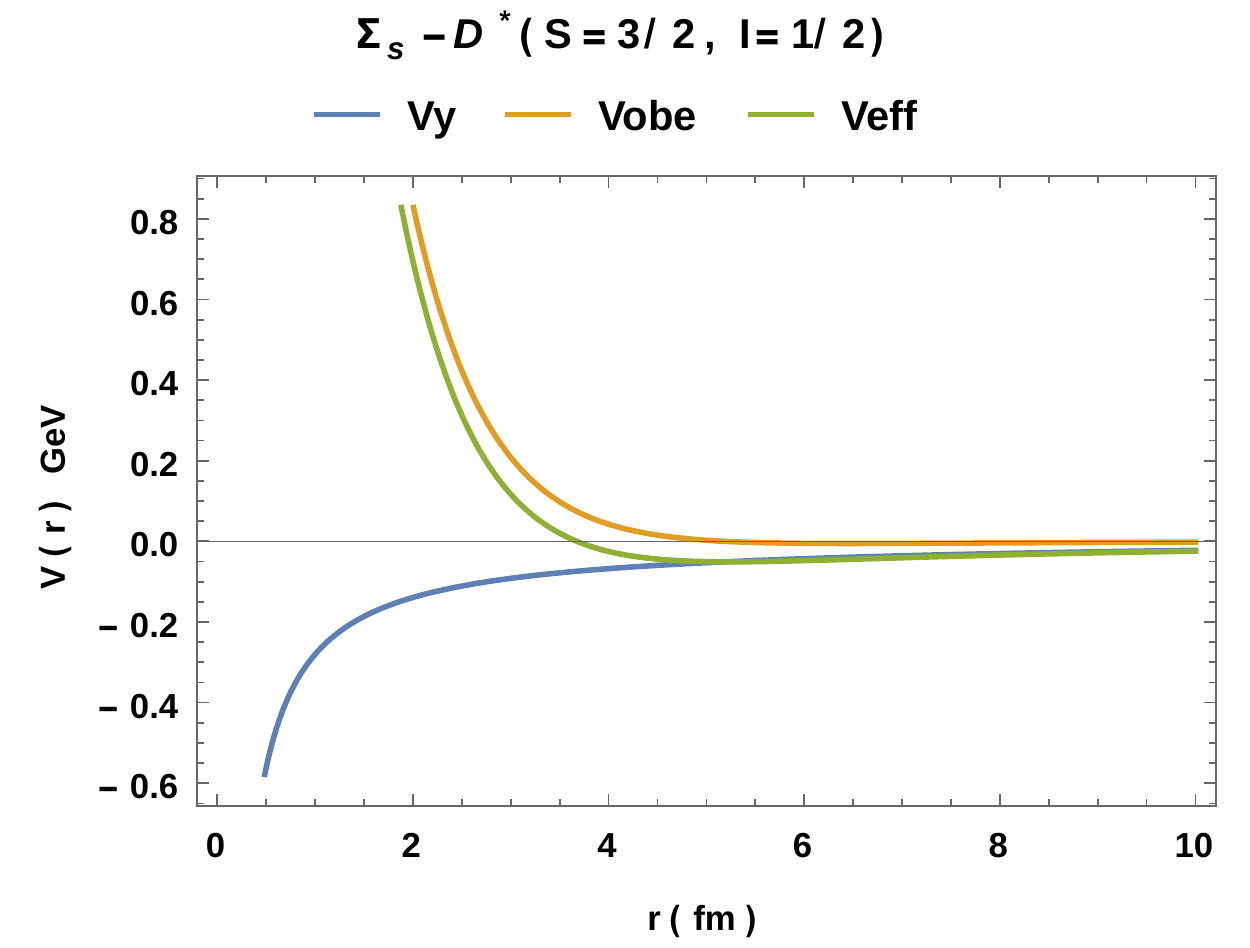}
\includegraphics[scale=0.67]{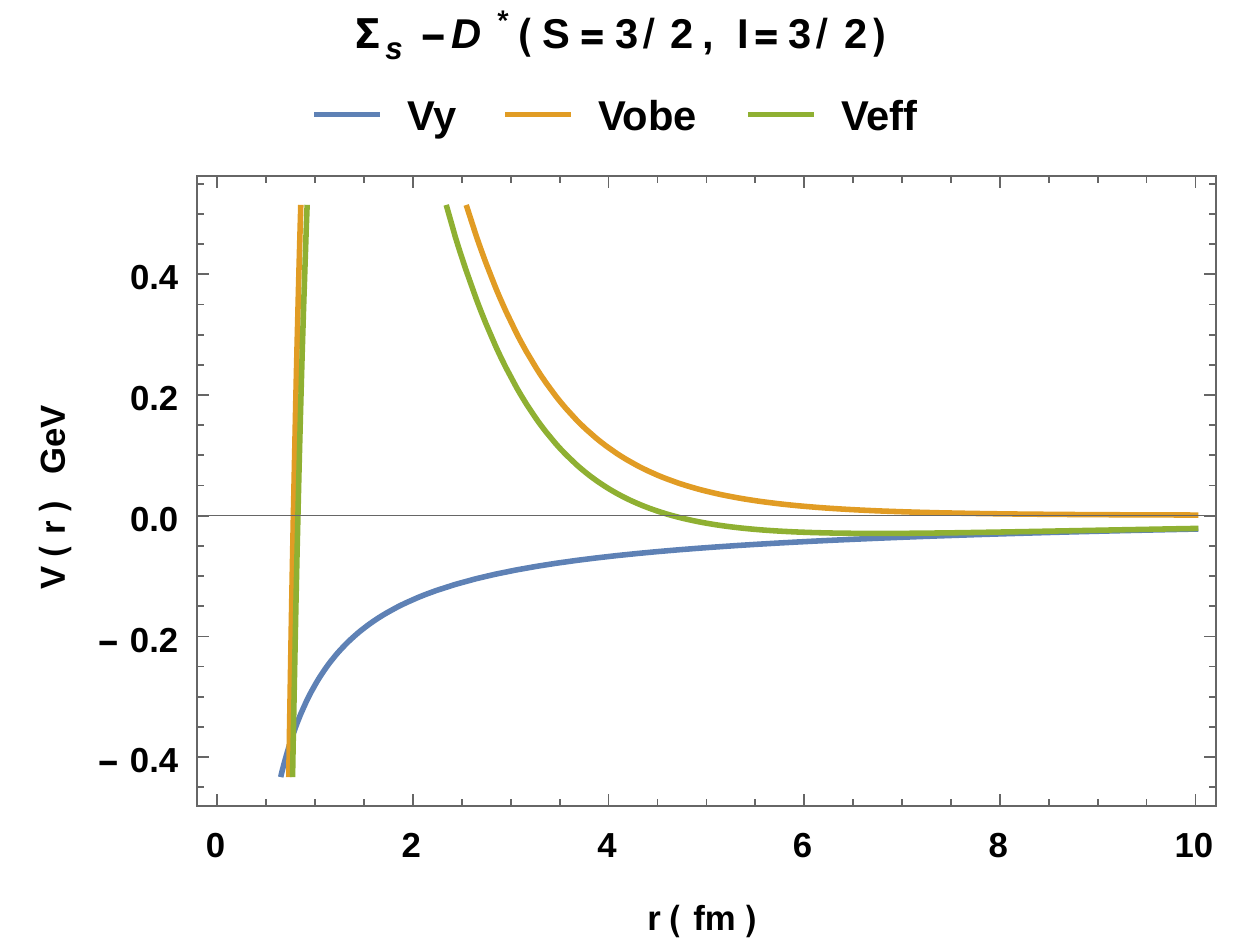}
\includegraphics[scale=0.67]{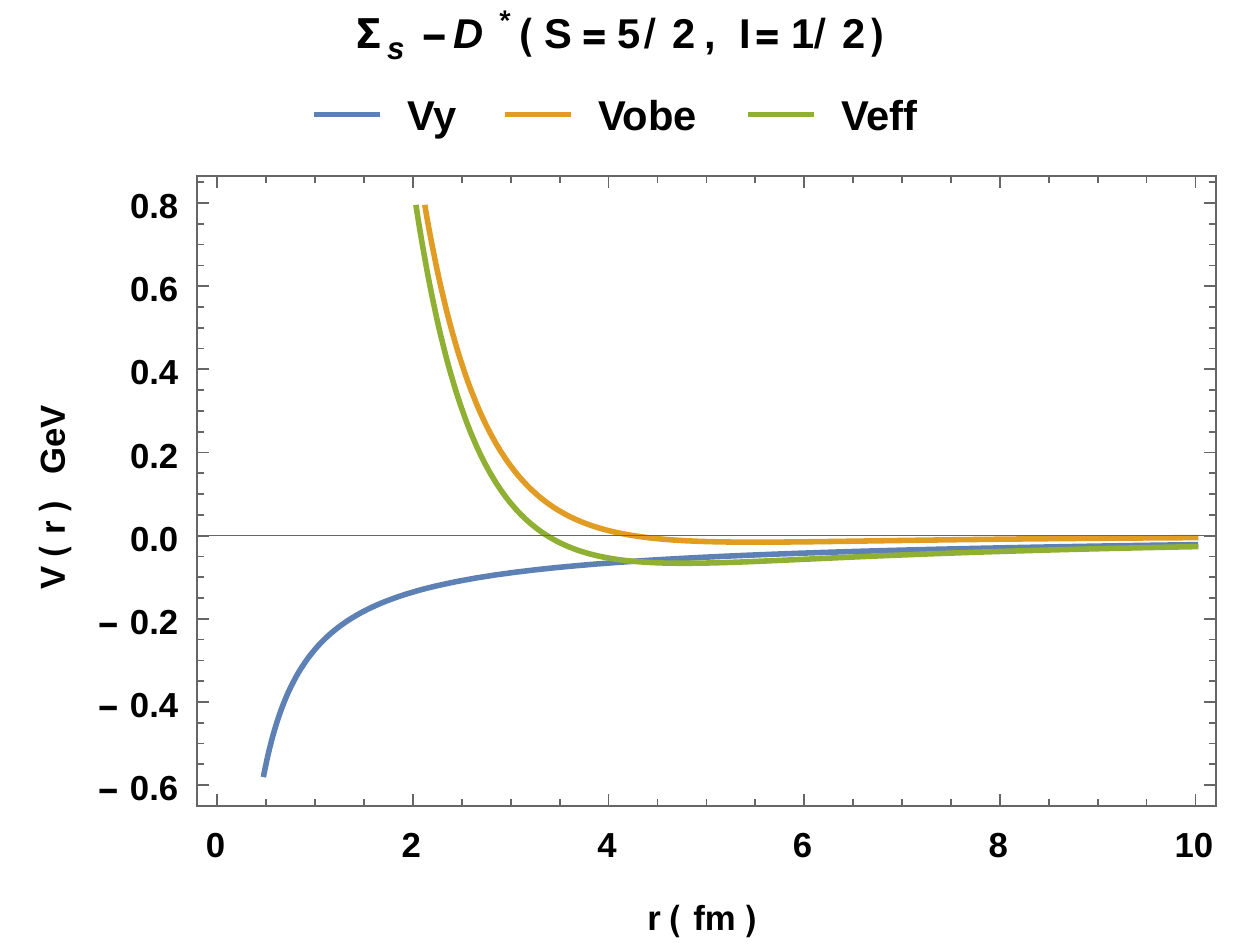}
\includegraphics[scale=0.67]{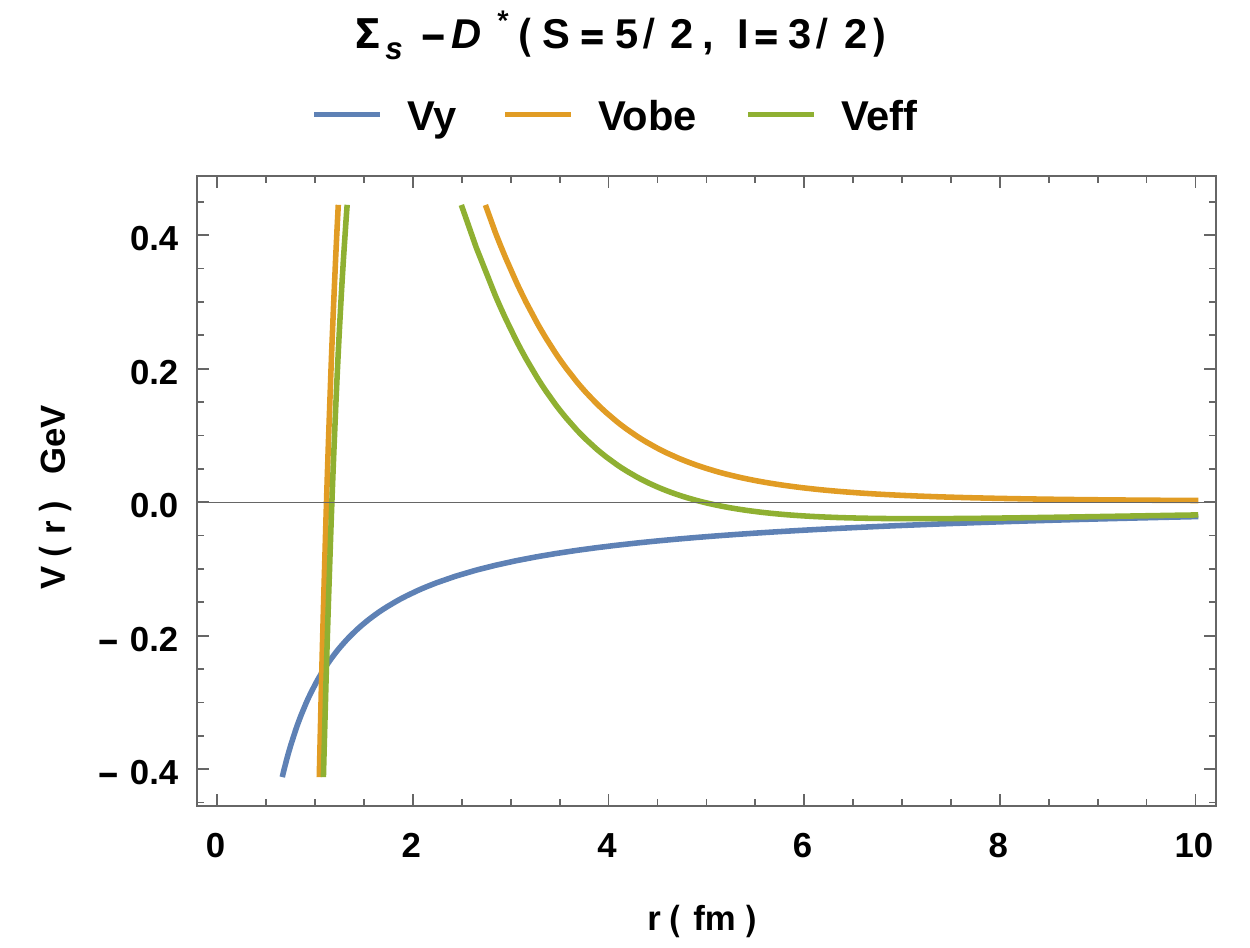}
\end{figure*}

\begin{figure*}[b]
\addtocounter{figure}{-1}
\caption{to be continued..}
\includegraphics[scale=0.67]{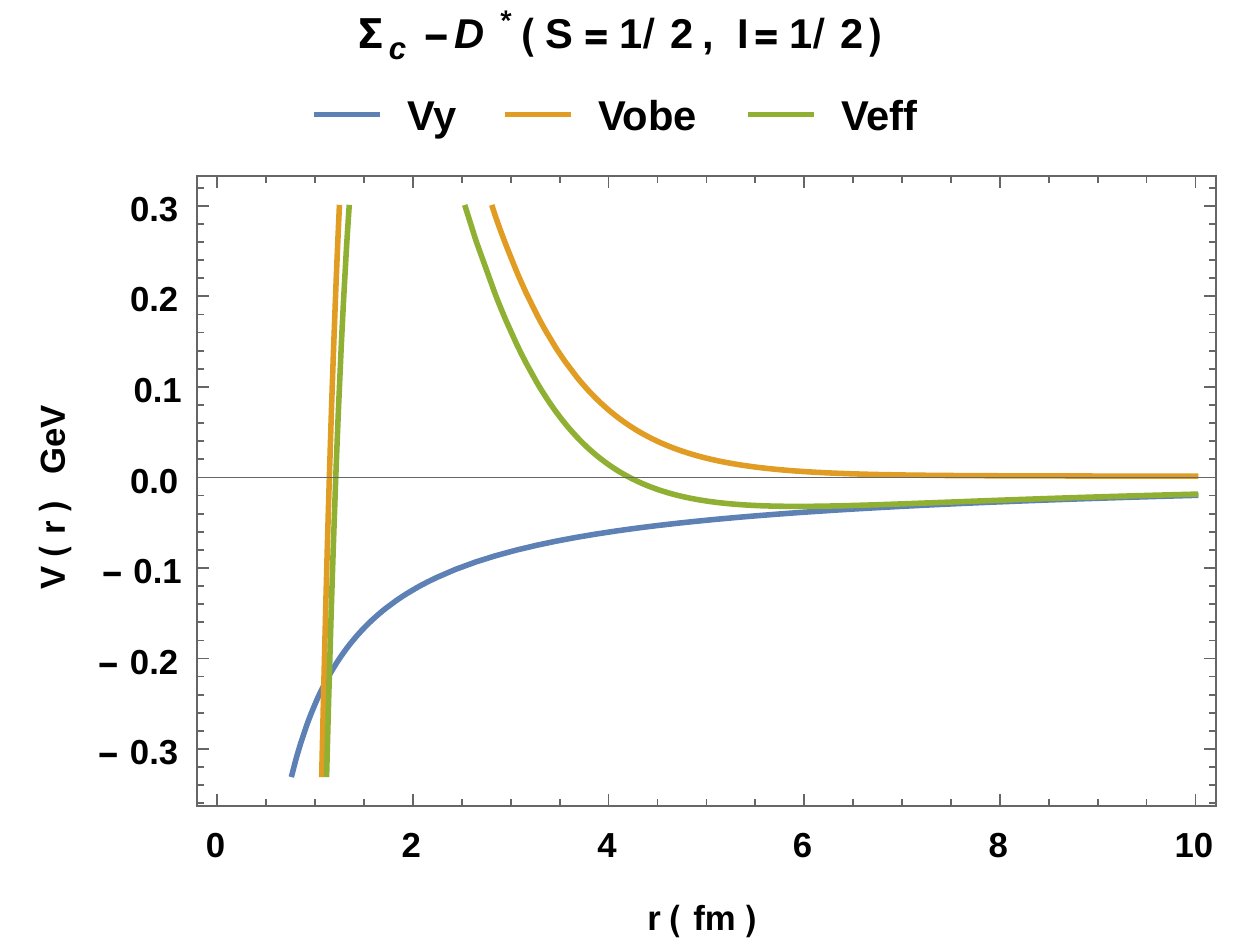}
\includegraphics[scale=0.67]{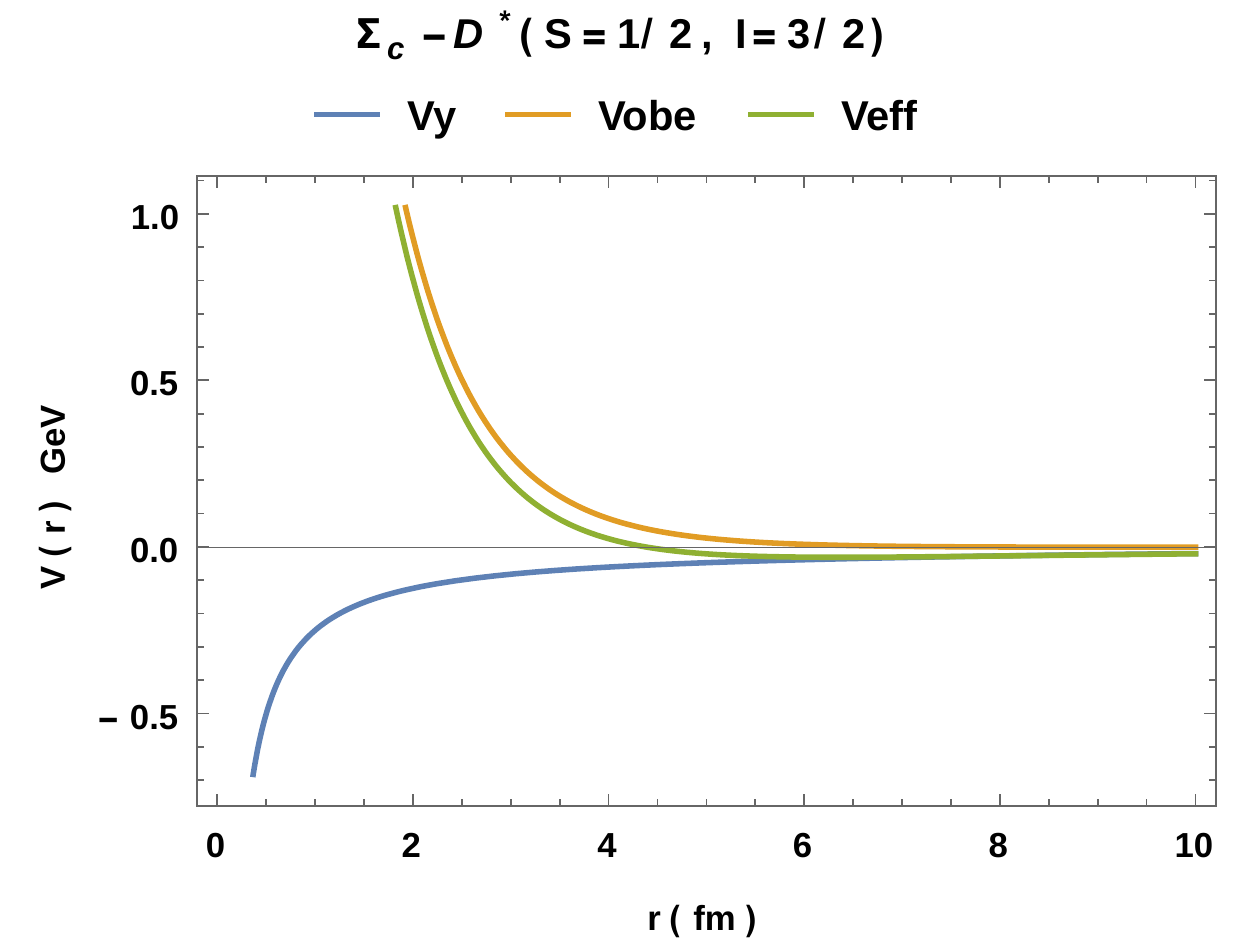}
\includegraphics[scale=0.67]{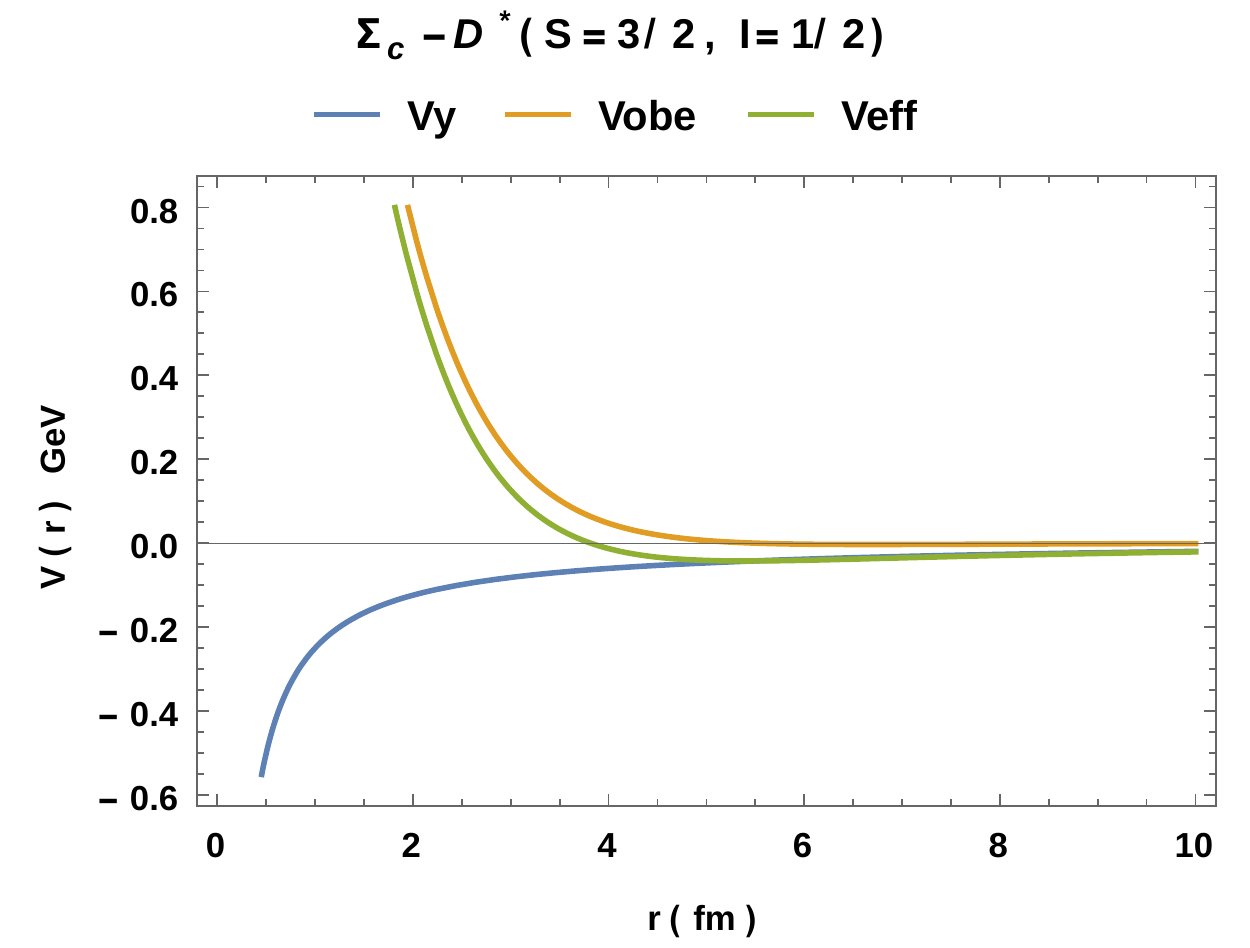}
\includegraphics[scale=0.67]{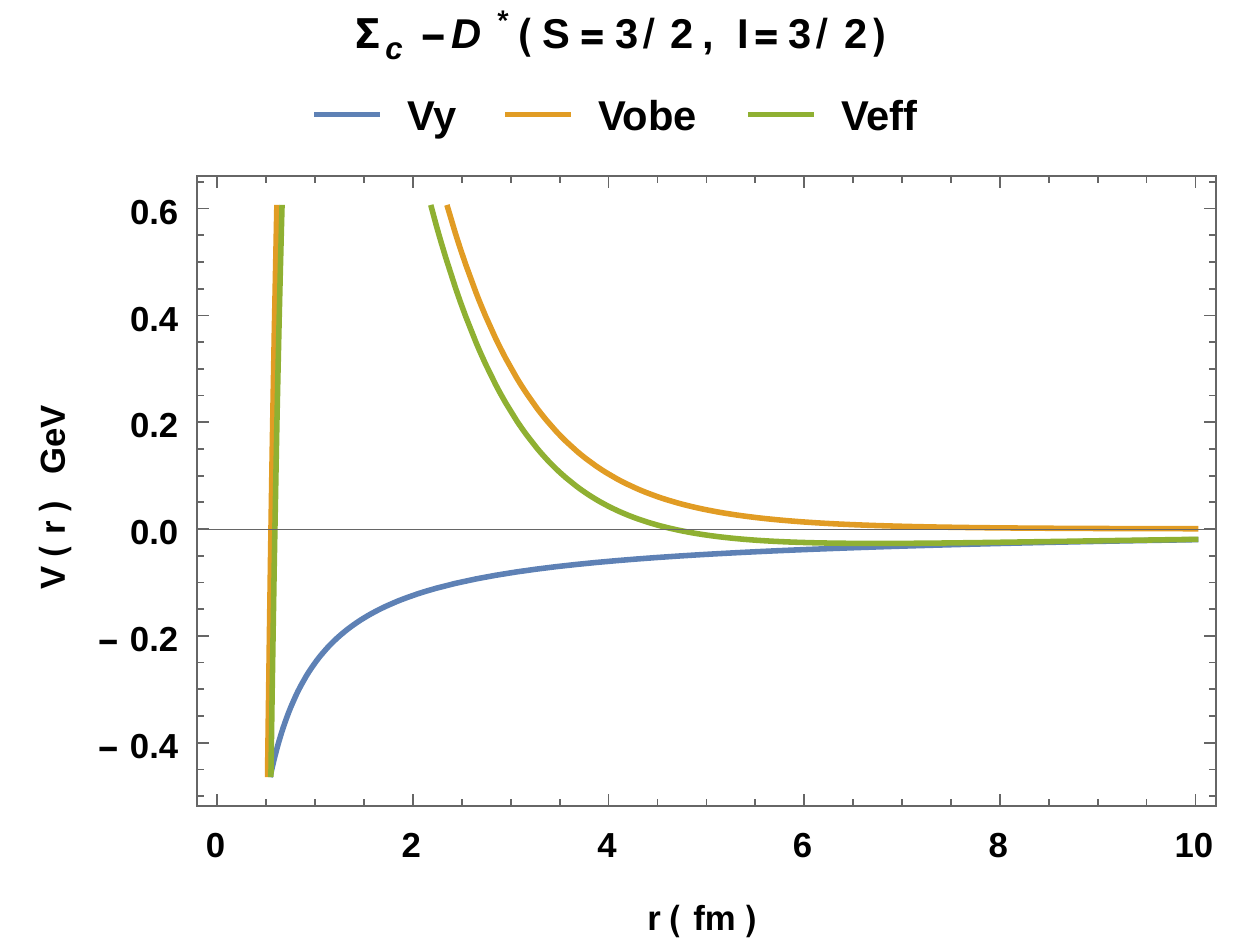}
\includegraphics[scale=0.67]{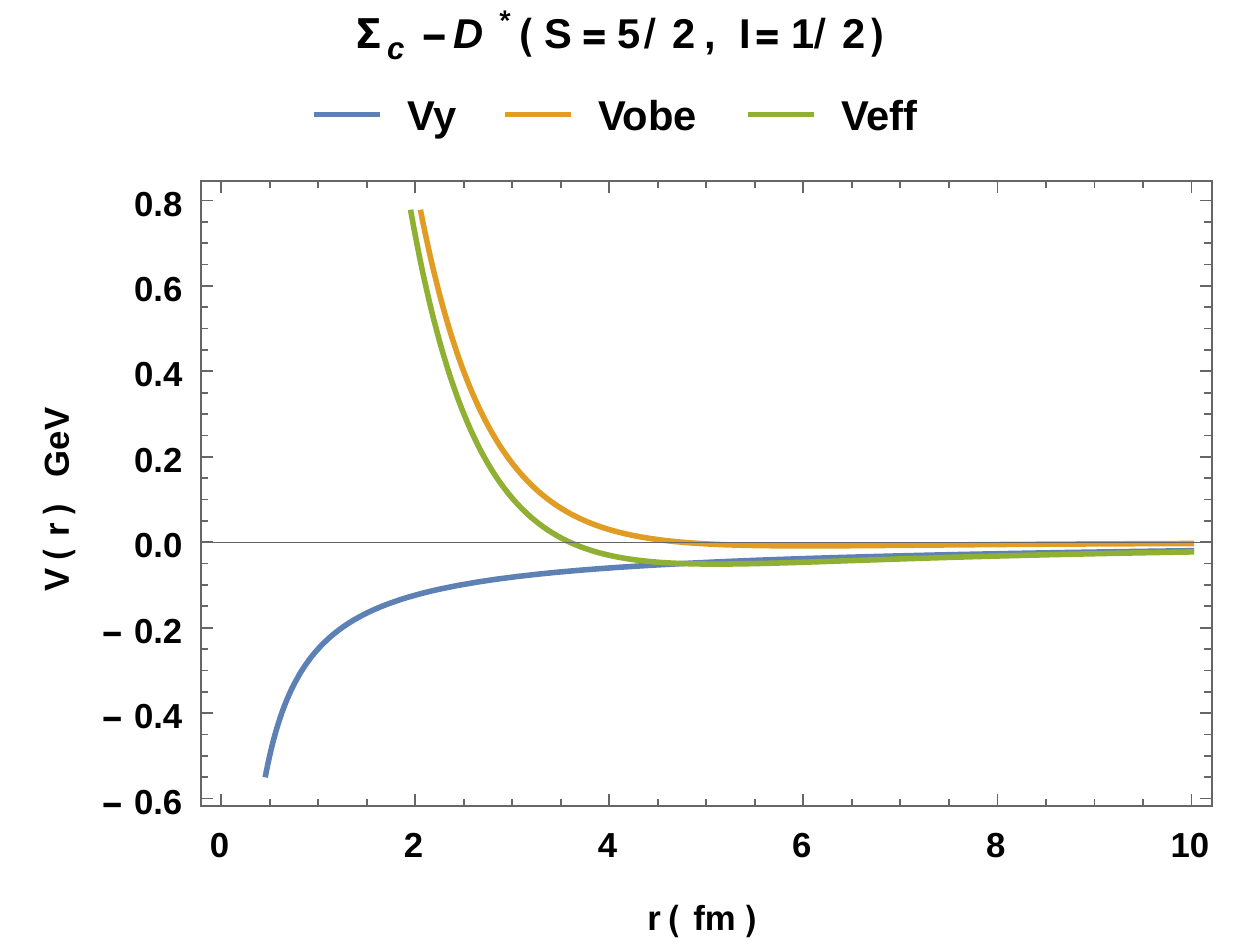}
\includegraphics[scale=0.67]{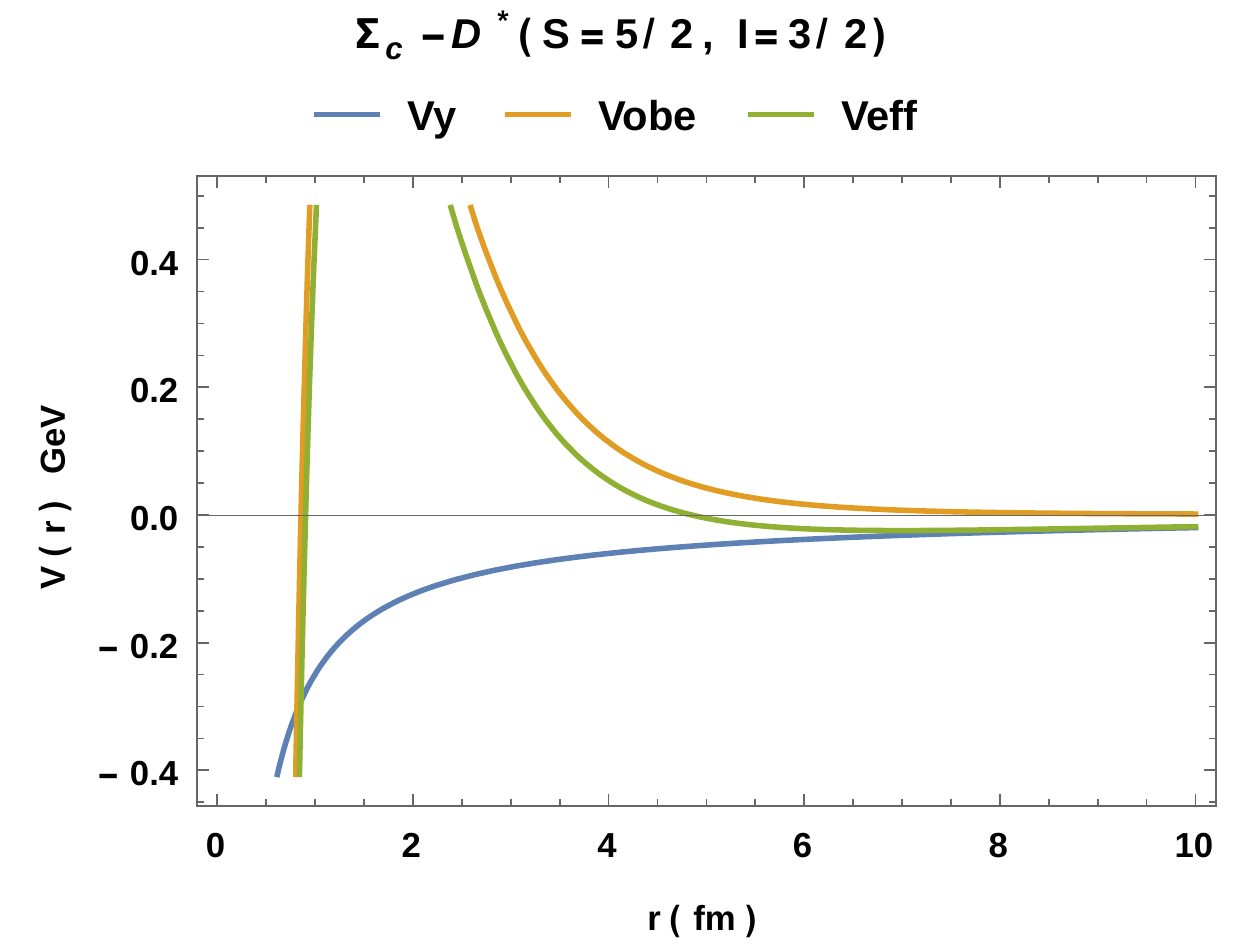}
\end{figure*}

\begin{figure*}[b]
\addtocounter{figure}{-1}
\caption{to be continued..}
\includegraphics[scale=0.67]{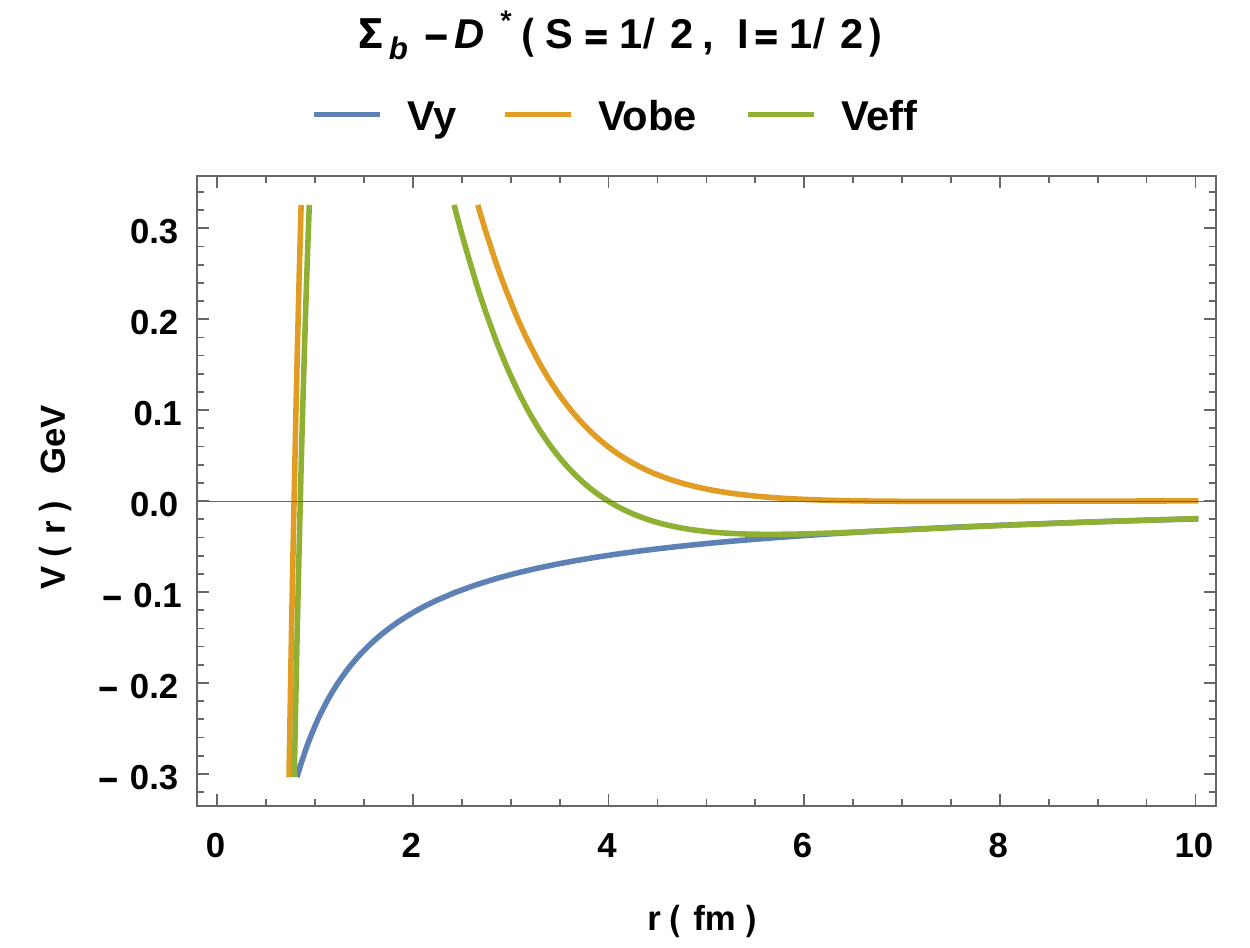}
\includegraphics[scale=0.67]{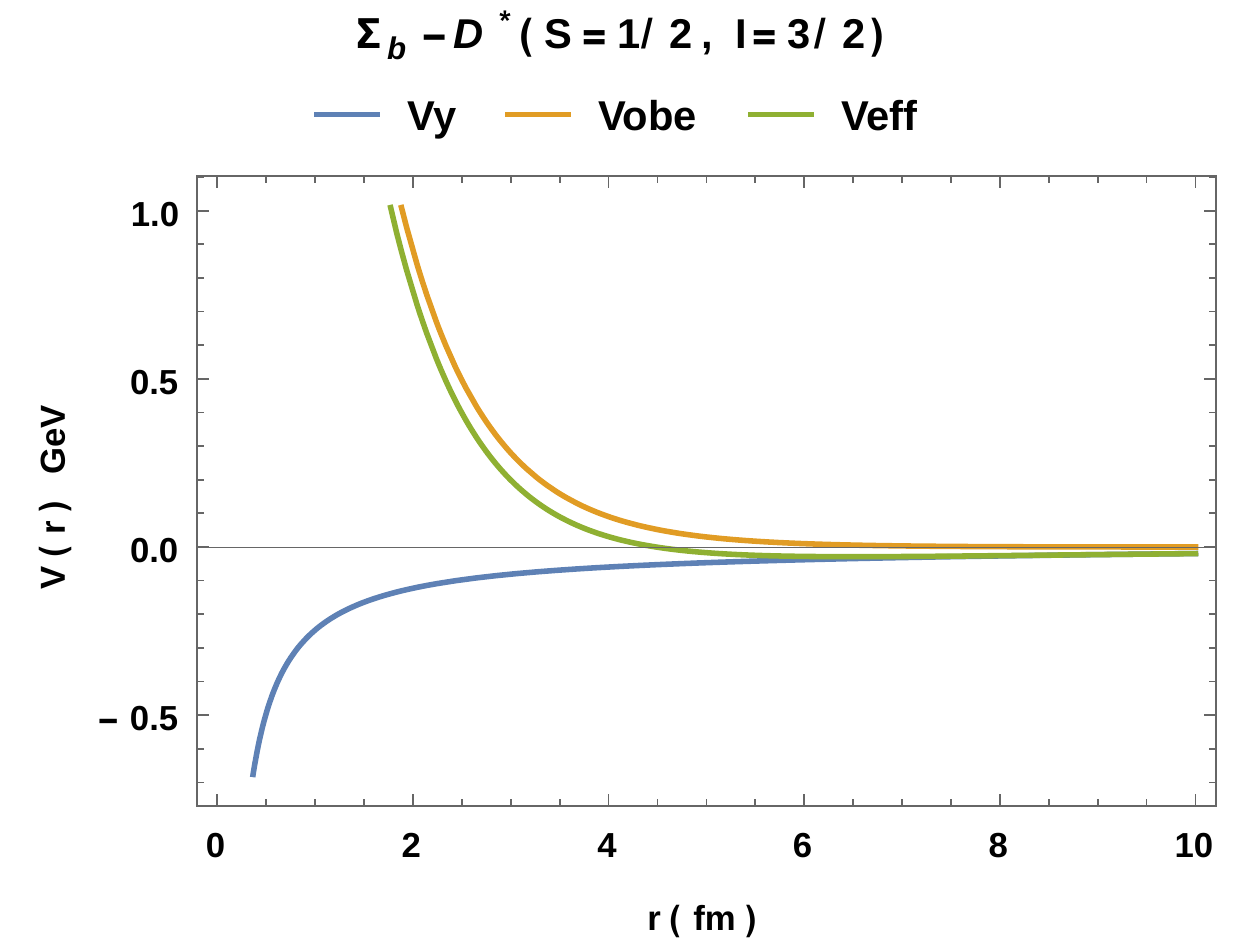}
\includegraphics[scale=0.67]{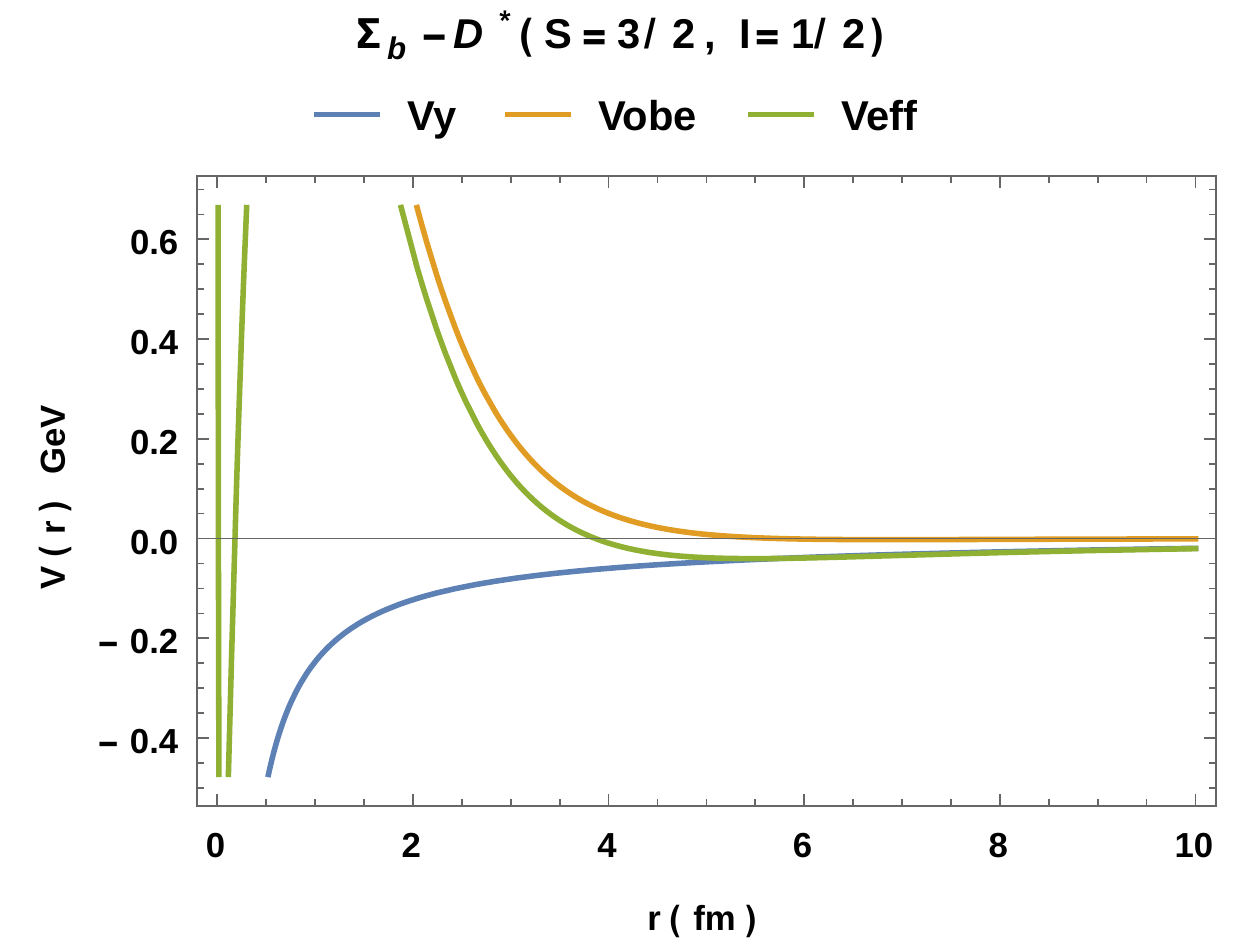}
\includegraphics[scale=0.67]{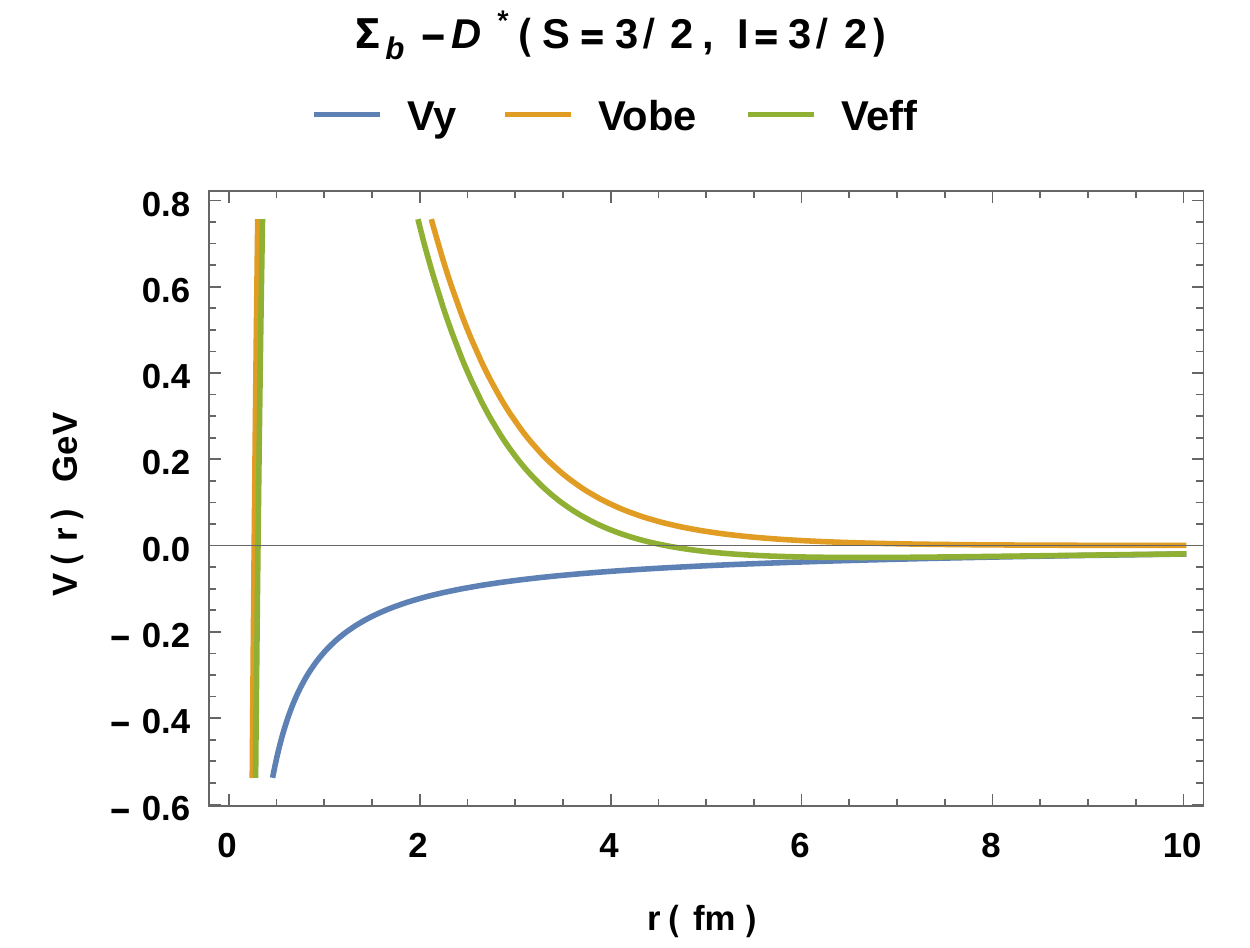}
\includegraphics[scale=0.67]{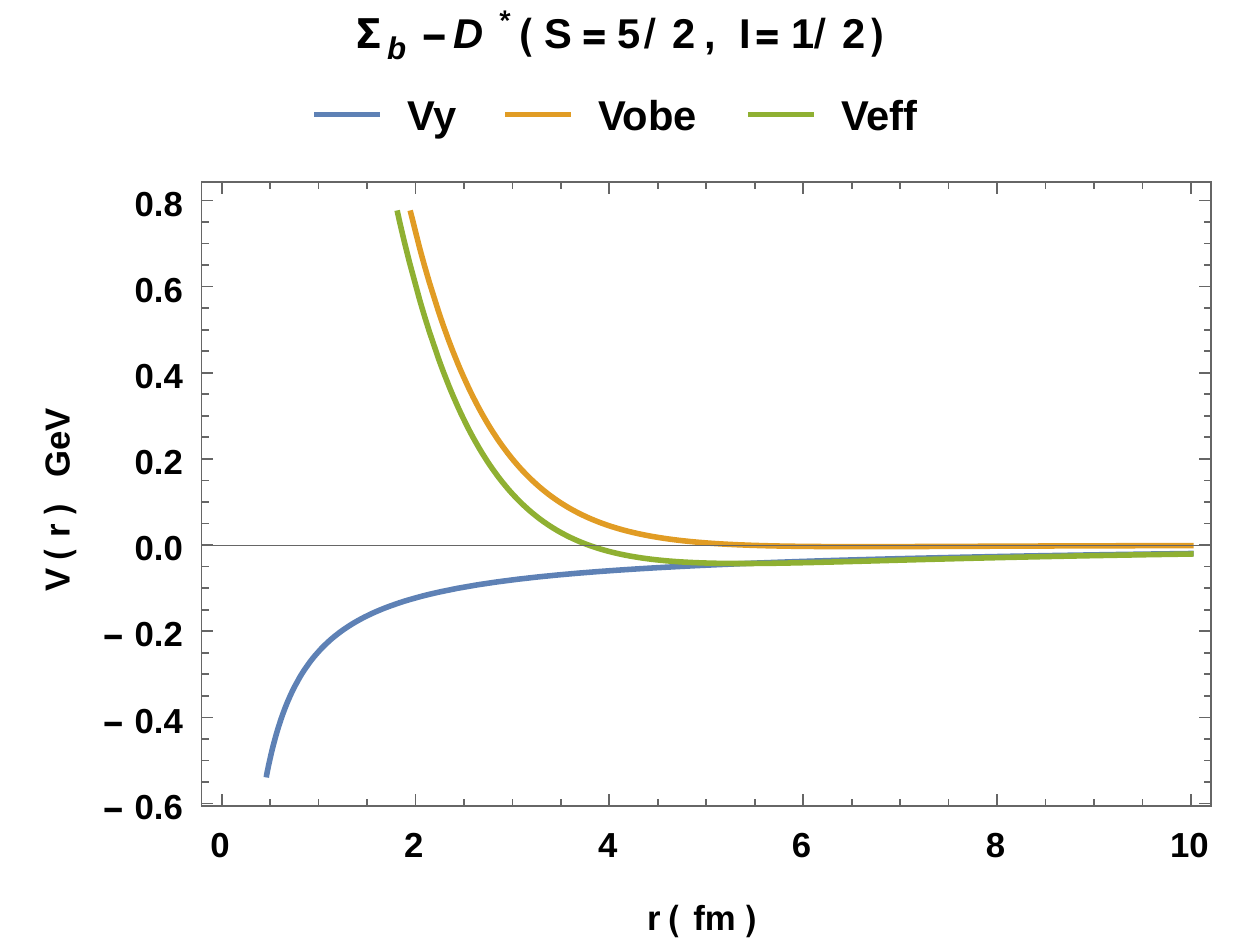}
\includegraphics[scale=0.67]{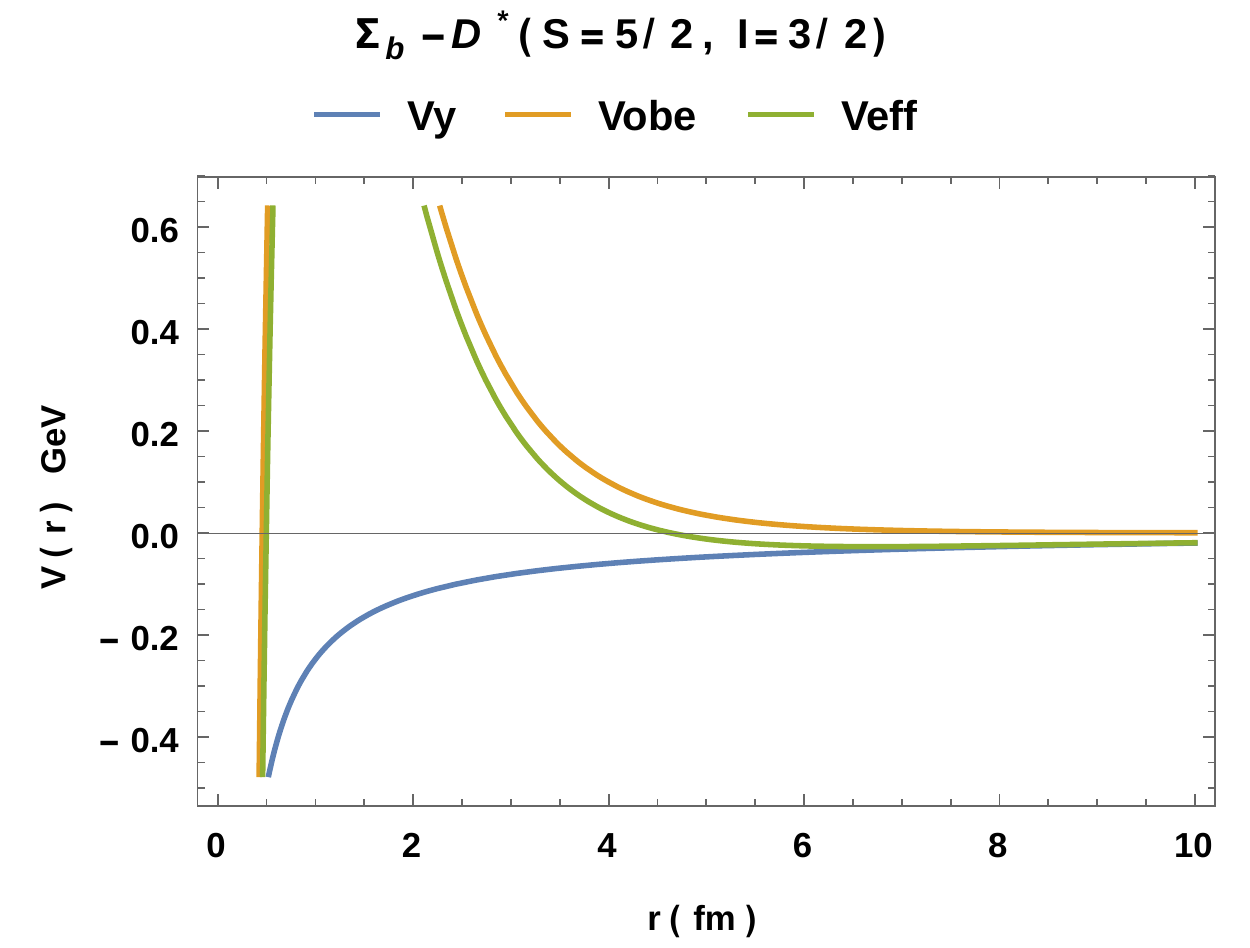}
\end{figure*}

%%%%%%%%%%%%%%%%%%%%%%%%%%% Figure %%%%%%%%%%%%%%%%%%%%%%%%%%%%%%%%%

%\bibliographystyle{revtex4-1}
\bibliographystyle{epj}
\bibliography{chapter3}% Produces the bibliography via BibTeX.

\begin{thebibliography}{51}

\bibitem{arxiv-Rathaud-penta}
D.P. Rathaud, A.K. Rai (2017), \texttt{1706.09323}

\bibitem{Aaij-prl2015-115-pentaquark}
R.~Aaij et~al. (LHCb Collaboration), Phys. Rev. Lett. \textbf{115}, 072001
  (2015)

\bibitem{Gell-Mann-physLett1964-8}
M.~Gell-Mann, Physics Letters \textbf{8}, 214  (1964)

\bibitem{Jaffe-prd1977-15}
R.J. Jaffe, Phys. Rev. D \textbf{15}, 267 (1977)

\bibitem{Strottman-prd1979-20}
D.~Strottman, Phys. Rev. D \textbf{20}, 748 (1979)

\bibitem{Lipkin-plb1987-195}
H.J. Lipkin, Physics Letters B \textbf{195}, 484  (1987)

\bibitem{Nakano-prl2003-91}
T.~Nakano et~al., Phys. Rev. Lett. \textbf{91}, 012002 (2003)

\bibitem{Barmin-phyAtnucl2003-66}
V.V. Barmin et~al., Physics of Atomic Nuclei \textbf{66}, 1715 (2003)

\bibitem{Stepanyan-prl2003-91}
S.~Stepanyan et~al. (CLAS Collaboration), Phys. Rev. Lett. \textbf{91}, 252001
  (2003)

\bibitem{Aktas-plb2004-588}
A.~Aktas et~al., Physics Letters B \textbf{588}, 17  (2004)

\bibitem{Abt-prl2004-93}
I.~Abt et~al. (HERA-B Collaboration), Phys. Rev. Lett. \textbf{93}, 212003
  (2004)

\bibitem{Knöpfle-jpg2004-30}
K.T. Knöpfle et~al., Journal of Physics G: Nuclear and Particle Physics
  \textbf{30}, S1363 (2004)

\bibitem{Bai-prd2004-70}
J.Z. Bai et~al. (BES Collaboration), Phys. Rev. D \textbf{70}, 012004 (2004)

\bibitem{Aubert-prd2007-76}
B.~Aubert et~al. (BABAR Collaboration), Phys. Rev. D \textbf{76}, 092004 (2007)

\bibitem{Link-plb2005-622}
J.~Link et~al., Physics Letters B \textbf{622}, 229  (2005)

\bibitem{Moritsu-prc2014-90}
M.~Moritsu et~al. (J-PARC E19 Collaboration), Phys. Rev. C \textbf{90}, 035205
  (2014)

\bibitem{Abelev-epjc2015-75}
B.~Abelev et~al. (ALICE Collaboration), The European Physical Journal C
  \textbf{75}, 1 (2015)

\bibitem{Li-jhep2015-2015}
G.N. Li, X.G. He, M.~He, Journal of High Energy Physics \textbf{2015}, 1 (2015)

\bibitem{Lebed-plb2015-749}
R.F. Lebed, Physics Letters B \textbf{749}, 454  (2015)

\bibitem{Maiani-plb2015-749}
L.~Maiani, A.~Polosa, V.~Riquer, Physics Letters B \textbf{749}, 289  (2015)

\bibitem{Wang-epjc2016-76}
Z.G. Wang, The European Physical Journal C \textbf{76}, 70 (2016)

\bibitem{Rui-Chen-npa2016-954-penta}
R.~Chen, X.~Liu, S.L. Zhu, Nuclear Physics A \textbf{954}, 406  (2016), recent
  Progress in Strangeness and Charm Hadronic and Nuclear Physics

\bibitem{Rui-Chen-cpc2017-41-penta}
R.~Chen, J.~He, X.~Liu, Chinese Physics C \textbf{41}, 103105 (2017)

\bibitem{Azizi-prd2017-96}
K.~Azizi, Y.~Sarac, H.~Sundu, Phys. Rev. D \textbf{96}, 094030 (2017)

\bibitem{Azizi-prd2017-95}
K.~Azizi, Y.~Sarac, H.~Sundu, Phys. Rev. D \textbf{95}, 094016 (2017)

\bibitem{Shen-arxiv2017}
C.W. Shen, Y.H. Lin (2017), \texttt{1710.09037}

\bibitem{Scoccola-prd2015-92}
N.N. Scoccola, D.O. Riska, M.~Rho, Phys. Rev. D \textbf{92}, 051501 (2015)

\bibitem{Guo-prd2015-92}
F.K. Guo et~al., Phys. Rev. D \textbf{92}, 071502 (2015)

\bibitem{Guo-epjc2016-52}
F.K. Guo et~al., The European Physical Journal A \textbf{52}, 318 (2016)

\bibitem{Meißner-plb2015-751}
U.G. Meißner, J.~Oller, Physics Letters B \textbf{751}, 59  (2015)

\bibitem{Xiao-Hai-plb2016-757}
X.H. Liu, Q.~Wang, Q.~Zhao, Physics Letters B \textbf{757}, 231  (2016)

\bibitem{Eides-epjc2018-78}
M.I. Eides, V.Y. Petrov, M.V. Polyakov, The European Physical Journal C
  \textbf{78}, 36 (2018)

\bibitem{Chen-physreport2016-639}
H.X. Chen et~al., Physics Reports \textbf{639}, 1  (2016), the hidden-charm
  pentaquark and tetraquark states

\bibitem{Swanson-PhysRep2006-429}
E.S. Swanson, Physics Reports \textbf{429}, 243  (2006)

\bibitem{Esposito-phys.Rep2017-668}
A.~Esposito, A.~Pilloni, A.~Polosa, Physics Reports \textbf{668}, 1  (2017)

\bibitem{Ali-Prog.Part.Phys2017-97}
A.~Ali, J.S. Lange, S.~Stone, Progress in Particle and Nuclear Physics
  \textbf{97}, 123  (2017)

\bibitem{Guo-RevModPhys2018-90}
F.K. Guo et~al., Rev. Mod. Phys. \textbf{90}, 015004 (2018)

\bibitem{Machleidt-PhysRep1987-149}
R.~Machleidt, K.~Holinde, C.~Elster, Physics Reports \textbf{149}, 1  (1987)

\bibitem{Lee-prd2011-84}
N.~Lee et~al., Phys. Rev. D \textbf{84}, 014031 (2011)

\bibitem{Ding-prd2009}
G.J. Ding, J.F. Liu, M.L. Yan, Phys. Rev. D \textbf{79}, 054005 (2009)

\bibitem{Rai-epjc2015}
A.K. Rai, D.P. Rathaud, Eur. Phys. J. C \textbf{75}, 462 (2015)

\bibitem{Rathaud-epjp2017-132}
D.P. Rathaud, A.K. Rai, The European Physical Journal Plus \textbf{132}, 370
  (2017)

\bibitem{Rathaud-IJP2016-90}
D.P. Rathaud, A.K. Rai, Indian J. Phys. \textbf{90}, 1299 (2016)

\bibitem{Ebert-prd2009}
D.~Ebert, R.N. Faustov, V.O. Galkin, Phys. Rev. D \textbf{79}, 114029 (2009)

\bibitem{Badalian-prd2004}
A.M. Badalian, A.I. Veselov, B.L.G. Bakker, Phys. Rev. D \textbf{70}, 016007
  (2004)

\bibitem{Machleidt-prc2001-63}
R.~Machleidt, Phys. Rev. C \textbf{63}, 024001 (2001)

\bibitem{Patrignani-PDG2016}
C.~Patrignani et~al. (Particle Data Group), Chin. Phys. \textbf{C40}, 100001
  (2016)

\bibitem{Paris-prc2000-62}
M.W. Paris, V.R. Pandharipande, Phys. Rev. C \textbf{62}, 015201 (2000)

\bibitem{Naghdi-PhysParNucl2014-45}
M.~Naghdi, Physics of Particles and Nuclei \textbf{45}, 924 (2014)

\bibitem{Wu-prc2011-84}
J.J. Wu et~al., Phys. Rev. C \textbf{84}, 015202 (2011)

\bibitem{Weinberg-RhysRev1965}
S.~Weinberg, Phys. Rev. \textbf{137}, B672 (1965)

\end{thebibliography}

\end{document}